\documentclass[fleqn,usenatbib]{mnras}

\usepackage{newtxtext,newtxmath}

\usepackage[T1]{fontenc}

\usepackage{amsmath,siunitx}
\DeclareRobustCommand{\VAN}[3]{#2}
\let\VANthebibliography\thebibliography
\def\thebibliography{\DeclareRobustCommand{\VAN}[3]{##3}\VANthebibliography}

\usepackage{graphicx}	
\usepackage{amsmath}	
\usepackage{longtable}

\usepackage[normalem]{ulem}

\title[Ram-pressure stripping candidates]{Diagnostic diagrams for ram-pressure stripped candidates}

\author[A. C. Krabbe et al. et al.]{
A. C. Krabbe,$^{1,2}$\thanks{E-mail:angela.krabbe@usp.br}
J. A. Hernandez-Jimenez,$^{2}$, C. Mendes de Oliveira,$^{1}$ Y.~L. Jaffe,$^{3,4}$ C.~B. Oliveira Jr.,$^{2}$
\newauthor{N. M. Cardoso,$^{1}$ A. V. Smith Castelli,$^{5,6}$ 
O. L. Dors,$^{2}$ A. Cortesi,$^{5,6}$ J. P. Crossett$^{3,4}$}
\\
$^{1}$Departamento de Astronomia, Instituto de Astronomia, Geofísica e Ciências Atmosféricas da USP, Cidade Universitária, 05508-900 São Paulo, SP, Brazil \\
$^{2}$Universidade do Vale do Para\'{\i}ba, Av. Shishima Hifumi, 2911, Zip Code 12244-000, S\~ao Jos\'e dos Campos, SP, Brazil\\
$^{3}$ Departamento de Física, Universidad Técnica Federico Santa María, Avenida España 1680, Valparaíso, Chile \\
$^{4}$ Instituto de F\'isica y Astronom\'ia, Universidad de Valpara\'iso, Avda. Gran Breta\~na 1111, Valpara\'iso, Chile
$^{5}$Instituto de Astrofísica de La Plata, CONICET–UNLP, Paseo del Bosque s/n, B1900FWA, Argentina \\
$^{6}$Facultad de Ciencias Astronómicas y Geofísicas, Universidad Nacional de La Plata (UNLP), Paseo del Bosque s/n, B1900FWA, Argentina \\
$^{7}$Observatório do Valongo, Universidade Federal do Rio de Janeiro, Ladeira Pedro Antônio 43, Rio de Janeiro, RJ, 20080-090, Brazil}
\date{Accepted XXX. Received YYY; in original form ZZZ}

\pubyear{2023}

\begin{document}
\label{firstpage}

\pagerange{\pageref{firstpage}--\pageref{lastpage}}
\maketitle

\begin{abstract}

This paper presents a method for finding ram-pressure stripped (RPS) galaxy candidates by performing  a morphological analysis of galaxy images obtained from the Legacy survey.  We consider a sample of about 600 galaxies located in different environments such as groups and clusters, tidally interacting pairs and the field. The sample includes 160 RPS 
previously classified in the literature into classes from  J1 to J5, based on the increasing level of disturbances. Our  morphological analysis was done using the {\sc astromorphlib} software followed by the inspection of diagnostic diagrams involving combinations 
of different parameters like  the asymmetry ($A$), concentration ($C$), Sérsic index ($n$), and bulge strength parameters $F(G,\,M_{20})$.  
 We found that some of those diagrams display a distinct region in which galaxies classified as J3, J4 and J5 decouples from isolated galaxies. We call this region as the morphological transition zone and we also found that tidally interacting galaxies in pairs are predominant within this zone.
Nevertheless,   after visually inspecting the objects in the  morphological transition zone to discard obvious contaminants, we ended up with 33 bonafide new RPS candidates in the studied nearby groups and clusters (Hydra, Fornax, and CLoGS sample), of which one-third show clear evidence of unwinding arms. Future works may potentially further increase significantly the samples of known RPS using such method.

\end{abstract}

\begin{keywords}
galaxies: clusters: general  -- galaxies: clusters: intracluster medium -- galaxies: evolution -- galaxies: interactions -- galaxies:irregular -- galaxies: structure 
\end{keywords}



\section{Introduction}
 In the hierarchical scenario of galaxy formation, galaxy interactions and mergers play a fundamental role, and they are the key to the growth of the structures observed in the Universe (e.g. \citealt{2006Natur.440.1137S}). 
 In fact,  the environment of a galaxy has a profound impact on its structural properties and star formation history. For instance, in high-density environments such as  cluster cores, the  quiescent early-type galaxies dominate, while the star-forming spirals are more common in the lower-density environments of groups (\citealt{1980ApJ...236..351D,2002MNRAS.334..673L}).  In clusters of galaxies, one of the most effective mechanisms in transforming galaxies is the so-called ram-pressure stripping (RPS, eg. \citealt{1972ApJ...176....1G,2015MNRAS.448.1715J, 2016AJ....151...78P}), which involves the interaction of galaxies with  the intra-cluster medium (ICM), as the galaxies fall into the cluster. The ram-pressure stripping can quenche star formation from the outside-in,  because the lost of gas outside the stripping radius, producing truncated H$\alpha$ profiles \citep{2016AJ....151...78P},  in turn this could potentially trigger a state of gas ``starvation'' leading a global decline in star formation,  since the supply of cold gas to these galaxies is  halted and they  continue to form stars until they run out of the fuel residing in their disc \citep{peng2015}.


Striking examples of galaxies undergoing strong
RPS are the so-called  jellyfish galaxies (\citealt{2009AJ....138.1741C, 2009MNRAS.399.2221B,2014ApJ...781L..40E, 2016AJ....151...78P,jaffe18}). These objects display long tails of ionized gas  that can extend up to 100 kpc in length \citep{2010MNRAS.408.1417S,2014ApJ...781L..40E, Poggianti_2017,bellhouse2017}. Evidence of galaxies undergoing RPS came mainly from neutral gas (HI) observations (e.g. \citealt{1984ARA&A..22..445H,1990AJ....100..604C,2004AJ....127.3361K,2001A&A...377..812V,2001A&A...369..432V,2009AJ....138.1741C,2012ApJ...756L..28J}).
However, candidates for galaxies undergoing RPS can be also identified using broad and narrow optical band  imaging.
For example, \citet{2016AJ....151...78P},  hereafter P16, searched for  this class of objects in  low and intermediate redshift  environments  inspecting visually  B and V-band images and identified a sample of 344 candidates, which was used  afterward as base of 
the European Southern Observatory large program GASP\footnote{\url{https://web.oapd.inaf.it/gasp/}} \citep[GAs Stripping Phenomena in galaxies with MUSE,][]{Poggianti_2017}.
P16 defined the  JClass classification, which is assigned  according to a visual inspection of the optical images based on 
evidence for stripping signatures. The JClass system includes five classes, ranging 
 from the most extreme cases (JClass 5) to progressively weaker cases, down to the weakest (JClass 1).  P16  suggest that the 
 JClass 5 and 4 are the most secure candidates and include the most striking classical jellyfish galaxies. Notably, in the subsequent GASP survey, it was found that approximately 85\% of the RPS candidates identified by P16 in clusters were indeed confirmed to be undergoing RPS \citep{vulcani22}.
 However, it is worth noting that this original classification method, which relies on the visual inspection of images, clearly presents some shortcomings:
 the classification is subjective and depends on the experience and interpretation of the classifier. Also, the JClass system only considers the optical properties of the galaxy, neglecting   potential stripping signatures in other wavelengths.

\citet{2016MNRAS.455.2994M} investigated the observational signatures and physical origin of RPS in 63 massive galaxy clusters at $z = 0.3 - 0.7$, based on images obtained with the Hubble Space Telescope (HST). 
 In addition to the visual screening process, their selection
of RPS candidates  involved the application of morphological criteria in diagrams of non-parametric parameters
 such as Concentration ($C$) versus Asymmetry ($A$),  Gini versus $M_{20}$ coefficient, and  $Sk_{0-1}$ versus $Sk_{1-2}$, which are  skeletal decomposition parameters.   By incorporating morphological criteria, the selection process becomes adept at identifying not only the visually apparent RPS candidates but also uncovering and classifying the less obvious cases of RPS. 
\citet{2020MNRAS.495..554R} used multi-band images from the Canada-France-Hawaii Telescope (CFHT) to identify a sample of galaxies in the Coma cluster which appear  to be experiencing stripping. These authors found  the RPS candidates as clear outliers in  quantitative morphological diagrams, such as C versus A and Gini versus $M_{20}$ coefficient, confirming the disturbed nature of these galaxies and the potential of  these combinations of photometric parameters as a tool to select RPS candidates.
 Recently, \citealt{bellhouse22} conducted a study in which they explored the morphological parameters of stripped and poststarburst galaxies. They investigated various parameters, including the C versus A ratio, Gini coefficient versus $M_{20}$ coefficient, and the outer centroid variance versus full centroid variance. Their findings revealed that the centroid variance was the most effective standalone measure of galaxy morphology, which can be a potential indicator for detecting disturbed morphologies.

Despite the diligent use of quantitative morphological parameters to objectively select RPS candidates in the literature, further refinement of the morphological criteria remains essential to enhance the search for new RPS candidates. Conducting a robust study will help optimize the selection process. This paper presents a comprehensive study on the morphological parameters of RPS candidates extracted from the P16 sample,  which is one of the most extensive and homogeneous samples at low redshift to date. To improve the selection process, a novel and refined methodology for identifying RPS galaxy candidates was employed, involving a meticulous morphological analysis of galaxy images acquired from the DESI Legacy Imaging Surveys  \citep{dey19}. This study aims to enhance the accuracy and efficiency of RPS candidate identification, contributing to a deeper understanding of these intriguing phenomena and their relevance in different galactic environments.
This manuscript is organized as follows: In Section \ref{data}, we provide a detailed description of the observational data used in our study, as well as the selection criteria applied to construct the samples. The methodology employed for the photometric analysis is presented in Section \ref{meto}. The results obtained from our analysis and their corresponding discussion are in Section \ref{results}. Finally, we summarize the main findings and conclusions in Section \ref{conclusions}.  For the sake of readability throughout this work, we interchangeably use the terms "ram pressure stripping" and "jellyfish" to refer to the same phenomenon. It is worth noting that some authors in the literature prefer to reserve the term "jellyfish" for particularly striking examples of galaxies undergoing ram pressure stripping, characterized by the presence of visible trailing "tentacles". Throughout this paper we adopted a value of Hubble constant, $H_0$, of 70\,km\,s$^{-1}$\,Mpc$^{-1}$.

\section{Data}
\label{data}

We used  archival imaging data in $g$, $r$ and $z$ bands  of  the DESI Legacy Survey\footnote{\url{https://www.legacysurvey.org/}.}, which covers 14\,000 deg² of the extragalactic sky and provides high-quality optical images \citep{dey19}. This  is a combination   of the DECam Legacy Survey (DECaLS) using the Dark Energy Camera on the Blanco 4\,m telescope, the Mayall $z$ band Legacy Survey (MzLS) using the MOSAIC instrument on the Mayall 4\,m telescope, and the Beijing-Arizona Sky Survey (BASS) using the 90Prime instrument at the Steward Observatory Bok telescope. The eighth and tenth  data releases of the Legacy Survey  were used in  this study as well as four distinct data sets  as described below.  

 \begin{enumerate}
     \item  P16 cluster  sample: it is composed of  galaxies that are  likely  experiencing ram pressure stripping. We constructed this  sample of objects from the atlas with stripping candidates of  \cite{2016AJ....151...78P}, which is based  on optical images of the survey of OMEGAWINGS and WINGS \citep{2006A&A...445..805F,2015A&A...581A..41G}.
     We selected only  galaxies  with determined redshift and belonging to galaxy clusters, along with different JClass classifications provided in \cite{2016AJ....151...78P}. This sample consists of 140 objects. 
     The redshift ($z$) of this sample is in the range of 0.017 to 0.099.

    \item Isolated sample: the control sample is composed of isolated galaxies  taken from 
    Data Release 15 \citep[DR15,][]{DR15}  of the Sloan Digital Sky MaNGA survey 
    \citep{bundy15, blanton17}.  The initial pool of galaxies considered in our analysis did not have any physical neighbours within a projected radius of 100\,kpc.  From the accompanying Value-Added Catalogues of DR15 we used MaNGA Morphology Deep Learning catalogue provided by Dom\'inguez-S\'anchez, Bentabol \& Bernardi  Fischer\footnote{\url{https://www.sdss.org/dr18/data_access/value-added-catalogs/?vac_id=73}} \citep{fisher19}.
    This catalogue contains the morphological classification T-Type \citep{ttype}  and the Galaxy Zoo attribute P-merger \citep{lintott08}. The deep learning models have been trained and tested on SDSS-DR7 images with great success \citep{dominguez18}. We have only selected disc galaxies, 2 $<$ T-Type $\leq 6$, with a P-merger $< 0.15$, and a  disc inclination $<70^{\circ}$ (to avoid edge-on galaxies). The final sample of isolated galaxies is composed of  213 objects  with  $z$  in the range of  0.012 to 0.068.

    \item Interacting sample: this sample  comprises close pairs of galaxies taken from the Catalogue of Southern Peculiar Galaxies and Associations\footnote{\url{https://ned.ipac.caltech.edu/level5/SPGA_Atlas/frames.html}}\citep{arp87}. We have selected pairs from the categories:
     (1) Galaxies with interacting companions, (2) 	Interacting doubles (galaxies of comparable size), (8) Galaxies with apparent companions
     and (9) M51-types (companion at end of spiral arm). 
      The selected pairs in this study exhibit various stages of interaction, specifically focusing on the period between the first pre- and post-pericenter passages. We have excluded systems in more advanced stages of interaction, such as those with double nuclei or highly disrupted galaxies. This ensures that our analysis primarily encompasses pairs that are in relatively earlier stages of interaction.
     After applying these selection criteria, the final sample consists of 86 objects,  with  $z$  in the range of 0.004 to 0.100. For practical reasons and ease of implementation of the analysis code, we have chosen to analyze only the main component of each pair. The images are centred on these main components, facilitating a simpler implementation of the analysis process.

     \item  Cluster Members: in pursuit of discovering new candidates of RPS, our research focused on galaxies from three distinct samples, all belonging to groups and clusters, with spectroscopy-confirmed members. One set of samples consisted of groups selected from the Comprehensive Local-Volume Groups Sample (CLoGS) - a project conducted by \citet{2017MNRAS.472.1482O}, comprising 53 optically selected groups in the nearby universe (with distances $D \leq $ 80 Mpc). Remarkably, some of these groups are known to be x-ray emitters with low velocity dispersions, which provide an ideal environment for hosting RPS galaxies. We also expanded our study to encompass galaxies from the Fornax and Hydra Clusters. The combined samples involved 94 galaxies from the CLoGS project, including 47 from the Hydra Cluster and 19 from the Fornax Cluster, resulting in a total of 160 objects within a $z$ range of 0.003 to 0.018.
  
 \end{enumerate}

It is important to note that all the samples used in this study are from the local Universe, with redshift values $z<0.1$. The measurement of morphological parameters is subject to the effects of cosmological redshift, which can lead to a degradation of image resolution and signal-to-noise quality as redshift increases. However, this issue becomes critical only for higher redshifts beyond $z>0.5$, as discussed in numerical simulations performed by \citet{ferrari18}. Given the limited redshift range of the samples in this study (up to a maximum of $z=0.1$), the determination of morphological parameters can be considered reliable and less affected by cosmological redshift effects \citep{ferrari18}.

\section{Morphological Analysis}
\label{meto}

The galaxy morphology analysis was conducted using the {\sc astromorphlib}\footnote{This code is publicly available at
\url{https://gitlab.com/joseaher/astromorphlib}.} software developed by \citet{hernandez22}. This software is capable of automatically downloading Legacy images and providing routines to perform the photometric analysis. One of the functions calculates a 2D sky background model of the images. Additionally, it generates accurate segmentation maps of the galaxies over the image field and it performs deblending in case of  overlapping sources, which is particularly useful for analyzing interacting galaxies. The non-parametric analysis of the galaxy images is performed using the  {\sc statmorph} {\sc python} package \citep{rodriguez19}.  In the following sections, we give details for each step of the analysis.

\subsection{Modelling the sky background}

\begin{figure*}
\includegraphics*[angle=0,width=0.28\textwidth]{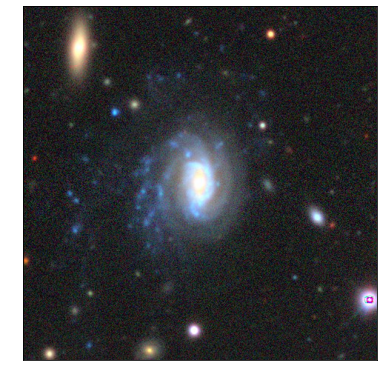}
\includegraphics*[angle=0,width=0.36\textwidth]{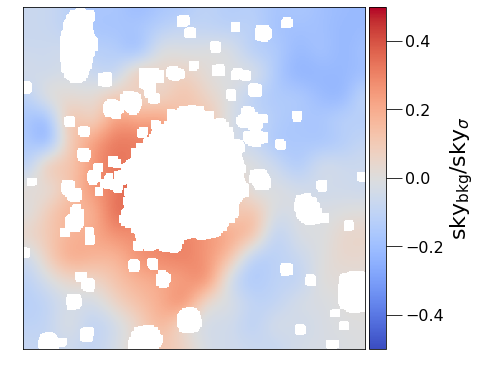}
\includegraphics*[angle=0,width=0.28\textwidth]{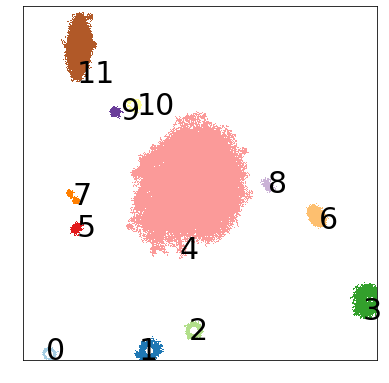}
\caption{A example of 2D sky background model and the segmentation procedure for the RPS galaxy JO201 (JClass 5). Left: the RGB image ($g$, $r$ and $z$ bands). Centre: the 2D background image
computed.  The intensity is the background signal (sky$_{{\rm bkg}}$) normalize by the standard deviation of background (sky$_{\sigma}$). Right: the segmentation maps of the identified sources, which are plotted by different colours and numbers. }
\label{sky_example}
\end{figure*}

The sky background in a given image can vary considerably from one region to the other, therefore it is necessary to generate a 2D model for sky subtraction. To do so, our first step was masking the sources,  such as foreground stars, background galaxies and the studied galaxy itself, over the image. We use the function {\sc make\_source\_mask} from {\sc photutils} \footnote{\url{https://photutils.readthedocs.io/}} package \citep{photutils} to detect the sources; the minimum number of connecting pixels for a source to be identified as such is set to 5 and the threshold  is 2 times above the median background level. A second step is performed with the {\sc Background2D } function from {\sc photutils}. This function first convolves the image with a Gaussian kernel  to smooth the noise, then computes the local sky background dividing the image into a grid - for each cell, sigma-clipping statistics is employed to calculate the median value of the sub-region. Afterwards, the  2D background model is generated by interpolating the previous low-resolution image. It is important to note that the pixel values of the identified sources were replaced  by the median sky background. Finally, this model is subtracted from the original image for its posterior analysis. An example of the 2D sky model computed  for the RPS galaxy JO201 is shown  in Fig.~\ref{sky_example}.

\subsection{Masking foreground stars}
In order to detect foreground stars in the background-subtracted image, the  {\sc astromorphlib} uses the  {\sc DAOStarFinder} function from the 
{\sc photutils}. This function is a {\sc python} implementation
of the {\sc DAOFIND} algorithm \citep{daophot}. Briefly,
this program searches local density maxima that have a peak amplitude greater than a certain threshold and have a size and shape similar to the defined 2D Gaussian kernel. The default threshold is defined as 10 times the sigma of the image background value - however, for some images this value was relaxed to detect more faint foreground
stars. The detected stars were masked using circular apertures. This
procedure was important, especially for galaxies that have overlapping foreground stars affecting the calculation of the physical and the  non-parametric parameters.

\subsection{Segmentation maps}
The  {\sc astromorphlib} uses the {\sc detect\_sources} function from {\sc photutils} to compute the segmentation maps. This routine
detects sources above a specified threshold value in the 
 background-subtracted image. The default threshold is defined as 2 times  the image background sigma value, nevertheless for some
 images this value was flexible to maximize the detection area 
 of the source or avoid some artifact structures in the image. 
In case of close interacting galaxies or background galaxies, 
the deblending procedure was performed by using  the {\sc deblend\_sources}
 from {\sc photutils}. The right panel of Fig.~\ref{sky_example}
 presents the segmentation maps calculated for the image of  RPS galaxy JO201. 

\subsection{Disc modelling}
\label{sec:modelling}
\begin{figure*}
\centering
\includegraphics*[angle=0,width=0.75\textwidth]{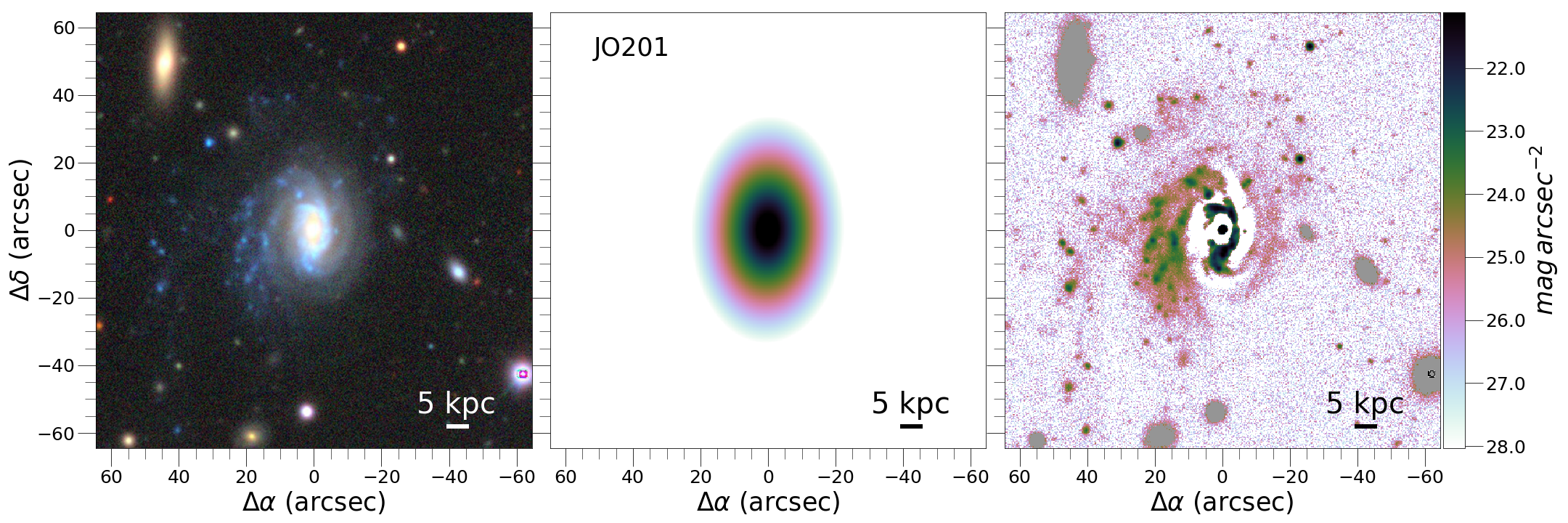}
\includegraphics*[angle=0,width=0.75\textwidth]{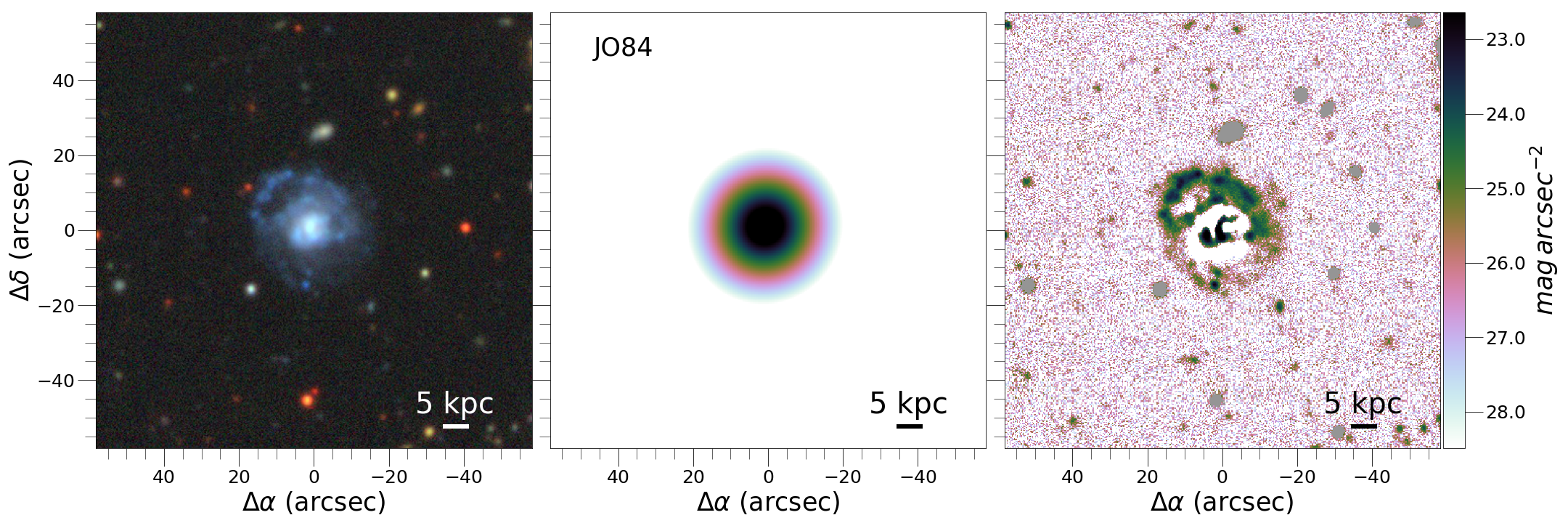}
\includegraphics*[angle=0,width=0.75\textwidth]{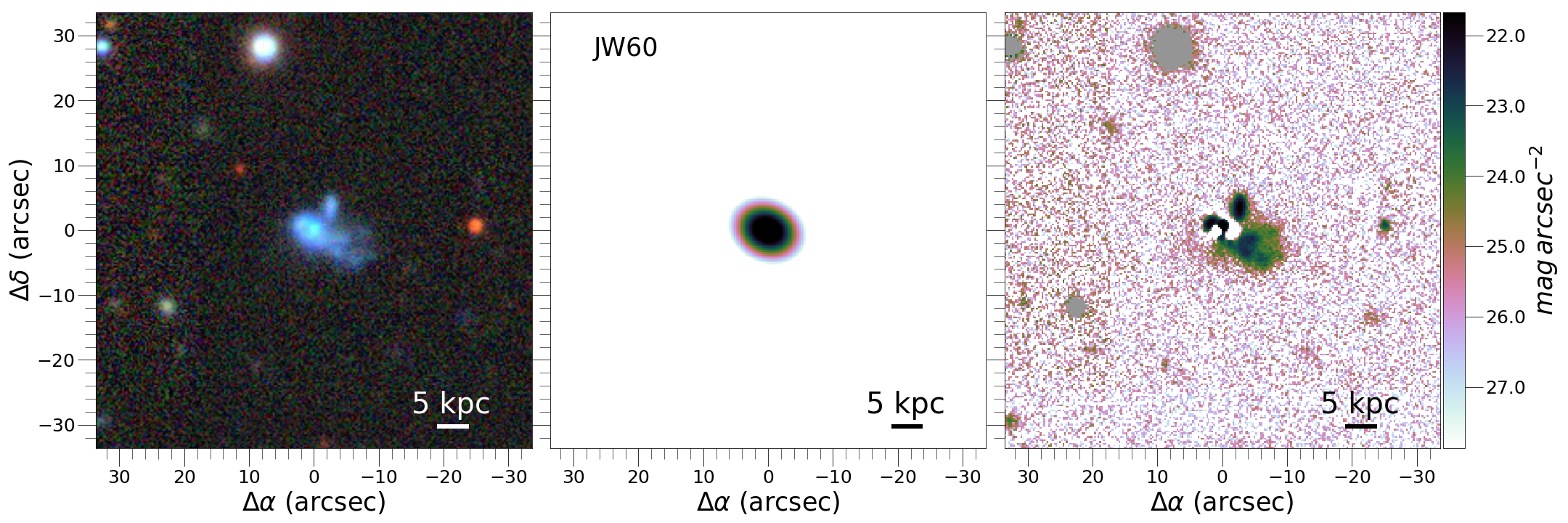}
\includegraphics*[angle=0,width=0.75\textwidth]{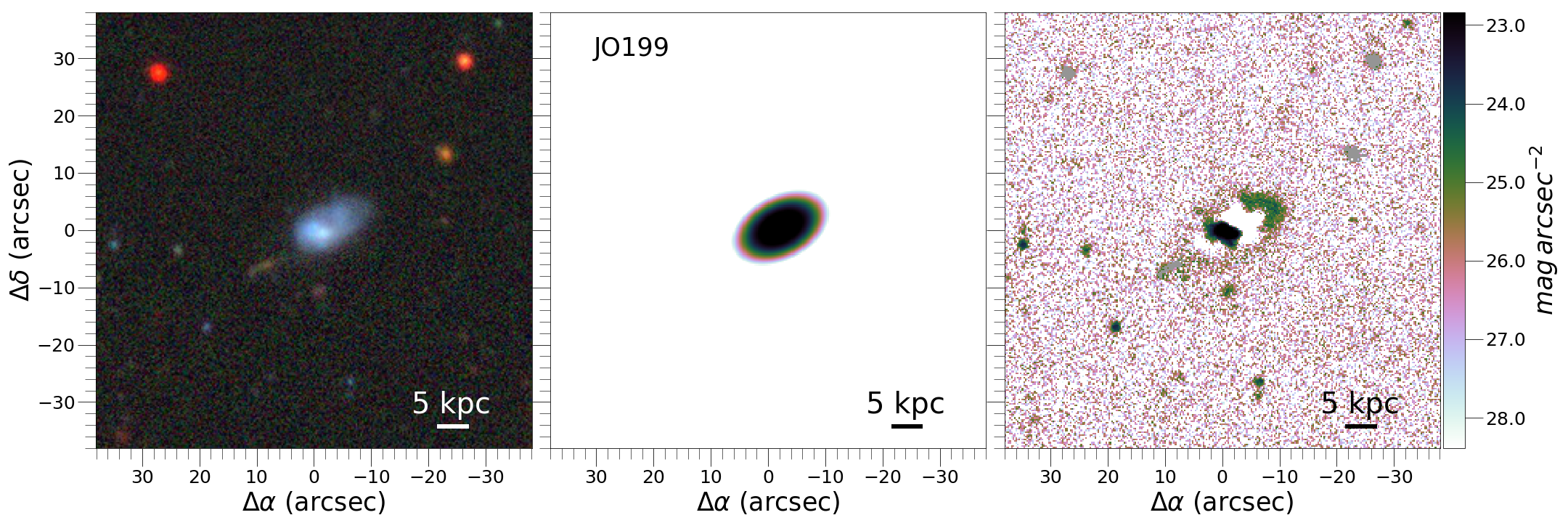}
\includegraphics*[angle=0,width=0.75\textwidth]{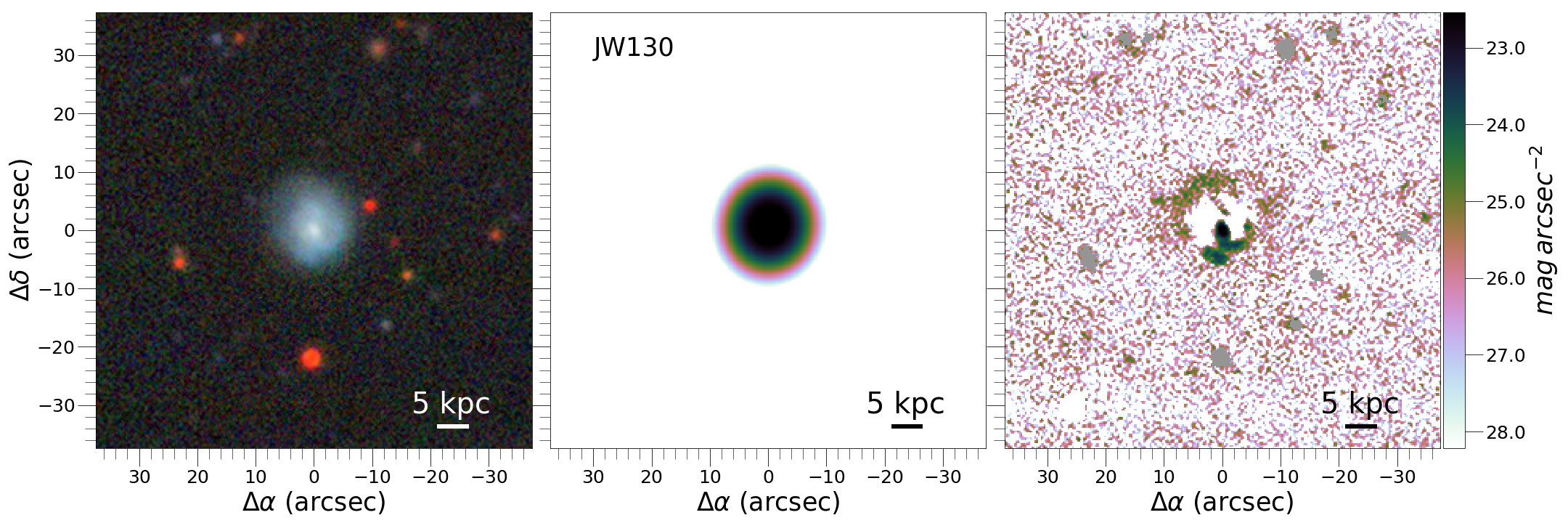}
\caption{Examples of the fitting results. Each column contains the RGB images (left), S\'ersic models (middle), and residual images  in $g$-band (right) for
JO201 (JClass 5), JO84 (JClass 4), JW60 (JClass 3), JO199 (JClass 2) and JW130 (JClass 1).  The grey regions visible in the residual maps represent areas that have been intentionally masked during the analysis. These masks are applied to distinguish between various components in the data, such as segmentation maps for background galaxies and foreground stars. 
The directions are   North ($\delta$) up and East ($\alpha$) left.}
\label{model_JC5}
\end{figure*}

A 2D S\'ersic model \citep{sersic} is computed to model the disc component of the galaxies. This modelling is an output of the function {\sc source\_morphology} from {\sc statmorph}.
This routine, in turn, uses {\sc Sersic2D} from  
{\sc astropy}\footnote{\url{http://www.astropy.org}} \citep{astropy:2013, astropy:2018, astropy:2022} and the fitting
functions from {\sc scipy}\footnote{\url{https://www.scipy.org}}
\citep{scipy}.
 The fitting algorithm is the Levenberg–Marquardt method and the modelling is computed over
the segmentation map of the galaxy.
The  S\'ersic computed parameters  are 
the radius containing half of the galaxy’s flux ($R_e$), the 
surface brightness ($I_e$) at  $R=R_{e}$, and the  S\'ersic
index ($n$).  These parametric parameters are used in our
statistical analysis. Fig.~\ref{model_JC5} shows one example of the fitting results  obtained for  each case  of the JClass classification.  The residual image is the difference between the original image  in $g$-band and the model image based on the best-fitting S\'ersic profile computed in $g$-band. 
 We opted for the g-band over the $r$ and $z$-bands due to its superior visibility of asymmetric star-forming structures, such as tails or bows, and the distinct shock morphologies associated with the RPS phenomenon
 \citep[e.g,][]{2016MNRAS.455.2994M}.

It is clearly seen that galaxies with higher  JClass have larger  residual. This  result is  expected because the galaxies with  higher JClass  present higher asymmetries or debris tails originating from unilateral external forces. For example,  the residuals are fairly small for JW130 (JClass 1) and much higher for JO84 (JClass 4) and  JO201 (JClass 5), which are the classes with the most striking jellyfish galaxies \citep{2016AJ....151...78P}.


\subsection{Stellar mass}

The  total stellar mass ($M_{\star}$) for the galaxies  was estimated using  the  relations between mass-to-light ratios and the colours
derived from \citet{bell03}. We computed
the stellar mass from the luminosity in the $z-band$ and the colour 
($r-z$) by using the coefficients given in Table 7 of
\citet{bell03}. The $M_{\star}$ was computed up to $2R_e$.
The luminosity in $z-band$ and the colour 
($r-z$) was used because the light in these bands is dominated by the old stellar population which builds up most of the stellar mass in the galaxy.   in addition, the determination of the mass-to-light ratios is
more accuracy for the red-end of the optical bands since
is less prone to uncertainties over a wide range of star
formation histories and the dust extinction \citep{bell03}.   The luminosity in $z-band$ and the colour ($r-z$) were determined  inside an ellipse with a semi-major axis  of  $2R_e$, using the position angle and ellipticity computed of the $r$-band 2D S\'ersic model, which is the deepest photometric band in our study.

\begin{figure*}
\centering
\includegraphics*[angle=0,width=\textwidth]{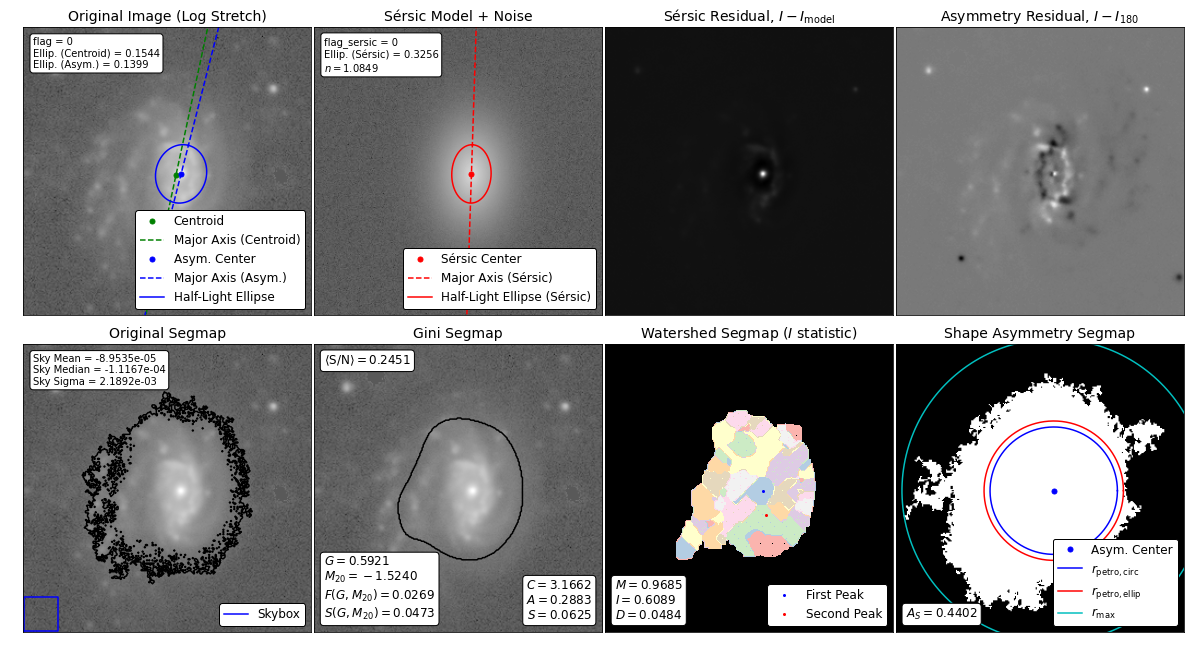}
\caption{An example of an output mosaic of  {\sc statmorph}  for JO201 (JClass 5)  ran over $g$-band image.  From left to right and from top to bottom: the input image, the Sérsic model, the Sérsic residual, the asymmetry residual, the original segmentation map, the Gini segmentation map, the watershed segmentation map, and the shape asymmetry segmentation map. }
\label{fig_statmorph}
\end{figure*}

\subsection{Non parametric parameters}

In addition to the photometric model for the  galaxies, we computed non-parametric parameters for
the studied galaxies  in $g$-band images.
As previously discussed in Section \ref{sec:modelling}, we selected this specific band for its heightened sensitivity in identifying the asymmetric structures produced by the RPS. We used the routine {\sc source\_morphology} from {\sc statmorph} \citep{rodriguez19}. This code is specifically
designed to calculate  non-parametric parameters such as Concentration, Asymmetry, and Gini index. The computation of these parameters was  done over $g$-band images.
Previous of computation of the parameters {\sc statmorph} determines the Petrosian radius  \citep[$r_{{\rm petro}}$,][]{petrosian76} . The parameters are computed over all pixels within $1.5r_{{\rm petro}}$.  For detailed technical information on how the morphological parameters are computed, we refer the reader to \citet{rodriguez19}. However, here,
we provide a concise summary of the statistics used in this work and calculated by {\sc statmorph}:

\begin{itemize}
\item  Gini - $M_{20}$: this classification system \citep{lotz04}  have been used to quantify galaxy morphology, including normal galaxies,  merger systems and irregulars.  Gini coefficient, $G$, measures the light ``inequality'' in the galaxy, while the $M_{20}$ statistic measures the second moment of a galaxy’s brightest regions. Associated to these parameters, the  {\sc statmorph} also calculates Gini–$M_{20}$ ``bulge statistic'' [$F(G,\,M_{20}$), \citet{snyder15b}], which is strongly correlated to the bulge strength, and Gini–$M_{20}$ ``merger statistic'' which is correlated to the merger ``strength'' [$S(G,\,M_{20}$), \citet{snyder15a}].

\item $CAS$: This statistics was popularized
by \citet{conselice03} and include the concentration index ($C$), the asymmetry index ($A$), and the  ``clumpiness'' or smoothness index ($S$).

\item  $MID$: These parameters were introduced by \citet{freeman13}and \citet{peth16} and  are more sensitive to weak morphological perturbations. The components of this statistic are the multimode statistic ($M$), which  measures the ratio between the areas of the two most ``prominent'' clumps within a galaxy, the intensity statistic ($I$), which measures the ratio between the two brightest sub-regions of a galaxy, and the  deviation statistic ($D$), which  measures the distance between the light centroid and brightest peak of the galaxy.

\end{itemize}

  Fig.~\ref{fig_statmorph} presents
an example of an output mosaic of  {\sc statmorph}  for JO201 (JC5). This mosaic shows
the input image, the Sérsic model, the residual image, the asymmetry residual image, the original segmentation map, the Gini segmentation map, the watershed segmentation map, and the shape asymmetry segmentation map.

\subsection{Running {\sc astromorphlib}}

The previous sections have described the entire sequence of analysis carried out for each galaxy in the studied samples. The initial step involved running the {\sc astromorphlib} tool automatically on each sample. Subsequently, we carefully examined all intermediate images and the final mosaic generated by {\sc statmorph}. A challenging aspect of the analysis was accurately determining the segmentation map for each object. If the segmentation map is not well-defined,  it is flagged in the determination of the Sérsic model and/or non-parametric parameters by {\sc statmorph}. To address this issue, we  ran {\sc astromorphlib} several times until we obtained the best input segmentation map,  without flags. This meticulous process ensures that the derived morphological parameters are not only accurate but also reliable, making meaningful and trustworthy posterior analysis.

\section{Results and Discussion}
\label{results}

\subsection{Correlations}

We conducted an analysis to investigate the potential correlation between the JClass classification and five measured physical parameters:  the S\'ersic model  $R_e$ and $n$, the color $g-r$, the stellar mass $M_{\star}$ and the absolute magnitude in the $z$ band $M_{z}$. Additionally, we also examined the correlation of the  JClass classification  with ten non-parametric parameters: $G$, $M_{20}$, $S(G,\,M_{20})$, $F(G,\,M_{20})$, $C$, $A$, $S$, $M$, $D$, and $I$. In total we have analysed fifteen parameters.
To assess the strength of the correlation between these variables, we employed the Spearman correlation coefficient, denoted as $r$. This coefficient quantifies the monotonic association between pairs of variables, providing insights into their level of correlation. Our focus was on identifying the most strongly correlated parameters, with a threshold of $|r|>0.2$. 

\begin{figure*}
\includegraphics*[width=0.9\columnwidth]{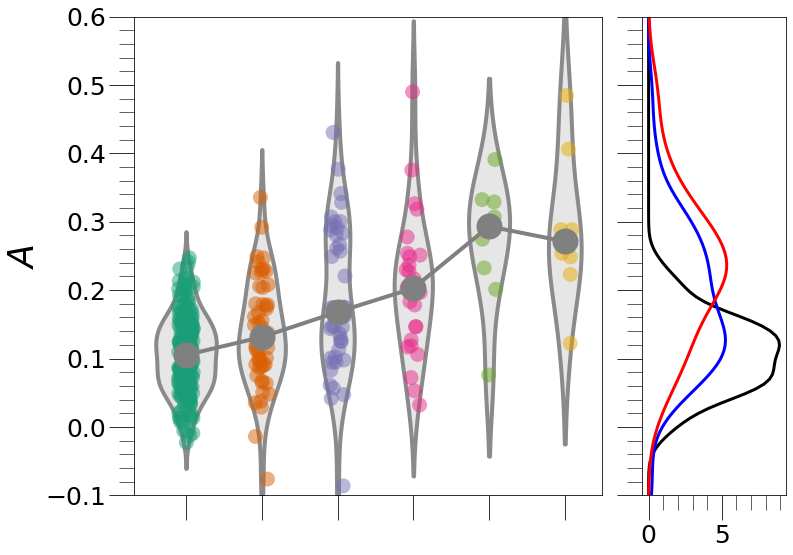}
\includegraphics*[width=0.9\columnwidth]{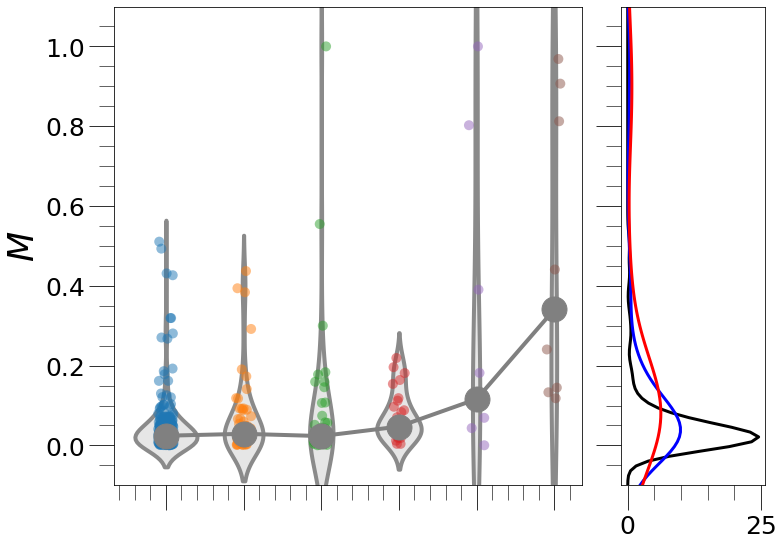}
\includegraphics*[width=0.9\columnwidth]{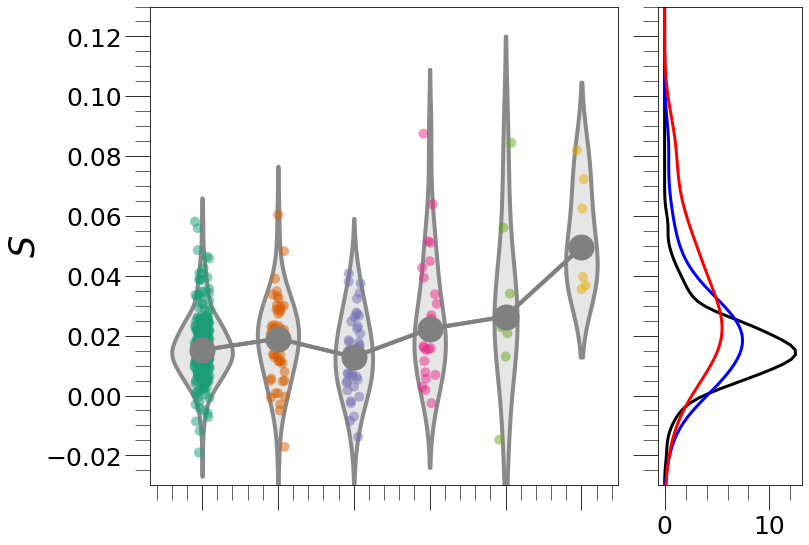}
\includegraphics*[width=0.9\columnwidth]{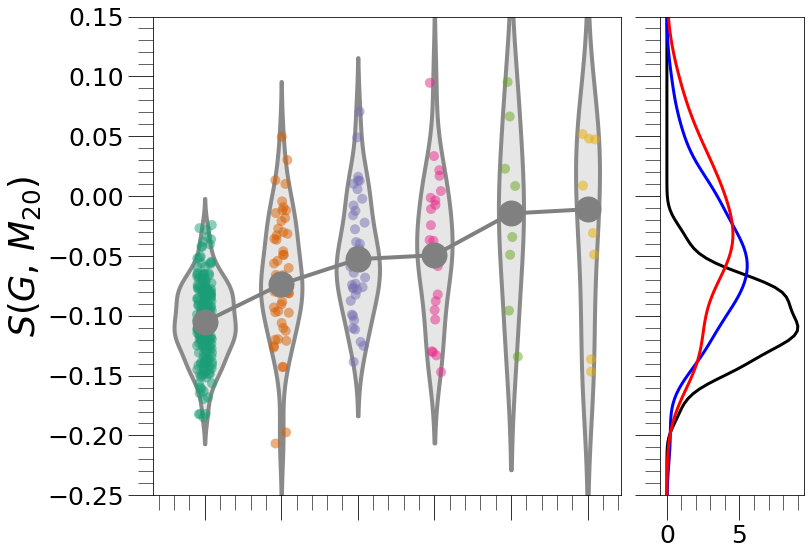}
\includegraphics*[width=0.9\columnwidth]{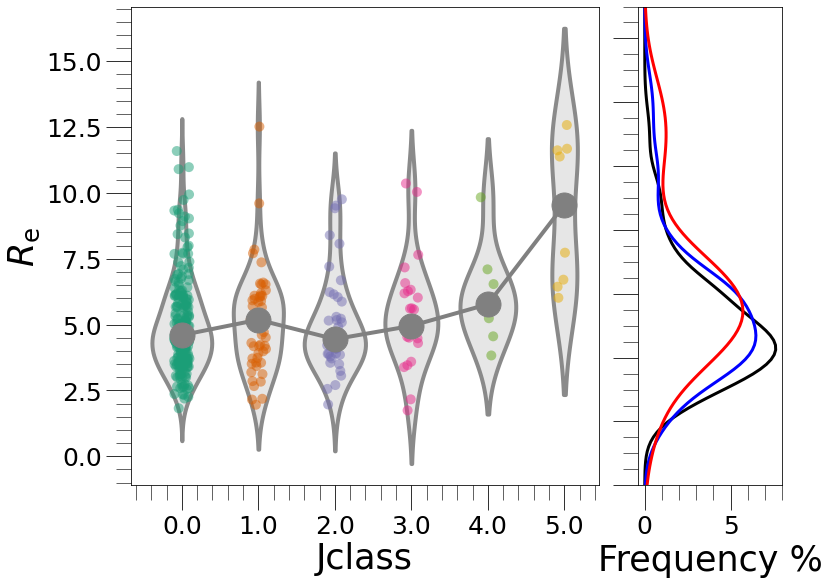}
\includegraphics*[width=0.9\columnwidth]{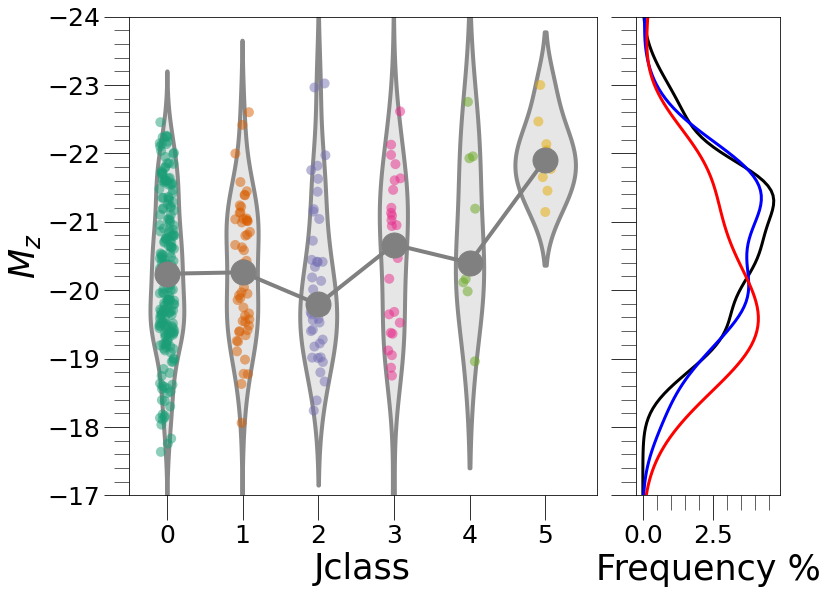}
\caption{Dependence of  $A$, $M$, $S$, $S(G,\,M_{20})$, $R_{\rm{e}}$ and $M_z$  on the JClass classification. The width of the violin plot along the $x$-axis indicates the normalized fraction at the corresponding $y$-axis value, while the black points represent the median values for a given JClass. There are side sub-panels with the density
histograms of each parameter for isolated (black lines) 
JClass $\geq3$ (red lines), and all JClass (blue lines) galaxies. The isolated galaxies are designated here as JClass equal to zero.}
\label{violinp}
\end{figure*}

Our analysis identified six significant correlations between parameters and the JClass classification. These parameters are   $A$ ($r=0.41$, p-values=6.3$\times10^{-7}$), $M$ ($r=0.32$, p-values=1.5$\times10^{-4}$), $S$ ($r=0.25$, p-values=2.9$\times10^{-3}$), $S(G,\,M_{20})$ ($r=0.23$, p-values=5.7$\times10^{-3}$),
$R_e$ ($r=0.22$, p-values=8.3$\times10^{-3}$), and $M_z$ ($r=0.22$, p-values=9.9$\times10^{-3}$).  In Fig.~\ref{violinp},  the six parameters as a function of the JClass classification are shown. These plots clearly show strong trends where an increase in JClass values is accompanied by higher parameter values, which are even more pronounced for JClass $\geq3$. Furthermore, if we separate the galaxies with JClass $\geq3$ into a single class and compare them with the isolated galaxies  (designated as JClass=0), we can see a bimodal distribution of values, particularly clear for the parameters $A$ and $S(G,\,M_{20})$, as shown in the density histograms in the
side sub-panels.

Out of the six strongly correlated parameters with JClass classification, the first four are related to the light distribution of the galaxy [$A$, $M$, $S$, and $S(G,\,M_{20})$], while the last two are scaling parameters ($R_{\rm{e}}$ and $M_z$). The first four parameters are associated with significant perturbations in the light distribution of the galaxy. For instance, the $A$ parameter is related to the presence of large asymmetric structures in the galaxy, while the $S$ parameter is associated with high level of clumpiness. Similarly, high values of the $M$ parameter are linked to substructures as bright as the nuclei of the host galaxy, and $S(G,\,M_{20})$ captures the merger statistics, increasing for objects with strong morphological perturbations. Hence, these parameters are effective in describing the characteristic features of RPS galaxies. 

On the other hand, the correlation between the JClass classification and physical properties such as effective radius and the absolute magnitude   has been previously noted by P16. According to their findings, higher JClasses correspond to more massive galaxies. This correlation could be attributed to the way gas is confined within galaxies  and the type of orbits in the cluster. The size and stellar mass of a galaxy play a crucial role in shaping its gravitational potential. This potential determines the strength of the gravitational forces acting on the gas within the galaxy. As a result, galaxies with larger radii and higher stellar masses may have lower gas densities and shallower potential wells.  Consequently,  the removal of gas from the host galaxy becomes easier, ultimately leading to the formation of long tails.   Notably, the formation of these ``tentacles'' is most likely under strong ram pressure conditions, which are typically encountered during the pericentre passage of galaxy orbits towards the centres of galaxy clusters.  In support of this explanation, a study conducted by \citet{jaffe18} focused on the phase-space analysis of the most striking examples from the P16 sample (higher JClass galaxies) and found that the most extreme cases of gas stripping occurred in galaxies with plunging orbits. These orbits take them close to the centre of the galaxy cluster at high speeds, experiencing intense ram pressure, which effectively strips away gas from the galaxy.


\subsection{The best diagnostic diagrams to select RPS and the morphological transition zone}
\label{jelly_zone}

\begin{figure*}
\centering
\includegraphics*[angle=0,width=0.9\columnwidth]{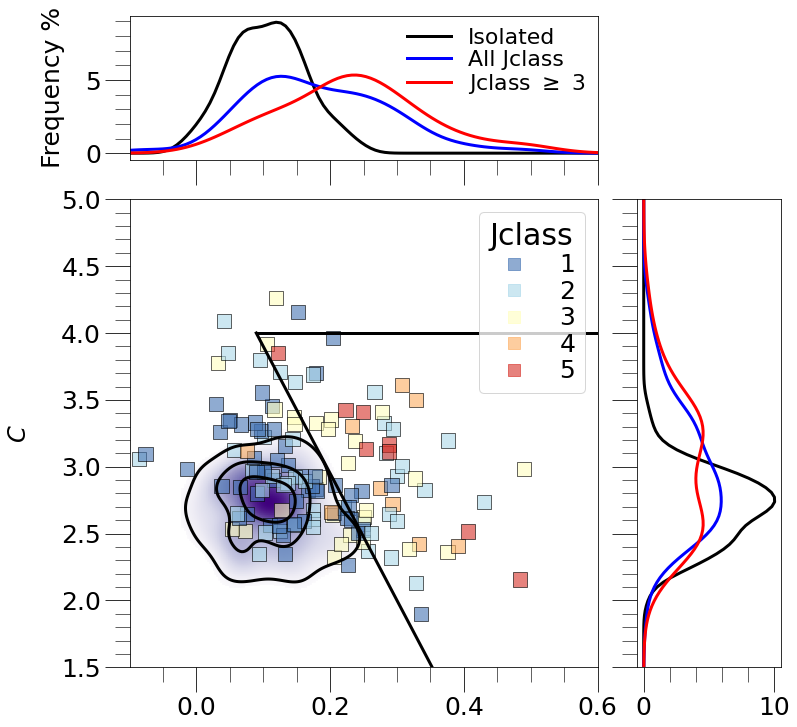}
\includegraphics*[angle=0,width=0.9\columnwidth]{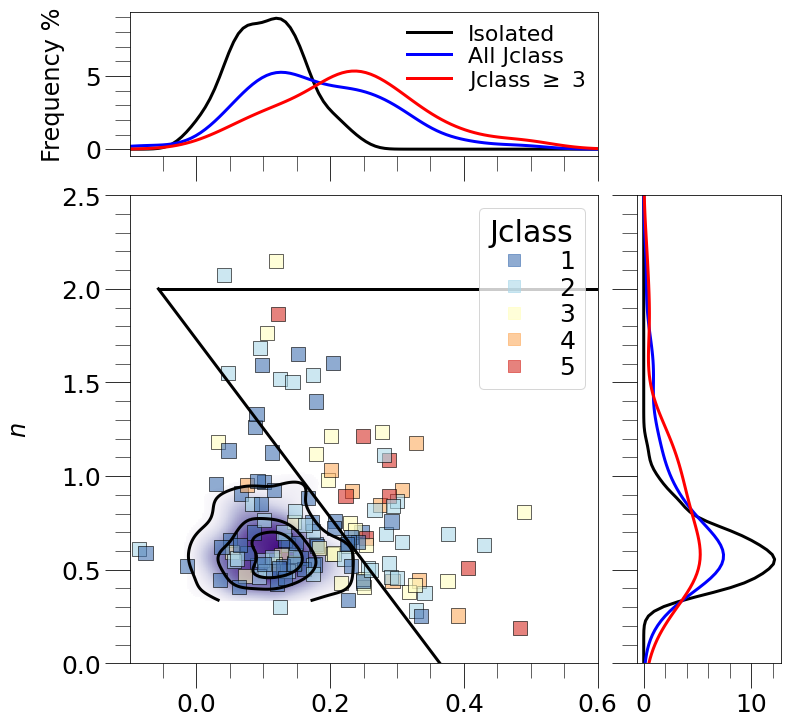}
\includegraphics*[angle=0,width=0.9\columnwidth]{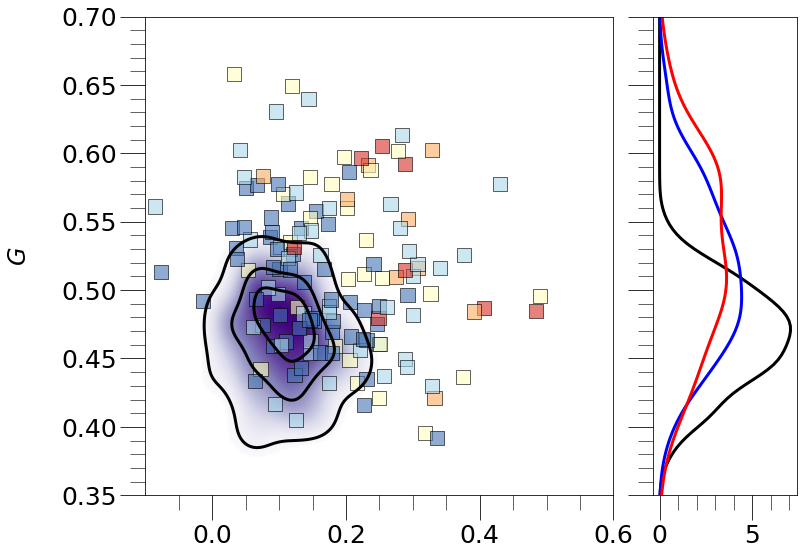}
\includegraphics*[angle=0,width=0.9\columnwidth]{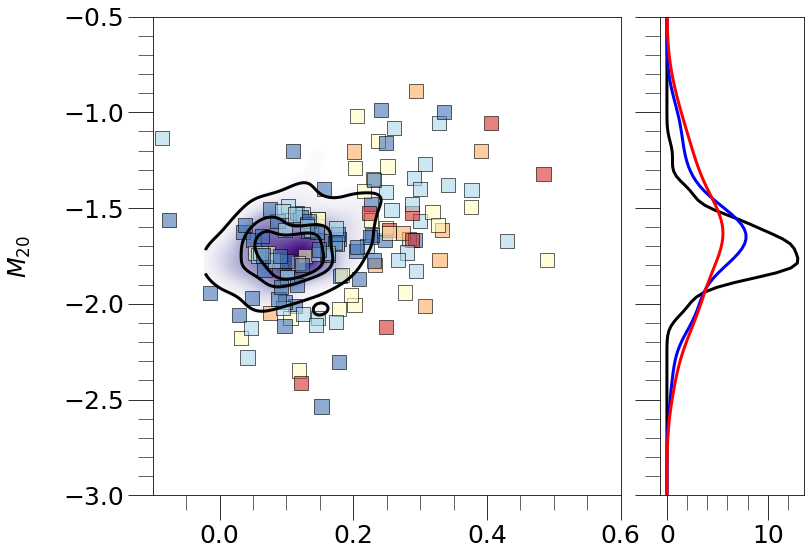}
\includegraphics*[angle=0,width=0.9\columnwidth]{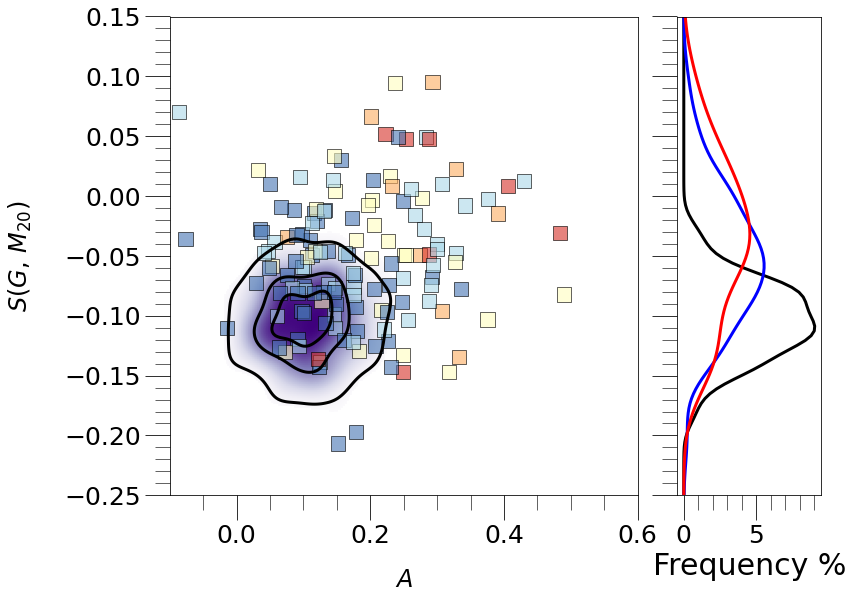}
\includegraphics*[angle=0,width=0.9\columnwidth]{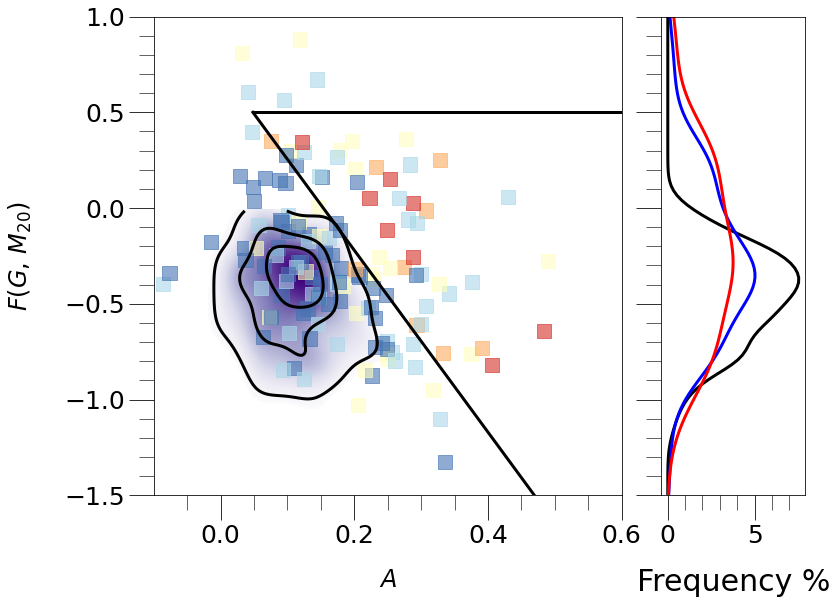}
\caption{Pairwise plots of  $A$  {\it vs} $C$ (top-left panel),  $A$ {\it vs} $n$ (top-right), 
$A$   {\it vs}  $G$ (medium-left), $A$   {\it vs} $M_{20}$ (medium-right), $A$ {\it vs} $S(G,\,M_{20})$ (bottom-left) and $A$  {\it vs} $F(G,\,M_{20})$ (bottom-right). The density maps in purple are the 
distribution of isolated galaxies and the iso-density contours  are 10, 50, and 80th percentiles (black lines). 
The solid black lines delineate the morphological transition zone boundaries in the diagnostic diagrams: $A$  {\it vs} $C$,  $A$ {\it vs} $n$ and
$A$  {\it vs} $F(G,\,M_{20})$.
The  top  and right sub-panels are  the density
histograms of each parameter for isolated (black lines), all JClass (blue lines), 
and JClass $\geq3$ (red lines) galaxies.}
\label{fig_jelly}
\end{figure*}

Identifying RPS candidates is a challenging task, and non-parametric parameters could be a crucial tool.  As seen in the previous section, there are strong correlations between these parameters and JClass classification. This raises the question of whether an optimal combination of non-parametric parameters exists that can clearly separate RPS and isolated galaxies. To seek out the best bi-dimensional plots where RPS galaxies (represented by the J3+J4+J5 classes) fall in a specific region away from isolated galaxies, we performed a Kolmogorov-Smirnov (KS) test for 105 possible pairs of the 15 computed parameters.  We used a {\sc python} routine with the numerical implementation of a multidimensional version of the KS test\footnote{\url{https://github.com/Gabinou/2DKS}} \citep{peacock83,fasano87,press96}. We then chose to investigate combinations with high KS test values ($> 0.7$), resulting in 6 pairs of combinations that obey this criterion: $A\, vs\,S(G,\,M_{20})$ (KS=0.79), $A\,vs\,C$ (KS=0.78), $A\,vs\,$Gini (KS=0.78), $A\,vs\,(G,\,M_{20})$ (KS=0.75), $A\,vs\,M_{20}$ (KS=0.74) and $A\,vs\,n$ (KS=0.73). 
It is evident that the key parameter is $A$ and  all optimal combinations invariably include it, which is reasonable 
 considering the prominent correlation it demonstrates with the JClass classification. Besides, isolated galaxies showcase a limited spectrum of $A$ values, specifically falling within the range of [0, 0.2]. Consequently, the inclusion of $A$ in the combinations proves highly effective in distinguishing between RPS and isolated galaxies.

The best pair combinations found are shown in Fig.~\ref{fig_jelly}. As expected, the plots show a good separation between  RPS galaxies (JClass $\geq3$)  and isolated galaxies, i.e., almost all JClass 3, JClass 4, and JClass 5 galaxies fall apart of the 
isolated area (represented as 80th percentile iso-density contour), while most  of the  J1 and J2 classes fall in the same area of isolated galaxies. We  identified the most effective plots, namely $A\,vs\,C$, $A\,vs\,n$, and $A\,vs\,F(G,\,M_{20})$, based on the criterion that no J5 galaxy should fall within the isolated regions\footnote{ Here, it is worth noting that we are favouring purity over completeness to enhance the efficiency of identifying compelling cases of RPS candidates. By focusing on purity, we aim to reduce false positives and ensure that the cases we identify as potential RPS candidates are more reliable.}. For each of these plots, we have defined boundaries that accurately differentiate between galaxies affected by RPS and isolated galaxies. 
To establish these boundaries, we have traced tangent lines to  the outermost iso-density contour of the 80th percentile of the isolated galaxy distribution as a reference.
 Hereafter, we will designate   the right side of these partition lines as morphological transition zones. However, to further refine this region, we have set maximum values for $C$, $n$, and $F(G,\,M_{20})$ at 4.0, 2.0, and 0.5, respectively. We have observed that no J5 objects are above these thresholds. The equations defining these lines for $A\,vs\,C$, $A\,vs\,n$, and $A\,vs\,F(G,\,M_{20})$ are as follows:

\begin{equation}
C_{lim} = -9.5\,A + 4.85,\,\,C_{{\rm max}}=4.0,
\end{equation}

\begin{equation}
n_{lim} = -4.75\,A + 1.73,\,\,n_{{\rm max}}=2.0, 
\end{equation}

\noindent and
\begin{equation}
F(G,\,M_{20})_{lim} = -4.75\,A +0.73,\,\,F(G,\,M_{20})_{{\rm max}}=0.5.
\end{equation}
 In summary, the areas defined by these lines are  effective boundaries to  select potential RPS candidates using images from the DESI Legacy Survey. Consequently, these three pair-wise plots can be considered as diagnostic diagrams. They will aid in identifying galaxies that exhibit morphological characteristics indicative of RPS, allowing for a systematic and reliable selection of potential RPS candidates.
 
\citet{2016MNRAS.455.2994M} used these kinds of planes of non-parametric parameters to select RPS candidates in  galaxy cluster samples at intermediate-redshifts ($z=0.3-0.7$). They  found that the RPS candidates are  outliers with respect to the distribution of normal galaxies in  clusters in  diagrams such  as  $C\,vs\,A$,   $G\,vs\,M_{20}$ coefficient, and  $Sk_{0-1}$ versus $Sk_{1-2}$.  Recently, \citet{bellhouse22} also studied
 jellyfish candidates in two galaxy clusters including the planes $C\,vs\,A$,   $G\,vs\,M_{20}$, and the outer centroid variance $vs$ full centroid variance. These authors found the last plane was the most effective standalone measure of galaxy morphology, which shows as promising tool to detect disturbed morphologies. The parameters of the planes $Sk_{0-1}$ versus $Sk_{1-2}$  and the outer centroid variance $vs$ full were not computed in our study, and the   Gini $vs\, M_{20}$ plot   was not included in the analysis due to the low KS test score of 0.57 (p-value=8.1$\times10^{-9}$).  On the other hand, the effectiveness of the $C,vs,A$ diagram to find potential candidates aligns with our findings. \citet{2016MNRAS.455.2994M} and \citet{bellhouse22} used a vertical line as a threshold to define "normal galaxies" zones based on the asymmetric axis in this plane.  However, as we  discussed earlier, we believe that the boundary formed by the tangential lines in the transition zones might be more effective in separating this group of galaxies. Furthermore, it is important to highlight the differences in the samples and data analysis between the mentioned studies and our current work. The samples used by both studies are at higher redshifts, ranging from $0.3-0.5$, and the photometric analysis was performed in a distinct band (HST/ACS,F606W). At these higher redshifts, the reliability of measuring photometric parameters decreases due to various factors, making a direct comparison with these works challenging \citep[e.g.,][]{ferrari18}. Another crucial point to note is that neither of the previous studies conducted a robust statistical analysis to identify the best combinations of non-parametric parameters for selecting RPS candidates, unlike our present study. Our comprehensive statistical analysis allows us to validate the effectiveness of the selected non-parametric parameters in identifying RPS candidates accurately.



Finally, as highlighted by \citet{2016MNRAS.455.2994M}, while these plots are effective in detecting galaxies experiencing RPS, they still exhibit significant contamination, requiring, in addition, a visual inspection of the systems, but with the great advantage of narrowing down the sample to a manageable size. 
 
\subsection{Interacting zone}

\begin{figure}
\centering
\includegraphics[angle=0,width=0.95\linewidth]{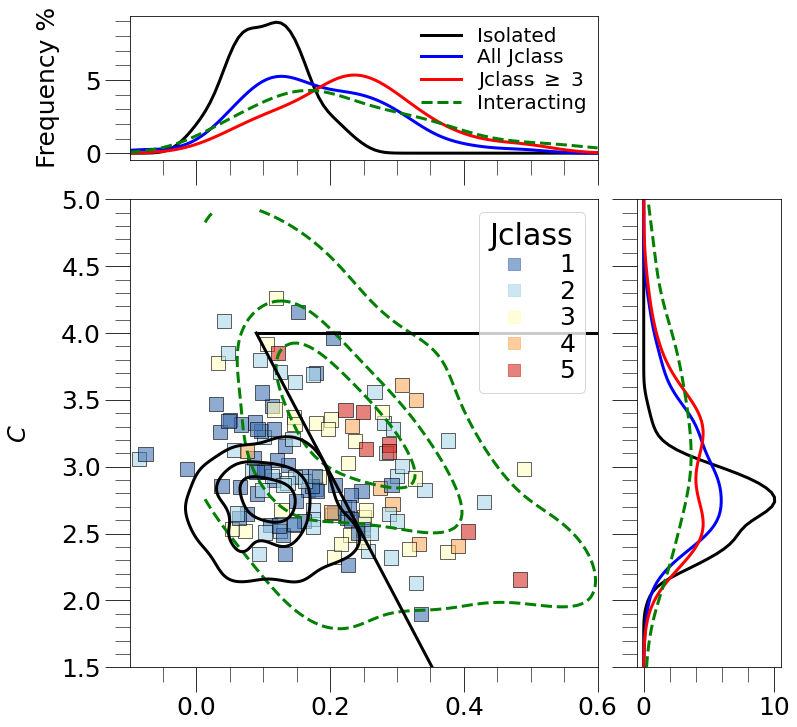}
\includegraphics[angle=0,width=0.95\linewidth]{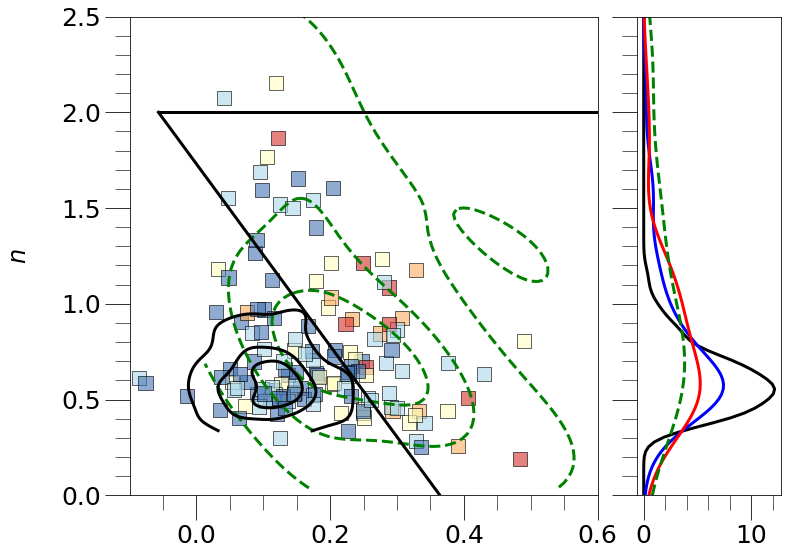}
\includegraphics*[angle=0,width=0.95\linewidth]{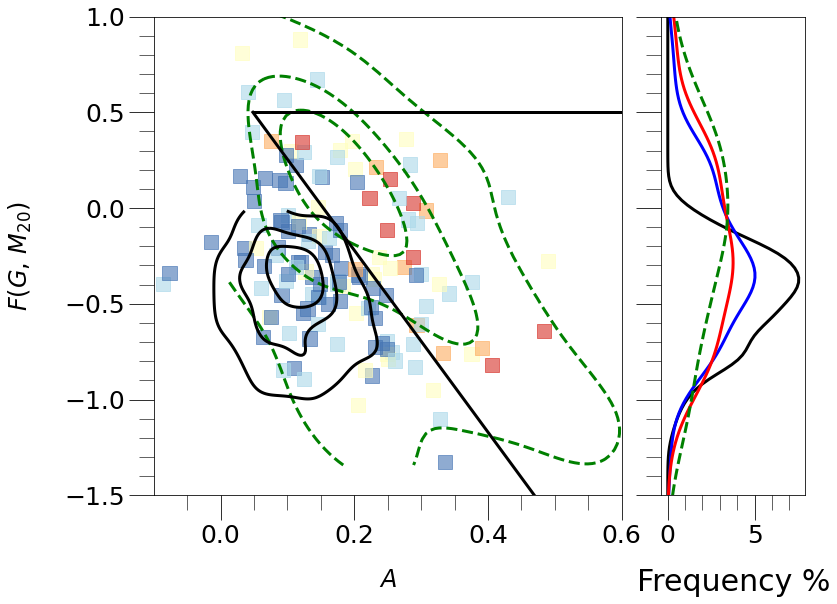}
\caption{The distribution of interacting galaxies in $A\,vs\,C$ (top panel), $A\,vs\,n$ (middle), and $A\,vs\,F(G,\,M_{20})$ (bottom) plots. The green density maps are the interacting galaxies (the iso-contour 
are dashed lines), while black iso-density contours (10, 50 and 80th percentiles) delineate the distribution of isolated galaxies. The solid black lines delineate the morphological   transition zone boundaries in each diagnostic diagram. The top and right  side sub-panels show
the density histograms for isolated, all J class, 
Jclass $\geq3$, and interacting galaxies of 
 $A$,  $C$, $n$, and  $F(G,\,M_{20})$ parameters.}
\label{fig_int}
\end{figure}

Distinguishing between the effects of galaxy tidal effects and ram pressure stripping mechanisms may be quite  difficult  because both mechanisms cause strong morphological perturbations \citep{kronberger08}. Then, to  assess  the effectiveness of the defined diagnostic diagrams in separating these two types of galaxies,  we plot in Fig.~\ref{fig_int} the distribution of the interacting galaxies. Interacting galaxies occupy a large area in diagnostic  plots, overlapping with both isolated and transition zones. However, the bulk of the interacting galaxy population falls within the transition zone, as seen from the iso-density contour of the 50th percentile.


Therefore,  while  these plots are not sufficient for distinguishing RPS and interacting galaxies,  they can  still be useful  to identify possible candidates
for galaxies undergoing RPS. However, further analysis is necessary to distinguish jellyfish galaxies from interacting galaxies. One possible way is by analyzing the type of morphological perturbations present in the galaxies. Tidal interactions between galaxies tend to produce bi-symmetric perturbations, whereas  ram-pressure stripping results in asymmetries that are preferentially aligned in one direction. 
 Another approach involves investigating the presence of nearby galaxy companions within a physically projected distance. If no obvious companions are found, it suggests that the ram pressure stripping (RPS) mechanism is likely the primary candidate. Conversely, the presence of physical companions increases the probability of gravitational interactions being the main driver of morphological perturbations.  However, there are important considerations to keep in mind. In high-density environments like galaxy clusters, galaxy-galaxy interactions occur at high velocities, resulting in short interaction times compared to the cluster crossing time. This means that the perturbing galaxy remains in close proximity to the perturbed galaxy for only a brief period. However, despite the brief interaction time, the effects of the interaction can be long-lasting. Therefore, the absence of nearby companions for a perturbed galaxy does not necessarily exclude the possibility of gravitational perturbations \citep{1998ApJ...495..139M}. Similarly, the presence of a close companion does not rule out the ram pressure mechanism as the primary perturbing force, as both mechanisms can act simultaneously \citep{kronberger08,moretti2018a}.   Therefore, conducting a meticulous analysis of the asymmetry patterns in morphological perturbations,  combined with an examination of the surrounding environment, can prove highly valuable in resolving the ambiguity in classifying galaxies located within the RPS zone.  In the following section  we will discuss this issue in detail.

\subsection{RPS candidates}

\begin{table*}
\caption{Galaxies in the morphological transistion zones}
\label{tab:Jelly_cand1}
\begin{tabular}{llcclcccccc}
\hline
\noalign{\smallskip}
Galaxy  & Cluster& RA & DEC  &  Morph  & $A$  & $C$  
  & $n$  &  $F(G,\,M_{20})$  &  intec  &  sep \\
  &  & [deg] & [deg]  &   &  &  &   
  &   &  & [kpc] \\
\hline
\noalign{\smallskip}
2MASX\,J01223182+0916534 &  LGG\,23	(CLoGS) & 20.6326  & 9.2815   & Early\_Tidal	     & 0.08 & 3.86  & 1.6   & 0.44  & -      & - \\
2MASX\,J10284923-3129509 &  Hydra	 & 157.2051 & -31.4975 & Early\_Blue\_Tidal  & 0.41 & 1.99  & 0.5   & -1.18 & -      & - \\
2dFGRS\,TGS471Z004       &  Fornax	 & 49.8540  & -32.6493 & Irr  	     & 0.32 & 2.15  & 0.3   & -1.06 & -      & - \\
6dFGS\,gJ102734.0-234711 &  Hydra	 & 156.8911 & -23.7857 & Jellyfish  	     & 0.22 & 2.80  & 0.7   & -0.38 & Major  & 51.6 \\
6dFGS\,gJ103502.9-293024 &  Hydra	 & 158.7620 & -29.5066 & Early\_Blue\_Tidal  & 0.08 & 3.71  & 1.6   & 0.38  & -      & - \\
6dFGS\,gJ104139.4-274638 &  Hydra	 & 160.4143 & -27.7773 & Jellyfish  	      & 0.33 & 2.73  & 0.7   & -0.55 & -      & - \\
ESO\,302-14              &  Fornax	 & 57.9204  & -38.4522 & Jellyfish  	     & 0.29 & 2.20  & 0.8   & -0.92 & -      & - \\
ESO\,436-29		 &  Hydra    & 157.597    & -30.393    & Jellyfish & 0.35 & 2.2 & 0.4 & -0.80 & -  & - \\
ESO\,437-043 &   Hydra & 160.414 & -27.777 & Jellyfish & 0.33 & 2.7 & 0.7 & -0.55 & -  & - \\
ESO\,437-30		 &  Hydra	 & 159.812    & -30.298    & Spiral\_Edge-on & 0.25 & 3.6 & 0.8 & 0.20 & -  & - \\
\noalign{\smallskip}
\hline
\end{tabular}
\end{table*}

\begin{figure*}
\centering
\includegraphics*[angle=0,width=0.32\textwidth]{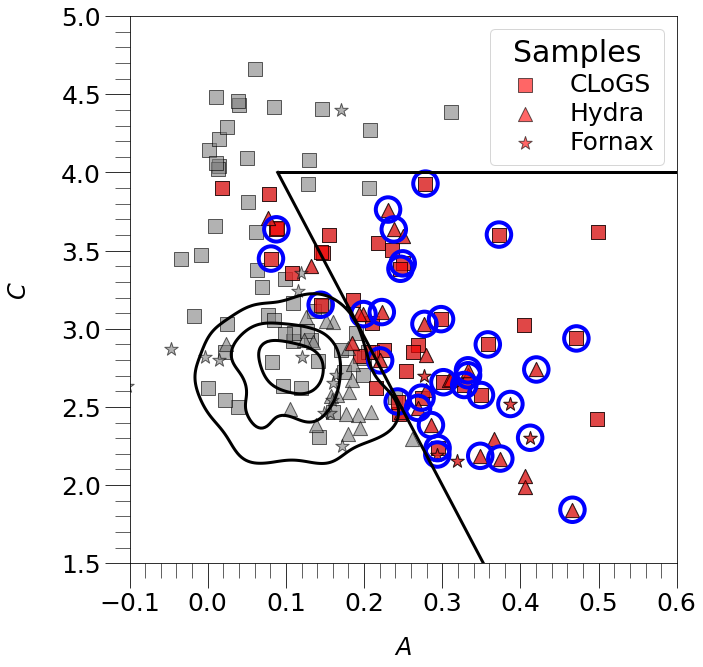}
\includegraphics*[angle=0,width=0.32\textwidth]{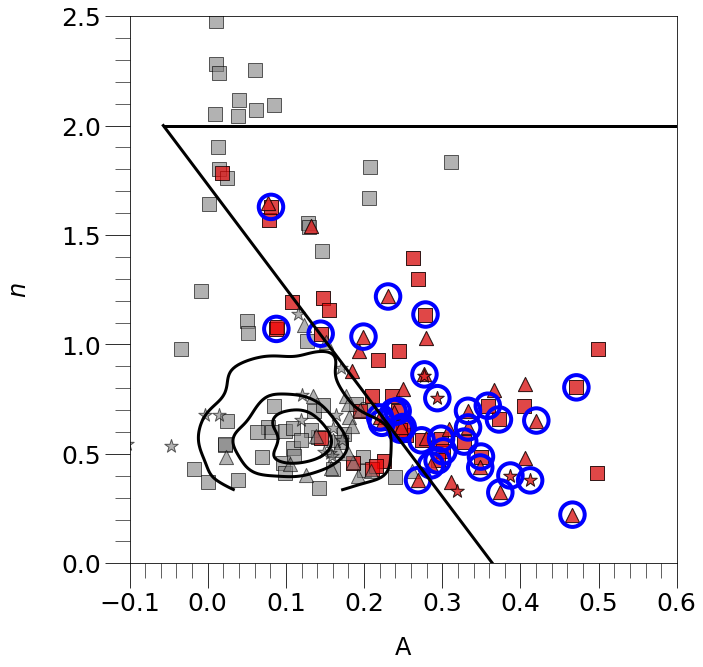}
\includegraphics*[angle=0,width=0.32\textwidth]{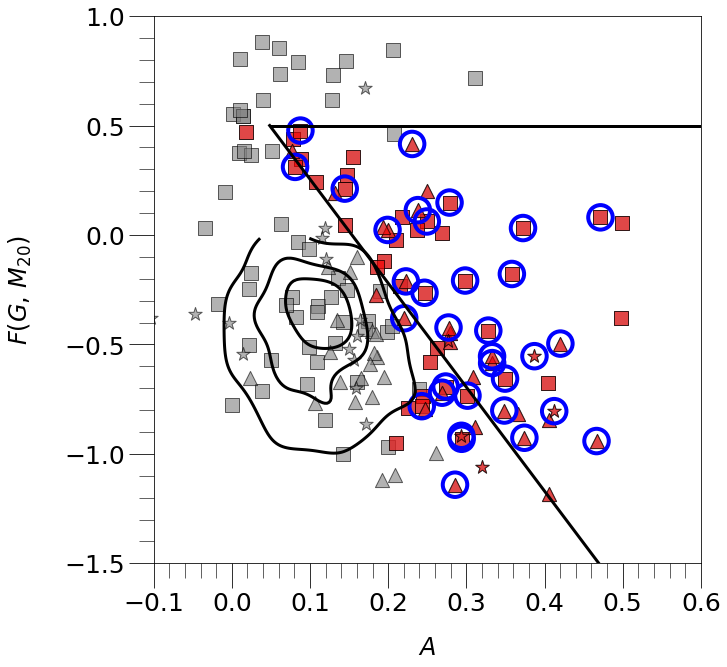}
\caption{The distribution of the RPS candidates (blue  open circles) from  Hydra,  CLoGS, and Fornax samples in the diagnostic diagrams: $A\,vs\,C$, $A\,vs\,n$, and $A\,vs\, F(G\_M_{20})$ plots. 
 The solid black lines delineate the  morphological transition zone boundaries in each diagnostic diagram.
 In red are the galaxies which fall inside the  morphological transition zones in the three plots, while in grey are those that do not. The purple
iso-density contours are 10, 50, and 80th percentiles of the
density map of the isolated sample.}
\label{fig_candidates}
\end{figure*}

In order to identify new RPS candidates in the  Hydra, CLoGS, and Fornax samples, we utilized the diagnostic diagrams established in Section \ref{jelly_zone}. The three diagnostic diagrams for these samples are displayed in Fig.~\ref{fig_candidates}, with the red markers indicating galaxies that fall within the morphological transition zones and the grey markers representing those that do not. Using these diagrams, we identified a total of 70  potential RPS  candidates.  It is important to mention that the three diagnostic diagrams selected almost the same fraction of galaxies, and all objects that felt in any of the RPS zones of these three diagrams were individually considered. The first 10 possible candidates are listed in Table \ref{tab:Jelly_cand1}, while the remaining ones are listed in {\bf the supplementary material (Table 1)}. These tables list the identification of the galaxy, the cluster to which the galaxy belongs, the equatorial coordinates (RA, DEC), our activity classification (i.e., Jellyfish, Tidal, etc.), and the computed values of $A$, $C$, $n$, and $F(G,\,M_{20})$. In addition, for interacting galaxies, there is a column (intec) indicating if they are either Major or Minor mergers, and a last column (sep) gives the projected distance between the interacting pair.

We carefully examine each potential RPS candidate to eliminate objects that possibly exhibit morphological perturbations that could be a result of gravitational interactions or another factor different from the stripping.  To achieve this, we conducted a visual methodical inspection and removed galaxies that displayed bi-symmetric perturbations.  We focused on analyzing the residual images of the galaxies after the subtraction of the 2D S\'ersic model (e.g., right panels of Fig.~\ref{model_JC5}).  In these maps is easier to look for jellyfish/tidal features, since the optical disc is subtracted,  while the internal asymmetric structures are uncovered and the outskirts structures are enhanced.

 Upon our examination, we identified 33 RPS candidates, resulting in a success rate of 47\%, which is at least twice the value reported by \citet{2016MNRAS.455.2994M} (less than 20\%). It is important to mention that we have removed the term ``potential'' and henceforth refer to them simply as candidates. Table \ref{tab:Jelly} lists these candidates.  Additionally, it is worth noting that if we had opted for a more restrictive selection approach, considering only the galaxies falling within the morphological transition zone in all three diagnostic diagrams, we would have lost 36\% of the classified RPS candidates. This outcome aligns with the findings reported by \citet{2016MNRAS.455.2994M}. Considering the results, it becomes evident that combining all three diagnostic diagrams provides the optimal approach for selecting RPS candidates.

Of the remaining possible candidates, 37  were false positives, with a failure rate of 53\%. Most of these false positives (17) were interacting galaxies. The identified interacting spirals (6)  exhibit long tidal tails, while interacting ellipticals (11) present shells, ripples, or other inter structures resulting from the disruption of galaxies (we referred to these as Early\_Blue\_Tidals). The other large group of false positives (14) were edge-on galaxies that fell within the transition zone due to dust lane structures, resulting in high values of asymmetry. We also identified 4 spirals with high star-forming activity (classified as Spiral\_SF), 1 Low Surface Brightness galaxy (LSB), and 1 Irregular galaxy (Irr).

It is crucial to emphasize the significance of a meticulous examination process to distinguish the RPS candidates from false positives. Additionally, it is worth noting that the delineated morphological transition zones not only serve as a valuable tool for selecting RPS candidates but also hold importance in identifying galaxies undergoing morphological transformations within the galaxy cluster. Consequently, these zones can be utilized to study the evolution of galactic morphology within clusters, providing valuable insights into the process of morphological transformation. 
 
 During this visual inspection, we not only identified RPS candidates but also analyzed their observed jellyfish features. These included (i) the classification of the geometry of the stripping and (ii) the structure of the spiral arms.  The details of these classifications are given in the next unnumbered subsections.  It is important to note the classification of these jellyfish features represents  ``instantaneous'' pictures of a transient phenomenon. For instance, \cite{2005A&A...433..875R} identified three distinct phases during the RPS process: the instantaneous stripping phase, the dynamic intermediate phase, and the quasi-stable continuous viscous stripping phase. For example,  in the first phase, which occurs over a timescale of 20 to 200 Myr,  the outer part of the gas disc is displaced but only partially unbound. Also,  the intensity of the wake and the final set-up of the unbounded gas depends on several factors \citep{2006MNRAS.369..567R}. These include the density of ICM, the velocity of galactic wind (Match number of the flow) and its variability, the potential of the host galaxy (mass, size), the amount of gas available and its density, and its distribution in the galactic disc (e.g. \citealt{1999MNRAS.308..947A, 2006MNRAS.369..567R, kronberger08, jachym09,kapferer2009, boselli22}). Therefore, the classification given here should be considered as a rough approximation of a complex phenomenon.

\begin{table*}
\caption{RPS candidates}
\label{tab:Jelly}
\begin{tabular}{llllllllllcl}
\hline
\noalign{\smallskip}
Galaxy  & Cluster & View & Field  &  Unwinding?  & $i$ & $\alpha_{{\rm inner}}$   
& $\alpha_{{\rm outer}}$  &  $\Delta\alpha$ &  ``{\it Effects}''  & Ref.\\ 
\hline
\noalign{\smallskip}
6dFGS\,gJ102734.0-234711 & Hydra & Edge-on & Face-on & -- & 67.2 & -- & --  & -- &&\\
ESO\,302-14 & Hydra  & Face-on & Face-on & -- & 51.9 & -- & --  & -- & Tidal & 1 \\
ESO\,436-29 & Fornax  & Face-on  & Face-on & Unwinding & 25.9 & 23.7 & 29.3  & 5.6 &&\\
ESO\,437-043 & Hydra & Edge-on & Face-on & -- & 63.2 & -- & --  & --&&\\
ESO\,437-37 & Hydra  & Face-on & Face-on & -- & 48.6 & -- & --  & --&&\\
ESO\,437-64 & Hydra  & Face-on & Face-on & Unwinding & 43.5 & 20.7 & 27.8  & 7.1&&\\
ESO\,444-12 & LGG\,351   & Face-on & Face-on & Unwinding & 43.3 & 12.7 & 28.9  & 16.2&&\\
ESO\,500-42 & Hydra  & Edge-on & Edge-on & -- & 64.7 & -- & --  & --&&\\
ESO\,501-22 & Hydra  & Edge-on & Face-on & -- & 73.1 & -- & --  & --&&\\
ESO\,501-65 & Hydra  & Face-on & Edge-on & Unwinding & 49.3 & 33.6 & 46.9  & 13.3&  RPS+Tidal & 2\\
ESO\,501-74 & Hydra  & Edge-on & Edge-on & -- & 65.8 & -- & --  & --&&\\
ESO\,501-9  & Hydra  & Edge-on & Face-on & -- & 60.5 & -- & --  & --&&\\
ESO\,507-28 & LGG\,310 (CLoGS) & Face-on & Face-on & -- & 55.2 & -- & --  & --&&\\
IC\,2537    & LGG\,185 (CLoGS) & Face-on & Face-on & Unwinding & 49.5 & 13.1 & 44.1  & 31.0&&\\
IC\,4248    & LGG\,351 (CLoGS) & Face-on & Edge-on & Unwinding & 40.6 & 4.2 & 40.6  & 36.4&  RPS+Tidal & 3 \\
IC\,4275    & LGG\,351 (CLoGS) & Face-on & Edge-on & Unwinding & 53.5 & 17.1 & 21.0  & 3.9&   ? &3\\
IC\,4397     & LGG\,383 (CLoGS)  & Face-on & Face-on & -- & 43.8 & -- & --  & --&&\\
LEDA\,753354 & Hydra  & Face-on & Edge-on & -- & 47.5 & -- & --  & --& RPS?& 4\\
MCG-01-10-036 & LGG\,103 (CLoGS) & Edge-on & Edge-on & -- & 75.1 & -- & --  & --&&\\
MCG-04-25-054 & Hydra & Edge-on & Edge-on & -- & 77.1 & -- & --  & --&&\\
NGC\,1376     & LGG\,103 (CLoGS) & Face-on & Face-on & Unwinding & 10.4 & 15.9 & 19.2  & 3.3&Tidal&5\\
NGC\,1418     & LGG\,103 (CLoGS) & Face-on & Face-on & -- & 54.2 & -- & --  & --&&\\
NGC\,2939     & LGG\,127 (CLoGS) & Edge-on & Edge-on & Unwinding & 72.6 & 17.5 & 31.1  & 13.6&&\\
NGC\,5635     & Hydra & Edge-on & Face-on & -- & 65.8 & -- & --  & --&&\\
NGC\,5735     & LGG\,383 (CLoGS) & Face-on & Face-on & Unwinding & 38.8 & 13.3 & 26.6  & 13.3&&\\
NGC\,5989     & LGG\,383 (CLoGS) & Face-on & Edge-on & -- & 33.2 & -- & --  & --&&\\
NGC\,1427A    & LGG\,402 (CLoGS) & Face-on & Face-on & -- & 54.4 & -- & --  & --& RPS+Tidal & 6\\

NGC\,1437A & Fornax & Face-on & Edge-on & Unwinding & 45.0 & 11.5 & 34.1  & 22.6& RPS+Tidal & 6\\
NGC\,3314A & Fornax & ? & ? & -- & ? & -- & --  & --&RPS & 4\\
UGC\,724   & LGG\,18 (CLoGS)   & Face-on & Edge-on & Unwinding & 41.4 & 16.0 & 28.7  & 12.7&&\\
UGC\,8199  & LGG\,329 (CLoGS) & Edge-on & Edge-on & -- & 61.9 & -- & --  & --&&\\
UGC\,8200  & LGG\,329 (CLoGS) & Face-on & Face-on & Unwinding & 50.7 & 33.8 & 43.4  & 9.6&&\\
UGC\,9302  & LGG\,383 (CLoGS) & Edge-on & Edge-on & -- & 77.2 & -- & --  & --&&\\
\noalign{\smallskip}
\hline
\noalign{\smallskip}
\end{tabular}
\begin{minipage}[r]{16.5cm}
References: (1) \citet{2020A&A...643A.147D};(2)\citet{2021ApJ...915...70W};
(3)  \citet{2019MNRAS.489.5723F};  (4) \citet{2022A&A...668A.184H};
(5) \citet{2001AJ....121.1413P};  (6)   \citet{2023arXiv230211895S}
\end{minipage}
\end{table*}

 \subsubsection*{Geometry}
 

 \begin{figure*}
\centering
\includegraphics*[width=0.45\textwidth]{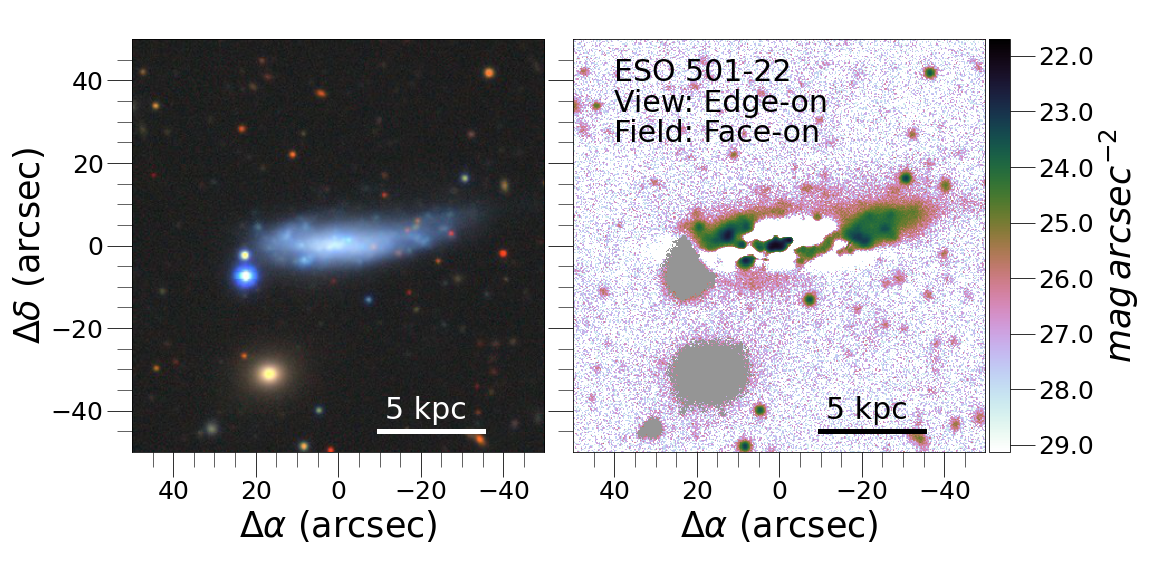}
\includegraphics*[width=0.45\textwidth]{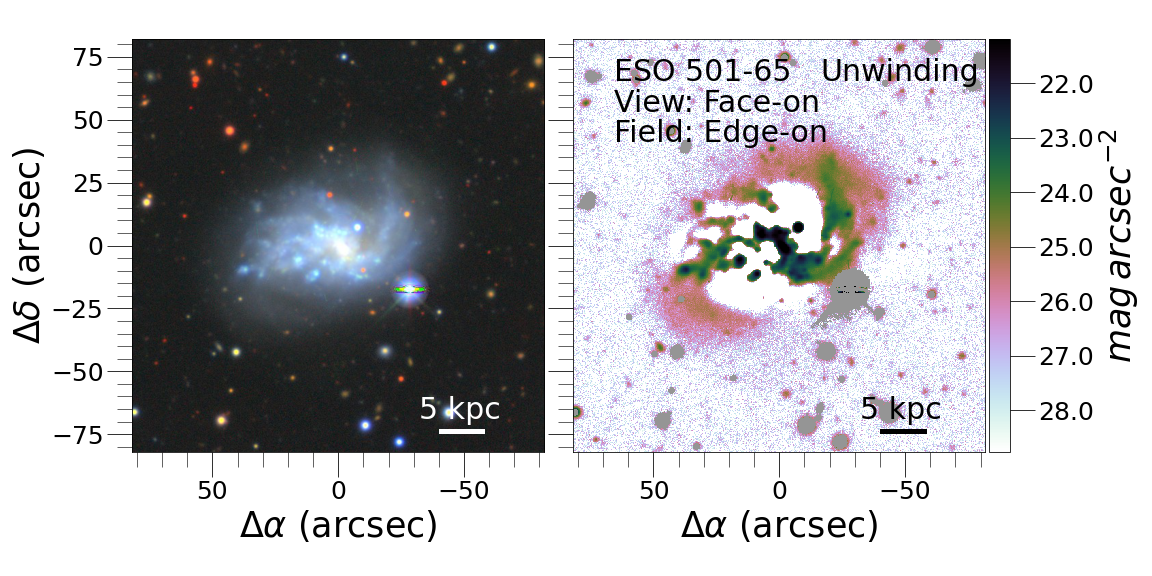}
\includegraphics*[width=0.45\textwidth]{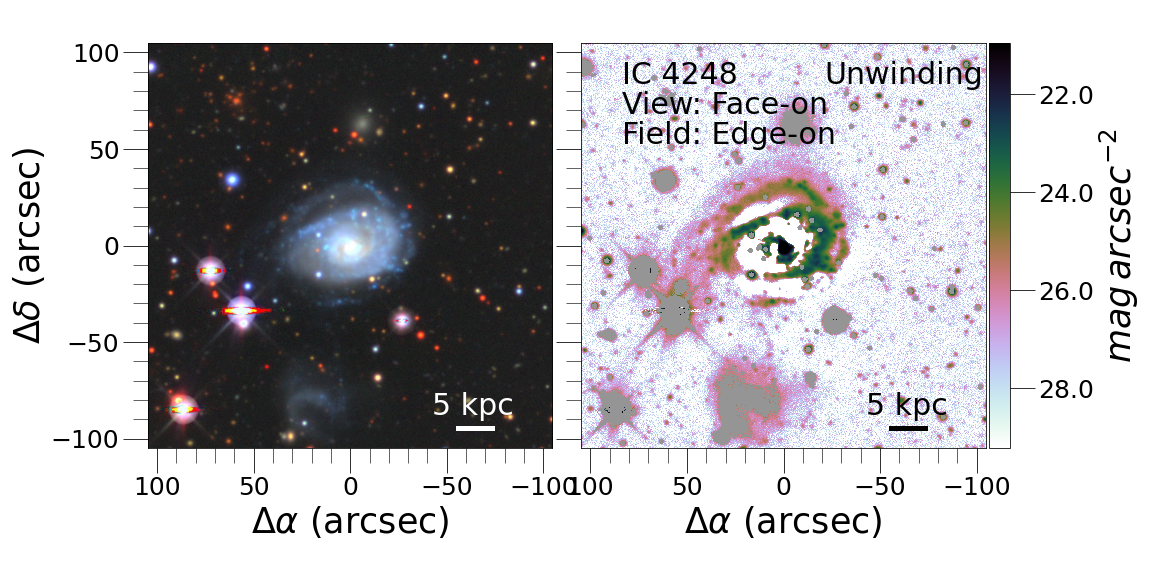}
\includegraphics*[width=0.45\textwidth]{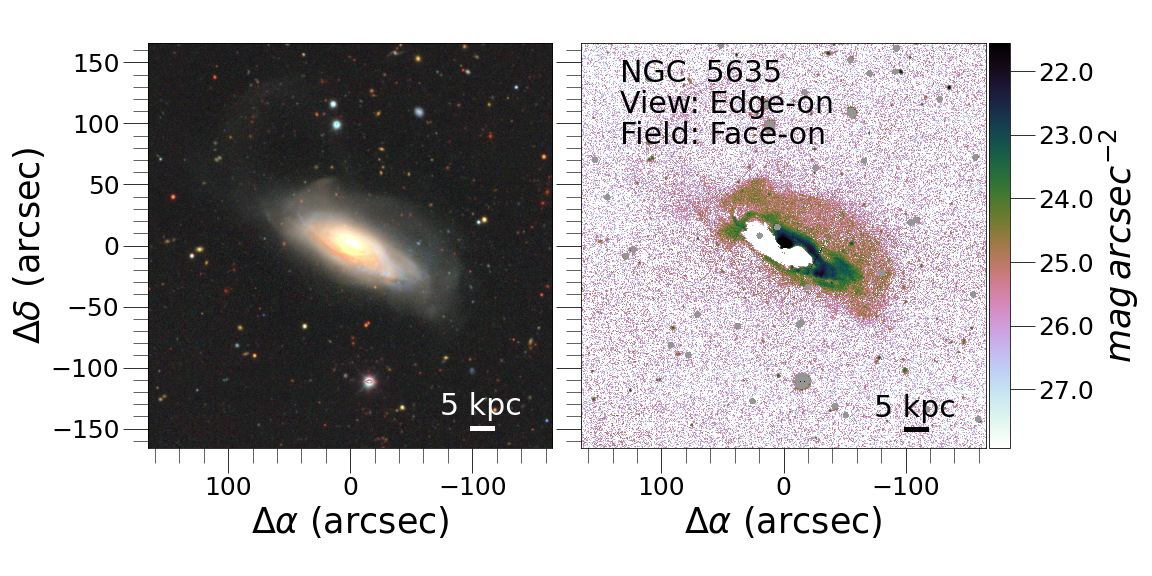}
\caption{Examples of new  RPS candidates found in this work. Each column contains the input image (left panel ) and residual image in $g$-band (right panel). 
The grey regions visible in the residual maps represent areas that have been intentionally masked during the analysis. These masks are applied to distinguish between various components in the data, such as segmentation maps for background galaxies and foreground stars. The directions are   North ($\delta$) up and East ($\alpha$) left.}
\label{fig_model_jelly}
\end{figure*}

The observed wake of the RPS galaxies is influenced by two critical angles: the observer view, which refers to the observer's line of sight, and the inclination angle between the galactic disc and the cluster wind direction, known as the stripping view. Thus, 
determining the geometry  of the galaxy is crucial in interpreting observational data accurately and gaining insights into the RPS effects on galaxy evolution
\citep{2006MNRAS.369..567R,jachym09}. For instance, the classic trailing tentacles in the jellyfish galaxies are produced when the stripping is face-on, but they are clearly visible in an edge-on observer view. When both the striping and observer view is face-on, the galaxies display an extended gas halo (see Fig.~5 of \citealt{2006MNRAS.369..567R}). On the other hand, galaxies suffering an edge-on stripping, when observed face-on, show an asymmetrical perturbation in the opposite direction of the ICM, whilst when they are seen edge-on exhibit a truncated disc in the leading side, and an extended and elongated galactic gas disc in the trailing side (see Fig.~6 of \citealt{2006MNRAS.369..567R}). To determine the observer view for the RPS candidates, we used the inclination angle derived from the fitted one-single Sérsic component model. Galaxies with a major/minor axis ratio greater than 0.5 ($i > 60 \deg$) were classified as edge-on, while those with a ratio less than or equal to 0.5 were considered face-on. Once the observer view was established, we analyzed the residual maps (e.g., left panel of Fig.~\ref{model_JC5}) to identify any jellyfish features and determine the most likely stripping view. This involved comparing the observed location and orientation of the features with the expected characteristics for different stripping scenarios as described previously.

Fig.~\ref{fig_model_jelly} shows the RGB images and residual maps of four RPS candidates, namely ESO\,501-22, ESO\,501-65, IC\,4248, and NGC\,5635. In each example, the residual map in the g-band clearly displays  RPS features. For instance, for ESO\,501-22, which is an edge-on galaxy, the residual map shows that it is undergoing face-on stripping. The stripped material is projected in the North direction and is escaping perpendicularly from the disc. On the other hand, ESO\,501-65 is a face-on spiral galaxy that is suffering edge-on stripping. The jellyfish "tentacles" are in the Northeast direction, and in this case, the leading side of the spiral (Southwest direction) is being compressed by the ram pressure. The stripping arms are also "unwinding" at larger radii, which will be analyzed in more detail in the next section. Similarly, IC\,4248 is another face-on spiral galaxy falling edge-on into the intergalactic medium (IGM) wind. The "tentacles" are trailing in the East direction and are opening at large radii as well. In the West direction, the leading side of the disc is also being compressed. Finally, NGC\,5635 is an edge-on galaxy that is undergoing face-on stripping. The stripping material is heading in a North-West direction and is being removed perpendicularly from the disc. We have displayed the RGB images and residual maps in the g-band for the remaining 29 candidates in {\bf the supplementary material (Fig. B1)}. The observer and stripping views are labeled in the residual maps, and this information is also listed in Table \ref{tab:Jelly}. It is worth noting that  the
RGB images, made from the combination of {\it g}, {\it r}, and {\it z} bands, proved to be essential along with the residual maps to identify the RPS featuring, in particular the patchy structures of the stripping tentacles.

 \subsubsection*{Unwinding galaxies }

\begin{figure*}
\centering
\includegraphics*[width=0.8\textwidth]{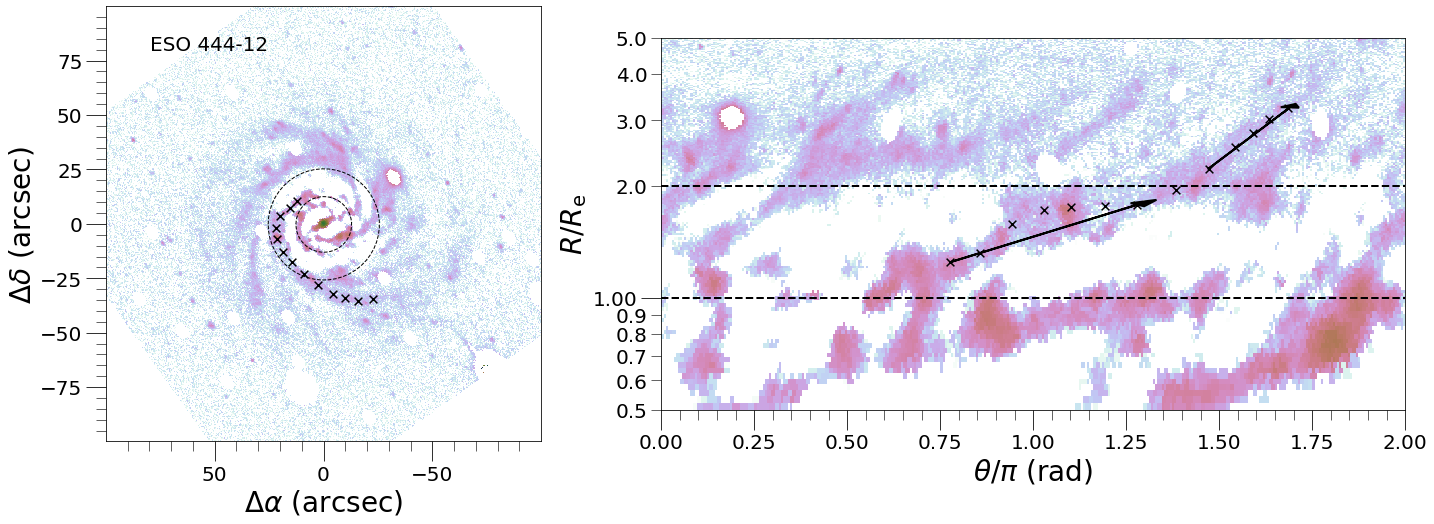}
\includegraphics*[width=0.8\textwidth]{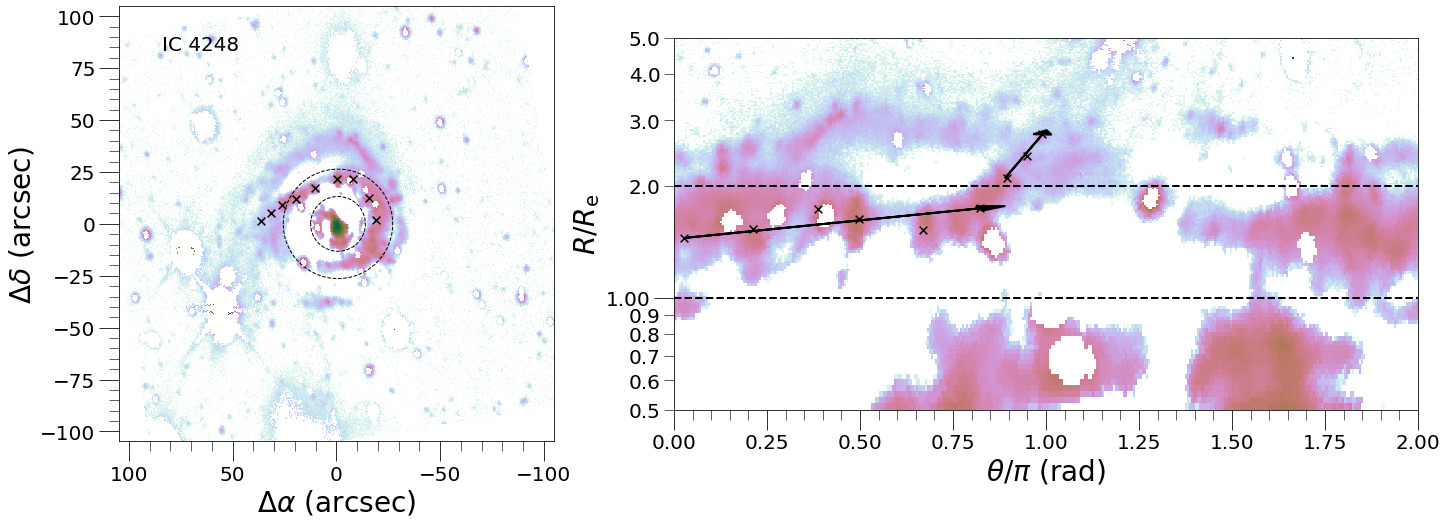}
\includegraphics*[width=0.8\textwidth]{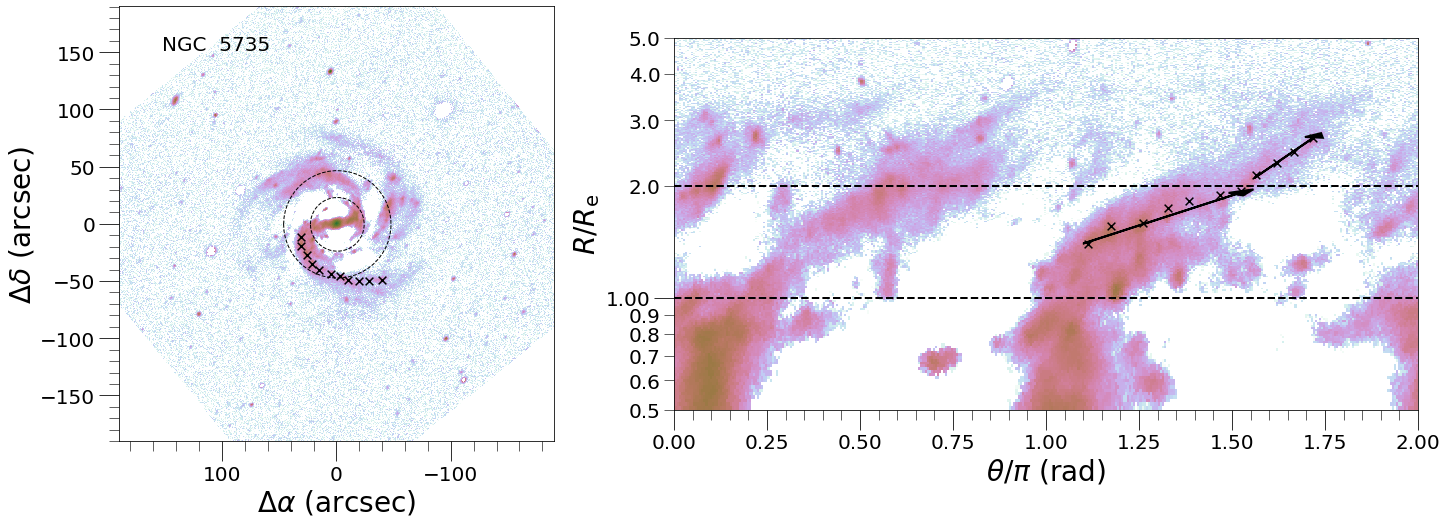}
\caption{ Examples of the unwinding RPS  candidates. The left panels contain the deprojected images in the $g$-bands and the right panels contain the radius {\it vs} $\theta$ plots. The ``x'' 's are the markers for the spiral arms for which were measured the aperture angles,  while the black arrows indicate the inner and outer ranges for the arms.  The directions are   North ($\delta$) up and East ($\alpha$) left.}
\label{fig_unwinding}
\end{figure*}

 In addition to the well-known characteristic features of jellyfish galaxies, ram pressure can also have the effect of "unwinding" the spiral arms of cluster galaxies \citep{2021MNRAS.500.1285B, vulcani22}. This phenomenon has been observed among face-on viewed galaxies with visible spiral arms in the GASP survey and studied in great detail by \citet{2021MNRAS.500.1285B}.  This work reveals that ram pressure can cause the outer parts of spiral arms to open up relative to the inner parts, and it was confirmed by measuring the pitch angle variation along the stripped spiral arms in a sample of 11 galaxies. Besides, the numerical simulations performed by \citet{2021MNRAS.500.1285B} demonstrate that this phenomenon can occur in both edge-on and face-on stripping scenarios.  In the former, due to the disc differential rotation, the effectiveness of stripping is also different resulting that the outer material being slowed down leading to the unwinding of the spiral arms. This unwinding is asymmetric, occurring only on the trailing side of the stripping.  In the face-on stripping, the gas removed from the outer edges falls to higher orbits with lower angular velocity, creating the appearance of unwinding. Unlike the edge-on scenario, the unwinding occurs around the disc in all directions.  It is worth noting that these effects are likely to be transient, lasting less than 0.5 Gyr and occurring only during the initial infall of the galaxy. Eventually, the effects disappear due to the action of ICM wind \citep{2021MNRAS.500.1285B}.

In our study, we identified RPS candidates that exhibit apparent unwinding arms. To confirm the presence of these features, we followed the same methodology as \citet{2021MNRAS.500.1285B}, which involved comparing the pitch angle measurements of the inner ($<2R_{\rm{e}}$) and outer ($>2R_{\rm{e}}$) parts of the stripping arms.
 we have performed this analysis over g-band images.
From these RPS candidates, we identified a total of 13 galaxies exhibiting these unwinding patterns.  The measured pitch angles for the inner ($\alpha_{{\rm inner}}$) and outer ($\alpha_{{\rm outer}}$) parts of the spiral arms, as well as their respective $\Delta\alpha$ are listed in Table \ref{tab:Jelly}.  We observed variations in pitch angles ranging from $\sim5^{\circ}$ to $\sim40^{\circ}$, similar to those found in the sample of \citet{2021MNRAS.500.1285B} (see their Table 2).  Out of the identified galaxies, seven are likely experiencing face-on stripping,  as indicated by the observed openness around the disc in all arms. Conversely,    six galaxies appear to be undergoing edge-on stripping, where unwinding is only observed in the trailing edges of their arms. Fig.~\ref{fig_unwinding} shows three of the unwinding galaxies: ESO\,444-12, IC\,4248, and NGC\,5735. The left panels present the deprojected images of these galaxies, generated by "stretching" the pixels along their projected minor axis using the same procedure as \citet{elme92}. From these images, we created $R$ {\it vs} $\theta$ plots (right panels), where the unwinding of the stripping arms is clearer, i.e., the slopes of the outer parts of the spirals are larger than the slopes of their inner parts.  To measure the pitch angle variations, we selected only one spiral arm in the deprojected images in which the unwinding is clear and manually placed "marks" along it (marked as "Xs" in the panels).  Then, the slopes were computed from the initial and end marks in each interval in the $R$ {\it vs} $\theta$ plots. For the inner parts of the arms, we considered marks until $<2R_{\rm{e}}$. Therefore, it is important to note that the values listed in Table 2 are rough estimations of the pitch angle variations. The deprojected images and $R$ {\it vs} $\theta$ plots for the remaining unwinding galaxies are shown {\bf in the   Fig. B2 of the supplementary material}, along with the markers used to measure the pitch angles. 

\subsubsection*{Comparison with previous studies}

Table \ref{tab:Jelly} also provides references for previously studied candidates and the corresponding scenarios (listed under the `` {\it Effects}'' column) proposed by 
these authors, which can be RPS, tidal interactions, or 
both.  ESO\,302-14, for example, was investigated by \cite{2020A&A...643A.147D} 
using \ion{H}{i} observations obtained with the MeerKAT radio telescope during the 
commissioning phase. They proposed an interaction scenario involving a low-mass 
companion that triggered recent star formation in the galaxy to explain the observed \ion{H}{i} morphology. However, the galaxy exhibits a one-sided trailing tail oriented towards the North-West direction (as observed in Fig.~2 and 10 of the corresponding work). The absence of a leading tail connecting both galaxies, if the small companion is the perturbing agent, contradicts the expected signature in the \ion{H}{i} contours. In contrast, within the context of the RPS scenario, the presence of a star-forming arc on the leading compressed side (as depicted {\bf in the Fig. B1 of the supplementary material}) lends greater plausibility to ram pressure stripping as the dominant mechanism, outweighing the possibility of tidal interactions.

\cite{2019MNRAS.489.5723F} investigated IC\,4248 and IC\,4275 within the context of the WALLABY survey \citep{wallaby}. They found that the \ion{H}{i} morphology of IC\,4248 exhibits a distorted structure with asymmetric features similar to the optical morphology, strongly suggesting RPS. However, the resolution of the data does not allow for kinematic confirmation. While gravitational effects from its companion galaxy NGC\,5135 were considered, the projected distance between them is approximately 250 kpc, making this interaction negligible and reinforcing the likelihood of RPS. As for IC\,4275, the \ion{H}{i} contour map (Fig. 9 in their work) reveals an asymmetric \ion{H}{i} distribution aligned with the detected unwinding arm (see Fig. B1 in the supplementary material), indicating a morphology consistent with RPS. However, they did not provide a specific scenario for this galaxy.

NGC\,1376 was included in the study conducted by \cite{2001AJ....121.1413P},
where a \ion{H}{i} asymmetric distribution was detected for this galaxy. Interestingly, this asymmetric distribution is in the same direction as the unwinding arms detected in our work (see  Figs. 3 and 8   of \cite{2001AJ....121.1413P}). They attributed the morphological perturbation to its compact dwarf companion, SBS\,0335-052. However, considering the projected distance between them (approximately 140 kpc), the RPS scenario is more likely for NGC\,1376 than a gravitational scenario.

A recent study by \citet{2023arXiv230211895S}  presented the results of 
 a deep and high-resolution MeerKAT survey of the Fornax cluster. The authors found that six galaxies within the cluster, including NGC\,1427A and NGC\,1437B, 
have long, one-sided, star-less \ion{H}{i} tails aligned radially within
the cluster.  As the ram pressure is expected to be relatively
weak in Fornax, the authors argue that \ion{H}{i} in the tails were pulled out of the stellar
body by tidal forces and, once there, became more susceptible to being further displaced by ram pressure.
That is, they supported the idea that there is a connection between tidal
interactions and the formation of ram-pressure \ion{H}{i} tails in Fornax. 

\cite{2022A&A...668A.184H} conducted  a multi-wavelength analysis of four galaxies (NGC\,3312, NGC\,3314A,  NGC\,3314B and 
LEDA\,753354) located in the center of the Hydra I galaxy cluster,  three of which are also included in our sample of RPS candidates. It should be noted that  NGC\,3314A  and NGC\,3314B are two overlapping spiral galaxies with a velocity difference of about 1850\,km/s, indicating that they are not physically related. In our study, it was only possible to estimate the non-parametric parameter for NGC\,3314A. These authors found evidence for RPS in  NGC\,3314A, and NGC\,3314B through the signature of  \ion{H}{i} contours.  Regarding LEDA\,753354, \cite{2022A&A...668A.184H} proposed that this particular galaxy may be at an early stage of experiencing ram pressure. The \ion{H}{i} contours (Fig. 4 in their study), revealed the presence of a faint tail extending in the trailing side and directed towards the North-East. Interestingly, in the opposite direction of this tail,  in what is the leading compressed side, 
there is a star-forming arc, which is observed {\bf in the residual maps shown in Fig. B1 in the supplementary material}. Then,
the optical morphology reinforces the  RPS scenario for 
 LEDA\,753354.

\subsubsection*{Caveats of the RPS selection}

The selected galaxies show promising indications of being candidates for RPS based on their morphological characteristics. However, in order to conclusively establish their physical origin, it is necessary to conduct spectroscopic follow-up observations.
  In particular, integral field unit (IFU) observations can be used to study the kinematics of both the gas and stellar components, providing crucial information to uncover the mechanisms behind the morphological perturbations, since the gravitational interactions act indistinctly on the  interstellar medium (ISM) and stars, whereas hydrodynamic interactions only affect the ISM \citep[e.g.][and reference therein]{Poggianti_2017,2021MNRAS.500.1285B,boselli22}. Additionally, observations of the ISM in the radio bands, tracing atomic and molecular gas, can  reveal the "invisible"  stripped material and its kinematics confirming the galaxy candidate as RPS  \citep[e.g.][]{2019MNRAS.489.5723F,2021ApJ...915...70W,2022MNRAS.510.1716R}. 
Nevertheless, these observations require a significant amount of telescope time, underscoring the importance of identifying good candidates to optimize observing time. Therefore, the diagnostic diagrams and the robustness of the visual inspection presented in this work could be useful in identifying compelling RPS candidates.

\section{Conclusions}
\label{conclusions}

Galaxies undergoing RPS present different  morphological signatures. In this paper, we present the results of the analysis of non-parametric and physical parameters in a sample of RPS galaxies by using $g$, $r$ and $z$ band images from  DESI Legacy Imaging Survey archive.  The sample of  RPS galaxies was taken from \citet{2016AJ....151...78P} and provides a classification into five disturbance classes (J1 to J5). A sample of interacting and isolated galaxies were included in this study, as well as a sample of galaxies belonging to clusters from the CLoGS project and of the Fornax,  and Hydra clusters. The latter samples of galaxy clusters  were considered to search for new candidates of RPS galaxies. The analysis of galaxy morphology was conducted using the {\sc astromorphlib} software developed by \citet{hernandez22}. The main findings are outlined below.

\begin{enumerate}

\item We analyzed the possible correlations between JClass classification and the measured physical and non-parametric parameters, using the Spearman correlation coefficient, $r$. We found that
the JClass values are correlated with six parameters, $A$, $M$, $S$, $R_{e}$ ,
$S(G,M_{20}$ and $M_{z}$ , mainly considering galaxies with JClass $\geq$ 3. The most stronger correlation is with the $A$ parameter $(r=0.41,$ p-values$6.3\times 10^{-7})$, being that the higher values of the JClass, the higher the asymmetry.

\item A Kolmogorov-Smirnov (KS) test  was performed for 105 possible pairs of the 15 computed parameters and then chosen to investigate combinations with high KS test values ($> 0.7$). The best pair combinations that show a good separation between RPS galaxies with JClass $\geq3$  and isolated galaxies are $C\,vs\,A$, $n\,vs\,A$, and $F(G,\,M_{20})\,vs\,A$. These pair combinations were called here as  diagnostic diagrams.

\item For each one of these diagnostic diagrams ($A\,vs\,C$, $A\,vs\,n$, and $A\,vs\,F(G,\,M_{20})$ ),   boundaries were defined that effectively differentiate between RPS  and isolated galaxies.  These boundaries were determined by utilizing the outermost iso-density contour of the 80th percentile of the isolated galaxy distributions as a reference.  The regions on the right side of these partition lines are referred to as morphological transition zones.

\item  We verified that the interacting galaxies occupy a large area in these diagnostic diagrams, overlapping with both isolated and morphological transition zones. However, the bulk of the interacting galaxy population falls within  the morphological transition zone.

\item The   diagnostic diagrams of $A\,vs\,C$, $A\,vs\,n$, and $A\,vs\,F(G,\,M_{20})$ were used  to identify new RPS candidates in the  Hydra, CLoGS, and Fornax samples, resulting  in a total of 70 new  potential RPS  candidates 

\item We  carefully visually examined   each  potential RPS candidate and  eliminated objects that could be exhibiting morphological perturbations as a result of gravitational interactions or another factor different from  stripping. A final sample of 33 new RPS candidates was  identified.

\item Our success rate of identifying RPS candidates is 47\% combining the three defined transition zones, which is at least twice the value reported by \citet{2016MNRAS.455.2994M} in similar analysis.  Considering the results, it is evident that combining all three diagnostic diagrams provides the optimal approach for selecting RPS candidates.

\item We found a total of 37  ``false'' positives (i.e., no RPS candidates) in the morphological transition zones. These ``false'' positives include mainly interacting galaxies but also edge-on galaxies and star-forming galaxies. Therefore, this morphological transition zones not only serve as a valuable tool for selecting RPS candidates but also hold importance in identifying galaxies undergoing morphological transformations within the galaxy cluster. Consequently, these zones can be utilized to study the evolution of galactic morphology within clusters.

\item Out of the total of 33 RPS candidates identified  in this final sample, 13 showed clear evidence of unwinding arms. These patterns were confirmed by  comparing the pitch angle measurements of the inner ($<2R_{\rm{e}}$) and outer ($>2R_{\rm{e}}$) parts of the stripping arms.

\item Out of the 33 selected candidates, 12 (36\%) are identified as edge-on galaxies, and among the 13 unwinding galaxies, 5 (38\%) exhibit an edge-on orientation. This observed proportion   in the distribution between edge-on and face-on galaxies does not appear to result from a selection bias. In the initial total sample of 160 galaxies and the potential sample of 70 galaxies within the transition zone, the edge-galaxies represent 54\% (87) and 56\% (39), respectively. Therefore, the observed proportion  could be attributed to random stripping-view variations in the sky rather than a selection bias. Nevertheless, it is crucial to emphasize that a more robust statistical analysis with a larger galaxy sample is necessary for a comprehensive understanding of this issue.

\end{enumerate}

As a final remark, the utilization of RGB images, generated by combining the {\it g}, {\it r}, and {\it z} band images, along with the incorporation of residual maps obtained through disc model subtraction, plays a vital role in effectively identifying features associated with ram pressure stripping. Moreover, it is crucial to consider the "geometry" of the galaxy, encompassing both the observer view and the orientation of the galactic disc relative to the ICM wind (stripping view). Additionally, the presence of unwinding signatures in the spiral arms serves as a key indicator. By taking all these factors into consideration, the objectivity of RPS identification is significantly enhanced.  Finally, the method presented in this paper provides an alternative and efficient approach to build large samples of interacting galaxies, complementing citizen science efforts like "Fishing for Jellyfish Galaxies" on Zooniverse \footnote{\url{https://www.zooniverse.org/projects/cbellhouse/fishing-for-jellyfish-galaxies}}.


\section*{Acknowledgements}

 We thank the anonymous referee for their helpful comments, questions, and suggestions on revising the manuscript. ACK thanks  FAPESP for the support grant 2020/16416-5 and the Conselho Nacional de Desenvolvimento Científico e Tecnológico (CNPq).  JAHJ acknowledges support from FAPESP, process number 2021/08920-8.  YJ acknowledges funding from  ANID BASAL Project No. FB210003. AC acknowledges the financial support provided by FAPERJ grant E-26/200.607 and 210.371/2022(270993). JPC acknowledges financial support from ANID through FONDECYT Postdoctorado Project 3210709. AVSC acknowledges financial support from Agencia I+D+i (PICT 2019-03299) and CONICET (PIP 1504).

The work made use the archive of The DESI Legacy Imaging Surveys.
The DESI Legacy Imaging Surveys consist of three individual and complementary projects: the Dark Energy Camera Legacy Survey (DECaLS), the Beijing-Arizona Sky Survey (BASS), and the Mayall z-band Legacy Survey (MzLS). DECaLS, BASS and MzLS together include data obtained, respectively, at the Blanco telescope, Cerro Tololo Inter-American Observatory, NSF’s NOIRLab; the Bok telescope, Steward Observatory, University of Arizona; and the Mayall telescope, Kitt Peak National Observatory, NOIRLab. NOIRLab is operated by the Association of Universities for Research in Astronomy (AURA) under a cooperative agreement with the National Science Foundation. Pipeline processing and analyses of the data were supported by NOIRLab and the Lawrence Berkeley National Laboratory (LBNL). Legacy Surveys also uses data products from the Near-Earth Object Wide-field Infrared Survey Explorer (NEOWISE), a project of the Jet Propulsion Laboratory/California Institute of Technology, funded by the National Aeronautics and Space Administration. Legacy Surveys was supported by: the Director, Office of Science, Office of High Energy Physics of the U.S. Department of Energy; the National Energy Research Scientific Computing Center, a DOE Office of Science User Facility; the U.S. National Science Foundation, Division of Astronomical Sciences; the National Astronomical Observatories of China, the Chinese Academy of Sciences and the Chinese National Natural Science Foundation. LBNL is managed by the Regents of the University of California under contract to the U.S. Department of Energy. The complete acknowledgments can be found at \url{https://www.legacysurvey.org/acknowledgment/}.

{\it Software:}   {\sc astromorphlib} \citep{hernandez22},
 {\sc statmorph} \citep{rodriguez19},  {\sc photutils} \citep{photutils}
{\sc astroplotlib} \citep{astroplotlib, hernandez13, hernandez15}, {\sc astropy} \citep{astropy:2013, astropy:2018, astropy:2022}, {\sc scipy} \citep{scipy}, {\sc numpy} \citep{numpy}, {\sc matplotlib} \citep{matplotlib}, {\sc pandas} \citep{pandas}, {\sc seaborn} \citep{seaborn},
{\sc jupyter} \citep{jupyter}, and {\sc bokeh} \citep{bokeh}.

\section*{Data Availability}

The data underlying this article will be shared on reasonable request
to the corresponding author.



\bibliographystyle{mnras}
\bibliography{krabbe} 




\appendix

\section{{\bf Extended Table  of galaxies in the morphological transition zone}.}

\begin{table*}
\caption{Galaxies in the morphological transition zones}
\label{tab:Jelly_cand2}
\begin{tabular}{llccllcccccc}
\hline
\noalign{\smallskip}
Galaxy  & Cluster &RA & DEC  &   Morph  & $A$  & $C$  
  & $n$  &  $F(G,\,M_{20})$   &  intec  &  sep \\
  &  &[deg] & [deg]  &  &  &  &   
  &   &  & [kpc] \\
\hline
\noalign{\smallskip}
2MASX\,J01223182+0916534 &  LGG\,23	(CLoGS) & 20.6326  & 9.2815   & Early\_Tidal	     & 0.08 & 3.86  & 1.6   & 0.44  & -      & - \\
2MASX\,J10284923-3129509 &  Hydra	 & 157.2051 & -31.4975 & Early\_Blue\_Tidal  & 0.41 & 1.99  & 0.5   & -1.18 & -      & - \\
2dFGRS\,TGS471Z004       &  Fornax	 & 49.8540  & -32.6493 & Irr  	     & 0.32 & 2.15  & 0.3   & -1.06 & -      & - \\
6dFGS\,gJ102734.0-234711 &  Hydra	 & 156.8911 & -23.7857 & Jellyfish 	     & 0.22 & 2.80  & 0.7   & -0.38 & Major  & 51.6 \\
6dFGS\,gJ103502.9-293024 &  Hydra	 & 158.7620 & -29.5066 & Early\_Blue\_Tidal  & 0.08 & 3.71  & 1.6   & 0.38  & -      & - \\
6dFGS\,gJ103704.4-312157 &  Hydra	 & 159.2685 & -31.3659 & Early\_Blue\_Tidal  & 0.19 & 3.10  & 1.0   & 0.04  & -      & - \\
ESO\,302-14              &  Fornax	 & 57.9204  & -38.4522 & Jellyfish 	     & 0.29 & 2.20  & 0.8   & -0.92 & -      & - \\
ESO\,436-29		 &  Hydra    & 157.597    & -30.393    & Jellyfish& 0.35 & 2.2 & 0.4 & -0.80 & -  & - \\
ESO\,437-043 &   Hydra & 160.414 & -27.777 & Jellyfish& 0.33 & 2.7 & 0.7 & -0.55 & -  & - \\
ESO\,437-30		 &  Hydra	 & 159.812    & -30.298    & Spiral\_Edge-on & 0.25 & 3.6 & 0.8 & 0.20 & -  & - \\

ESO\,437-35		 &  Hydra	 & 160.018    & -30.267    & Spiral\_Egde-on & 0.31 & 2.7 & 0.4 & -0.88 & -  & - \\
ESO\,437-37		 &  Hydra	 & 160.129    & -29.270    & Jellyfish& 0.20 & 3.1 & 1.0 & 0.02 & -  & - \\
ESO\,437-50		 &  Hydra	 & 160.879    & -30.772    & Spiral\_SF & 0.31 & 2.7 & 0.6 & -0.65 & Major  & 34.0 \\
ESO\,437-64		 &  Hydra	 & 162.273    & -29.375    & Jellyfish& 0.28 & 3.0 & 0.9 & -0.42 & -  & - \\
ESO\,444-12		 &  LGG\,351 (CLoGS)	 & 200.209    & -29.480    & Jellyfish& 0.14 & 3.2 & 1.0 & 0.21 & -  & - \\
ESO\,500-42		 &  Hydra	 & 156.834    & -23.805    & Jellyfish& 0.23 & 3.8 & 1.2 & 0.42 & Major  & 50.8 \\
ESO\,501-22		 &  Hydra	 & 158.840    & -27.696    & Jellyfish& 0.22 & 3.1 & 0.6 & -0.21 & -  & - \\
ESO\,501-32		 &  Hydra	 & 159.092    & -25.377    & Spiral\_Tidal & 0.28 & 2.6 & 0.6 & -0.49 & -  & - \\
ESO\,501-61		 &  Hydra	 & 159.524    & -25.094    & Early\_Blue\_Tidal & 0.37 & 2.3 & 0.8 & -0.82 & -  & - \\
ESO\,501-65		 &  Hydra	 & 159.639    & -27.737    & Jellyfish& 0.42 & 2.7 & 0.7 & -0.50 & -  & - \\

ESO\,501-74		 &  Hydra	 & 160.205    & -24.668    & Jellyfish& 0.37 & 2.2 & 0.3 & -0.93 & -  & - \\
ESO\,501-9		 &  Hydra	 & 158.242    & -27.670    & Jellyfish& 0.29 & 2.4 & 0.4 & -1.14 & -  & - \\
ESO\,501-96		 &  Hydra	 & 161.698    & -23.328    & Early\_Blue\_Tidal & 0.13 & 3.4 & 1.5 & 0.19 & -  & - \\
ESO\,507-28		 & LGG\,310	(CLoGS) & 192.941    & -26.090    & Jellyfish& 0.09 & 3.6 & 1.1 & 0.48 & Minor  & 29.2  \\
ESO\,507-35		 & LGG\,310 (CLoGS)     & 193.234    & -26.697    & Spiral\_Tidal & 0.24 & 2.5 & 1.0 & -0.73 & -  & - \\
ESO\,507-42		 & LGG\,310 (CLoGS)     & 193.400    & -26.294    & Spiral\_Edge-on & 0.15 & 3.5 & 1.2 & 0.27 & -  & - \\
ESO\,507-62		 & LGG\,310 (CLoGS)     & 194.846    & -27.427    & Spiral\_Edge-on & 0.25 & 2.7 & 0.6 & -0.58 & -  & - \\
IC\,401 		 & LGG\,126  (CLoGS)    &  76.082    & -10.077    & Spiral\_Edge-on & 0.50 & 3.6 & 1.0 & 0.05 & -  & - \\	   
IC\,2537		 & LGG\,185 (CLoGS)      &  150.966   & -27.571    & Jellyfish& 0.27 & 2.5 & 0.4 & -0.72 & -  & - \\		   
IC\,4248		 & LGG\,351	(CLoGS) &  201.697   & -29.881    & Jellyfish& 0.37 & 3.6 & 0.7 & 0.03 & -  & - \\

IC\,4275		 & LGG\,351	(CLoGS) &  202.964   & -29.732    & Jellyfish& 0.24 & 2.5 & 0.7 & -0.78 & -  & - \\		   
IC\,4397		 & LGG\,383  (CLoGS)    &  214.495   & 26.413     & Jellyfish& 0.30 & 3.1 & 0.6 & -0.21 & -  & - \\		   
LEDA\,722049		 & Hydra	 &  160.790   & -30.050    & Spiral\_SF & 0.28 & 2.8 & 1.0 & -0.45 & -  & - \\	   
LEDA\,753354		 & Hydra	 &  159.343   & -27.545    & Jellyfish& 0.47 & 1.8 & 0.2 & -0.94 & INDEF  & 31.0 \\	   
MCG-01-10-036		 & LGG\,103 (CLoGS)     &  56.741    & -3.462     & Jellyfish& 0.25 & 3.4 & 0.6 & 0.06 & -  & - \\		   
MCG-03-34-014		 & LGG\,338	(CLoGS) &  198.148   & -17.542	   & Spiral\_Edge-on & 0.23 & 2.9 & 0.5 & -0.79 & -  & - \\	 
MCG-04-25-054		 & Hydra	 &  159.858   & -23.755	   & Jellyfish& 0.24 & 3.6 & 0.7 & 0.11 & -  & - \\  		 
ESO\,500-14          & Hydra     & 156.131    & -23.553    & Early\_Tidal & 0.18 & 2.9 & 0.9 & -0.27 & -  & - \\
NGC\,489 		 & LGG\,23 (CLoGS)	 & 20.475     & 9.207	   & Spiral\_Edge-on & 0.09 & 3.6 & 1.1 & 0.35 & -  & - \\
NGC\,518 		 & LGG\,23 (CLoGS)	 & 21.068     & 9.325	   & Early\_Edge-on & 0.22 & 3.5 & 0.9 & 0.08 & -  & - \\

NGC\,586 		 & LGG\,27 (CLoGS)	 & 22.904     & -6.894     & Early\_Edge-on & 0.16 & 3.6 & 1.2 & 0.35 & Minor  & 37.1 \\
NGC\,600 		 & LGG\,27 (CLoGS)      & 23.272     & -7.312     & Spiral\_SF & 0.21 & 2.6 & 0.4 & -0.23 & -  & - \\
NGC\,1341		 & Fornax     & 51.992     & -37.149    & Spiral\_Tidal & 0.28 & 2.7 & 0.9 & -0.48 & -  & - \\
NGC\,1376		 & LGG\,103 (CLoGS)      & 54.273     & -5.044     & Jellyfish& 0.35 & 2.6 & 0.5 & -0.66 & -  & - \\
NGC\,1417		 & LGG\,103 (CLoGS)     & 55.489     & -4.705     & Spiral\_Tidal & 0.24 & 3.5 & 0.8 & 0.02 & -  & - \\
NGC\,1418		 & LGG\,103 (CLoGS)     & 55.567     & -4.731     & Jellyfish& 0.30 & 2.7 & 0.5 & -0.73 & -  & - \\
NGC\,1449		 & LGG\,103 (CLoGS)      & 56.513     & -4.138     & Early\_Tidal & 0.02 & 3.9 & 1.8 & 0.47 & -  & - \\
NGC\,2939	         & LGG\,127	(CLoGS)  & 144.533 & 9.524   & Jellyfish& 0.47 & 2.9 & 0.8 & 0.08 & -  & - \\
NGC\,3383		 & Hydra	  & 161.830 & -24.438  & Spiral\_SF & 0.25 & 2.5 & 0.6 & -0.80 & -  & - \\
NGC\,3625		 & LGG\,232	(CLoGS)  & 170.130 & 57.781  & Spiral\_Edge-on & 0.19 & 2.8 & 0.7 & -0.12 & -  & - \\
NGC\,3652		 & LGG\,236	(CLoGS)  & 170.663 & 37.765  & Spiral\_Tidal & 0.50 & 2.4 & 0.4 & -0.38 & -  & - \\
NGC\,3669		 & LGG\,232 (CLoGS) 	  & 171.362 & 57.721  & Spiral\_Edge-on & 0.21 & 2.9 & 0.4 & -0.95 & -  & - \\
NGC\,5635		 & LGG\,383	(CLoGS)  & 217.132 & 27.409 & Jellyfish& 0.28 & 3.9 & 1.1 & 0.15 & -  & - \\
NGC\,5653		 & LGG\,383	(CLoGS)  & 217.543 & 31.215 & Early\_Tidal & 0.27 & 2.9 & 1.3 & 0.01 & -  & - \\
NGC\,5735		 & LGG\,383	(CLoGS)  & 220.639 & 28.726 & Jellyfish& 0.33 & 2.6 & 0.6 & -0.44 & -  & - \\
NGC\,5989		 & LGG\,402	(CLoGS)  & 235.387 & 59.755 & Jellyfish& 0.29 & 2.2 & 0.5 & -0.93 & -  & - \\
NGC\,1427A		 & Fornax	  & 55.038 & -35.626  & Jellyfish& 0.41 & 2.3 & 0.4 & -0.80 & INDEF  & 47.9 \\
NGC\,1437A		 & Fornax	  & 55.758 & -36.273  & Jellyfish& 0.39 & 2.5 & 0.4 & -0.55 & -  & - \\
NGC\,3314A & Fornax & 159.304 & -27.684 &  Jellyfish& 0.33 & 2.7 & 0.6 & -0.59 & -  & - \\
UGC\,724		 & LGG\,18	  & 17.498 & 32.368 & Jellyfish& 0.08 & 3.4 & 1.6 & 0.31 & -  & - \\
\noalign{\smallskip} 
\hline
\noalign{\smallskip}

\end{tabular}
\end{table*}

\begin{table*}
\contcaption{Galaxies in the jellyfish zone}
\begin{tabular}{llccllcccccc}
\hline
\noalign{\smallskip}
Galaxy  & Cluster &RA & DEC  &   Morph  & $A$  & $C$  
  & $n$  &  $F(G,\,M_{20})$   &  intec  &  sep \\
  &  &[deg] & [deg]  & &  &  &  
  &   &  & [kpc] \\
\hline
\noalign{\smallskip}
UGC\,743		 & LGG\,18	(CLoGS)  & 17.827 & 31.889 & Early\_Tidal & 0.11 & 3.4 & 1.2 & 0.24 & -  & - \\
UGC\,803		 & LGG\,23 (CLoGS)       & 18.845 & 8.099  & LBS & 0.26 & 2.9 & 1.4 & -0.52 & -  & - \\
UGC\,1980		 & LGG\,66	(CLoGS)  & 37.640 & 32.176 & Spiral\_Edge-on & 0.21 & 3.0 & 0.8 & -0.02 & -  & - \\
UGC\,2032		 & LGG\,61	(CLoGS)  & 38.352 & 20.579 & Spiral\_Edge-on & 0.14 & 3.5 & 0.6 & 0.05 & -  & - \\
UGC\,6433		 & LGG\,236	(CLoGS)  & 171.383 & 38.061 & Early\_Blue\_Tidal & 0.40 & 3.0 & 0.7 & -0.68 & -  & - \\
UGC\,8199		 & LGG\,329 (CLoGS)       & 196.688 & 35.101 & Jellyfish& 0.27 & 2.6 & 0.6 & -0.69 & Major  & 59.9 \\
UGC\,8200		 & LGG\,329	(CLoGS)  & 196.733 & 35.132 & Jellyfish& 0.36 & 2.9 & 0.7 & -0.18 & Major  & 57.8 \\
UGC\,9253		 & LGG\,383	(CLoGS)  & 216.744 & 31.517 & Spiral\_Edge-on & 0.19 & 3.2 & 0.5 & -0.15 & -  & - \\
UGC\,9302		 & LGG\,383 (CLoGS)       & 217.291 & 31.799 & Jellyfish& 0.25 & 3.4 & 0.6 & -0.26 & -  & - \\
WPVS\,78		 & Hydra	  & 159.672 & -25.592& Spiral\_Tidal & 0.41 & 2.1 & 0.8 & -0.84 & -  & - \\

\noalign{\smallskip} 
\hline
\noalign{\smallskip}

\end{tabular}
\end{table*}

\section{Residual images of the RPS candidates.}

\begin{figure*}
\centering
\includegraphics*[width=0.45\textwidth]{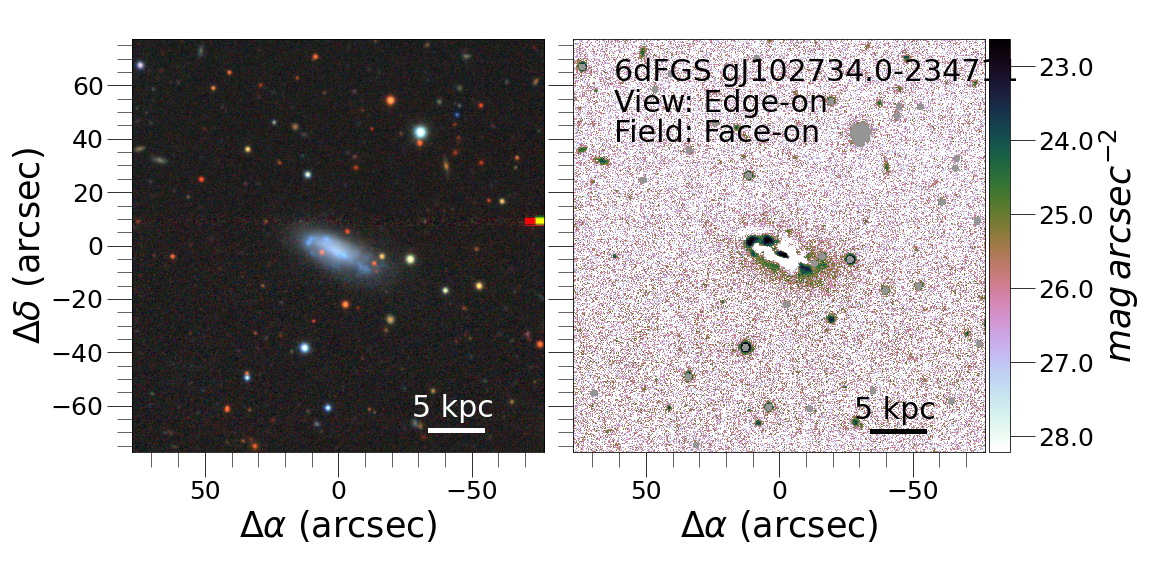}
\includegraphics*[width=0.45\textwidth]{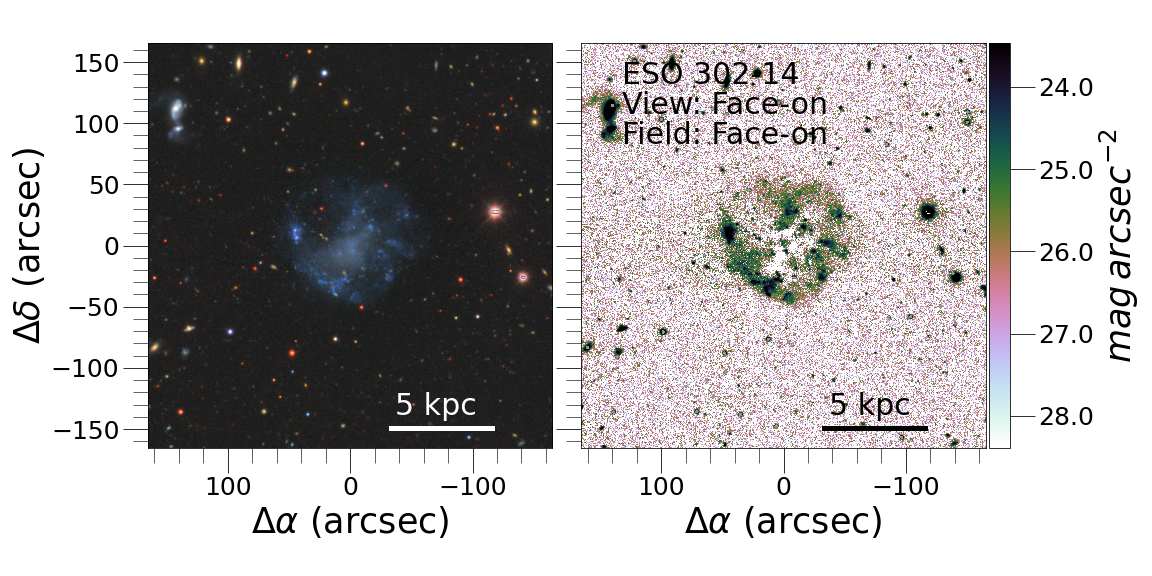}
\includegraphics*[width=0.45\textwidth]{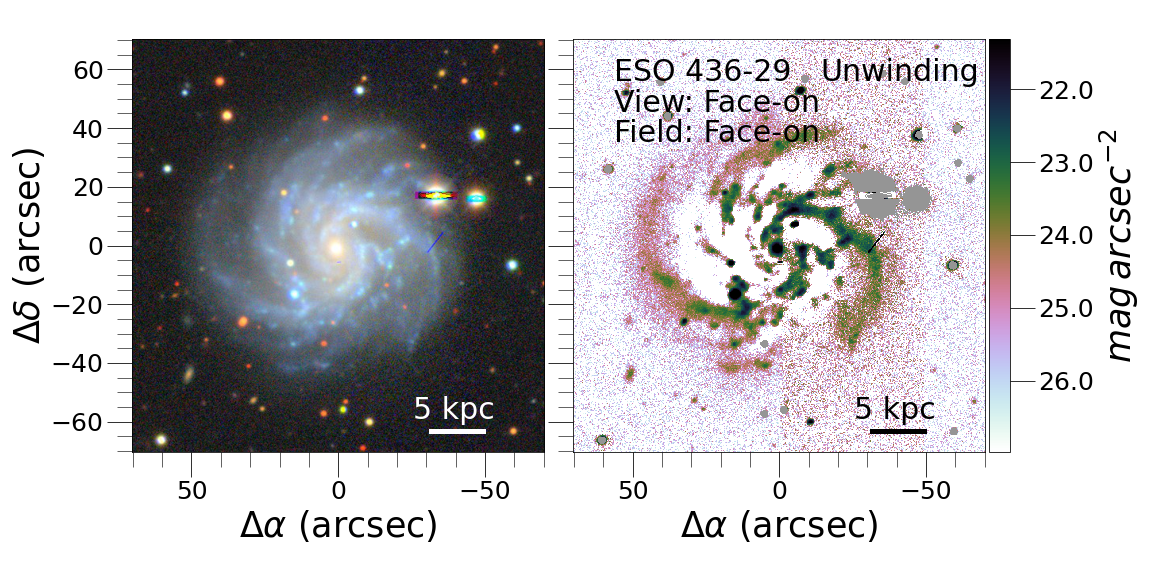}
\includegraphics*[width=0.45\textwidth]{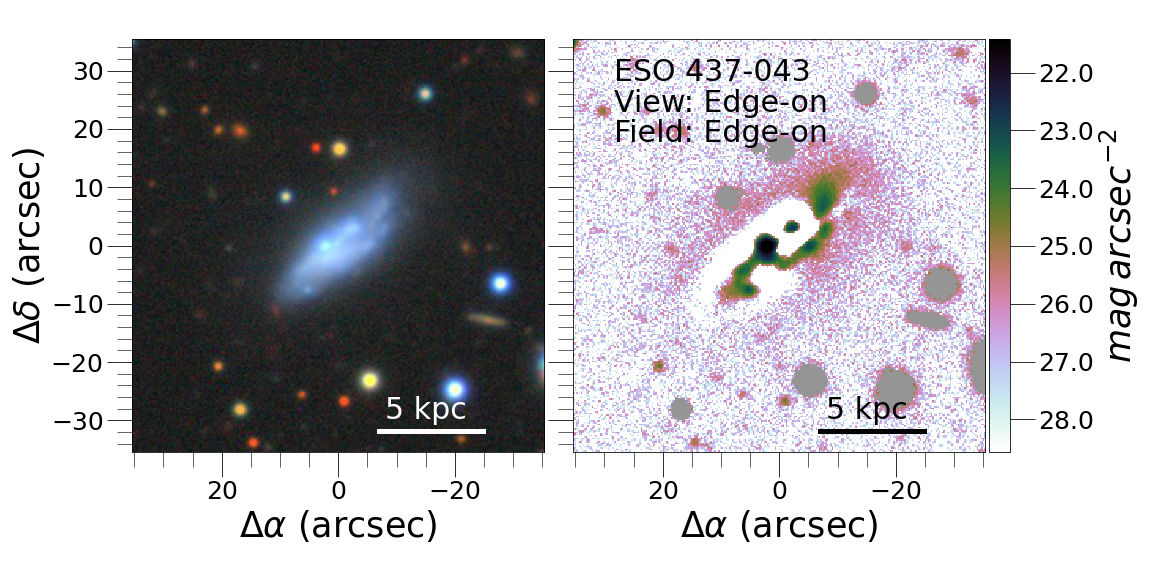}
\includegraphics*[width=0.45\textwidth]{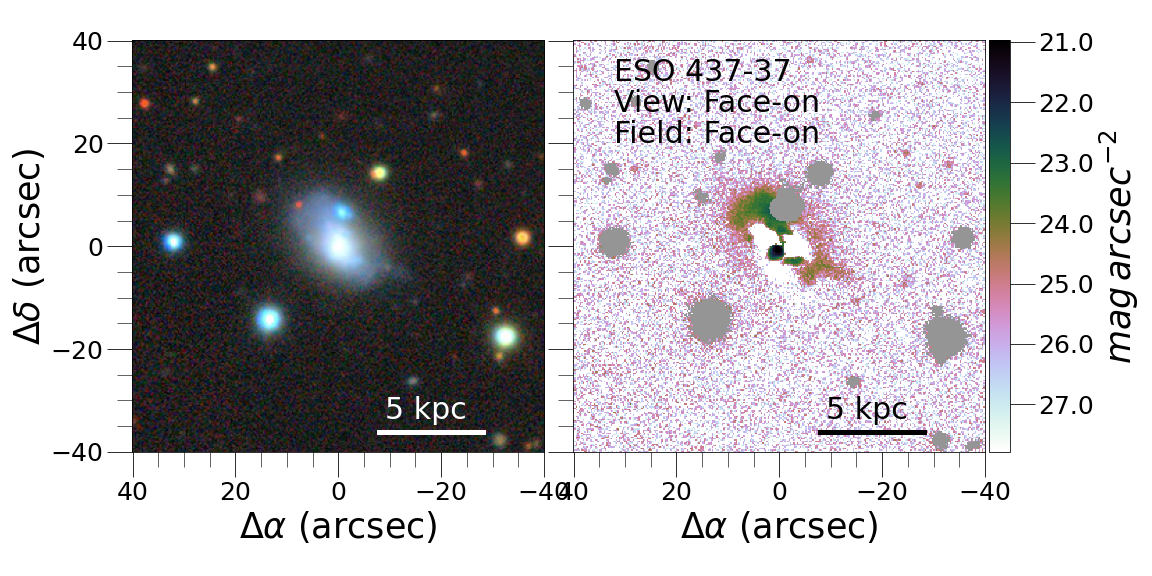}
\includegraphics*[width=0.45\textwidth]{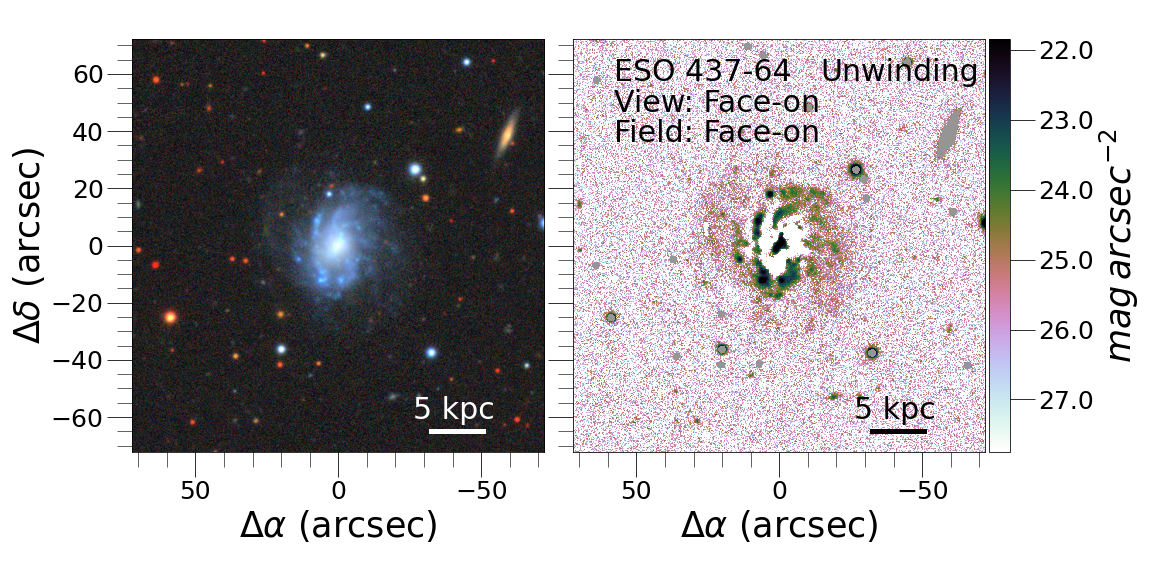}
\includegraphics*[width=0.45\textwidth]{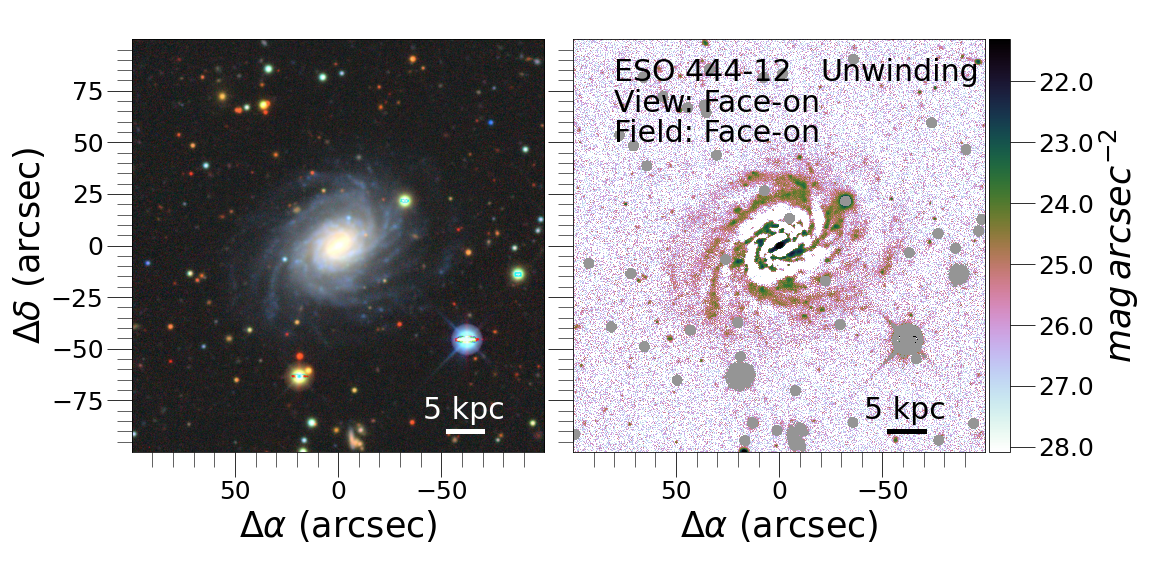}
\includegraphics*[width=0.45\textwidth]{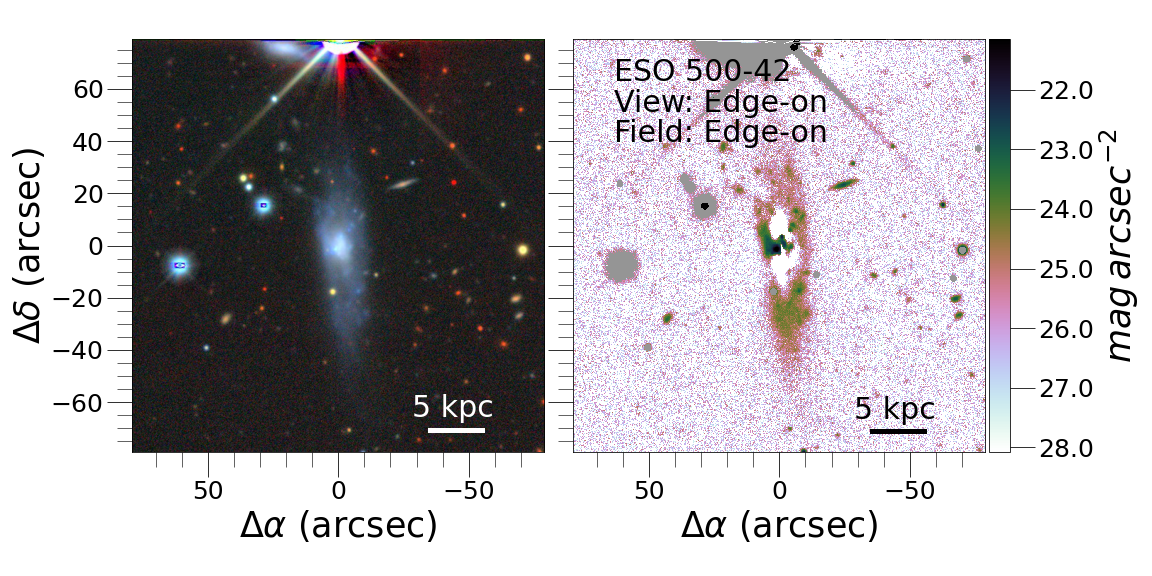}
\includegraphics*[width=0.45\textwidth]{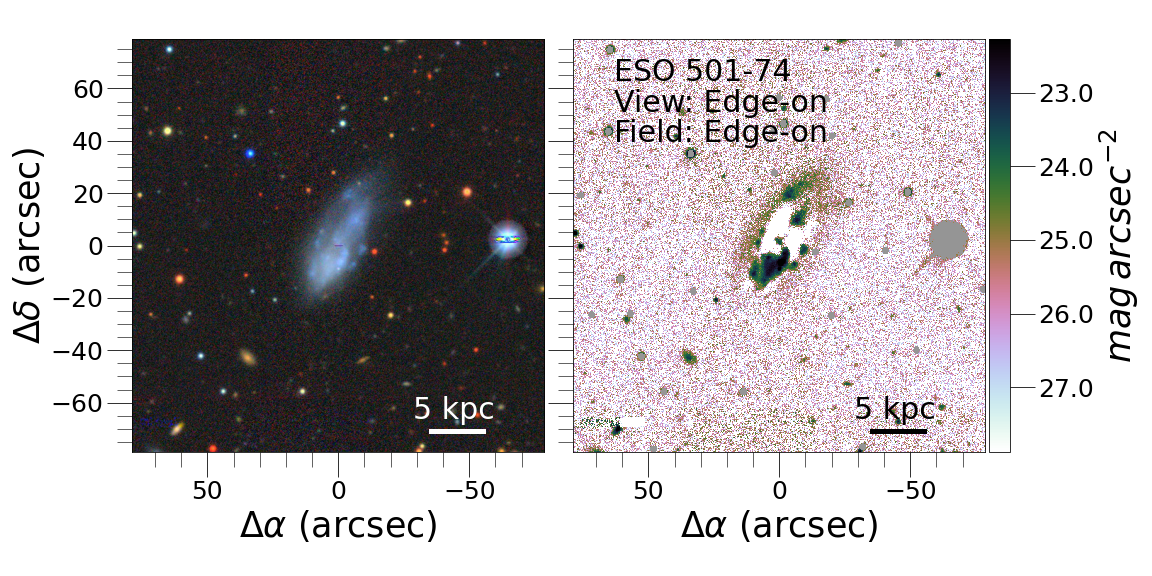}
\includegraphics*[width=0.45\textwidth]{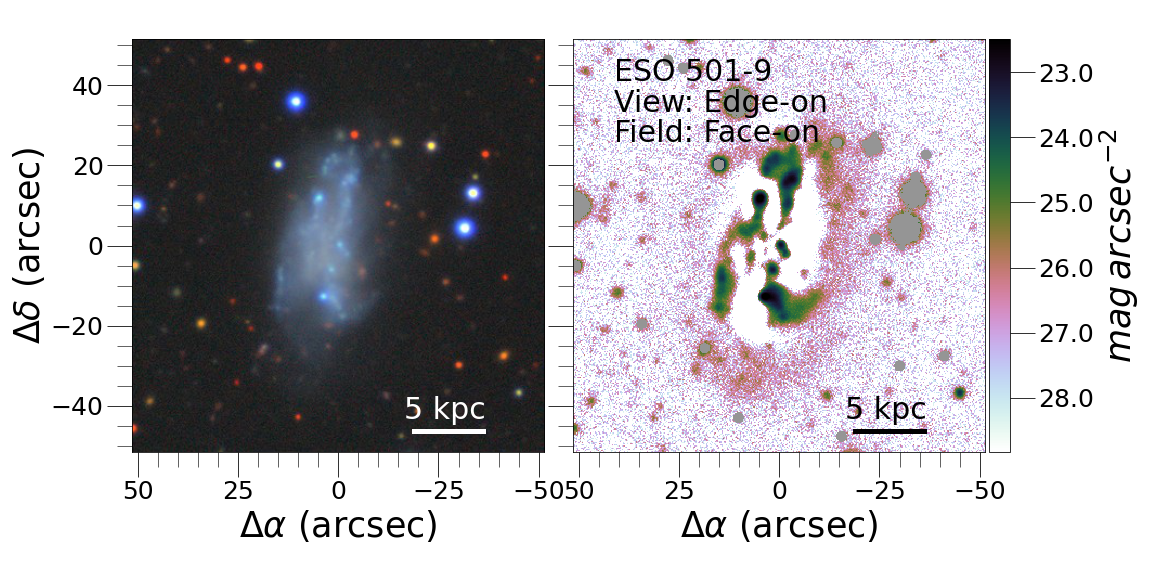}
\caption{Residual images of the RPS candidates. Each column contains the input image (left) and residual image
{\bf in $g$-band} (left) for each RPS candidate. The grey regions visible in the residual maps represent areas that have been intentionally masked during the analysis. These masks are applied to distinguish between various components in the data, such as segmentation maps for background galaxies and foreground stars. The North ($\delta$) and East ($\alpha$) directions are aligned in the conventional way.}
\label{fig_model_jelly_01}
\end{figure*}

\begin{figure*}
\contcaption{}
\includegraphics*[width=0.45\textwidth]{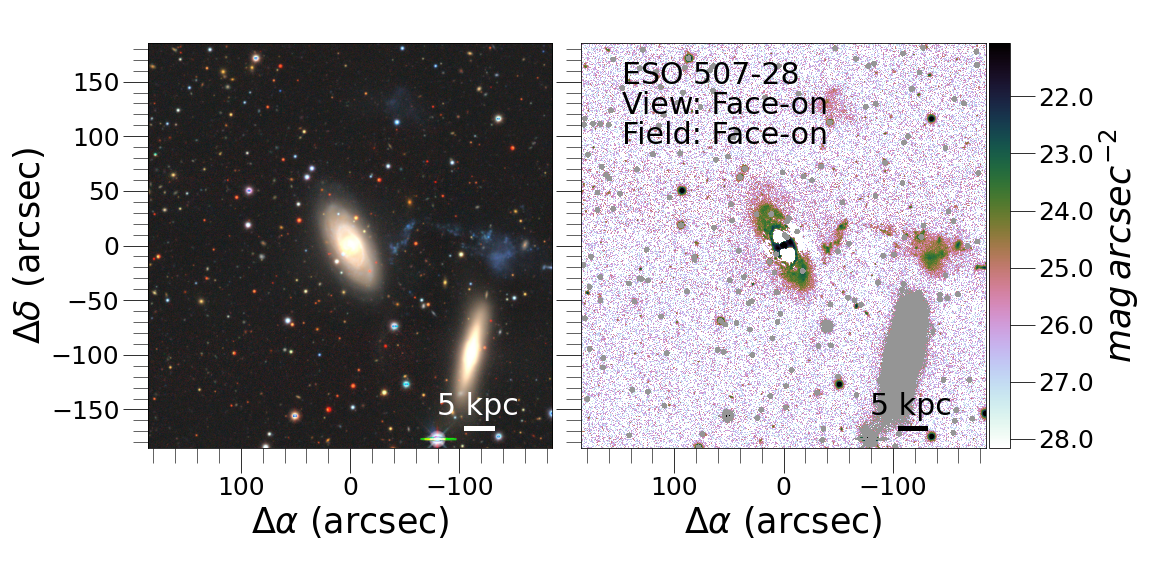}
\includegraphics*[width=0.45\textwidth]{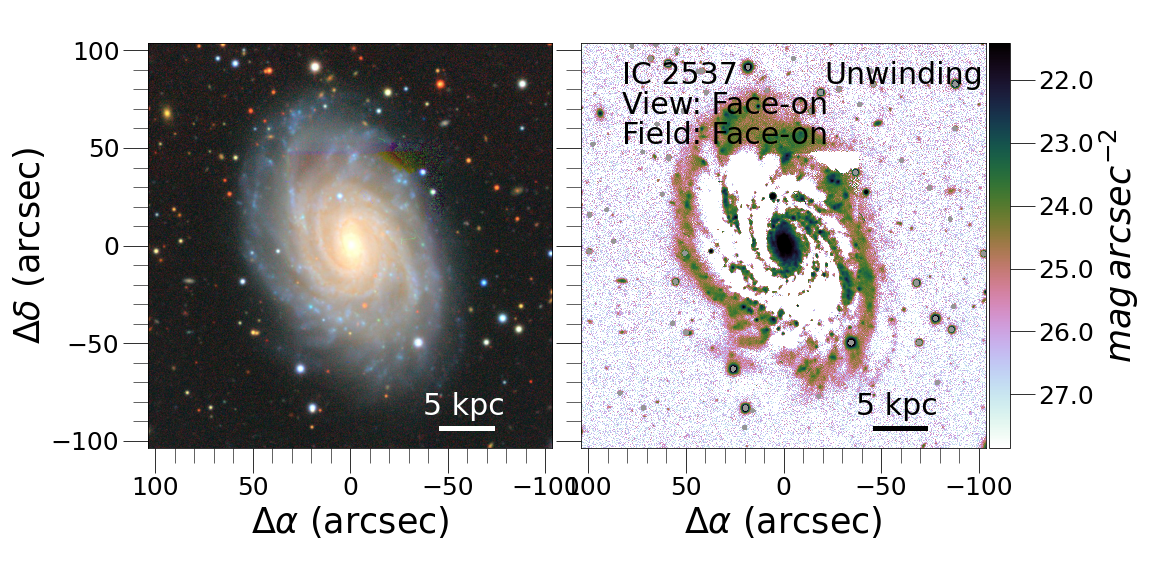}
\includegraphics*[width=0.45\textwidth]{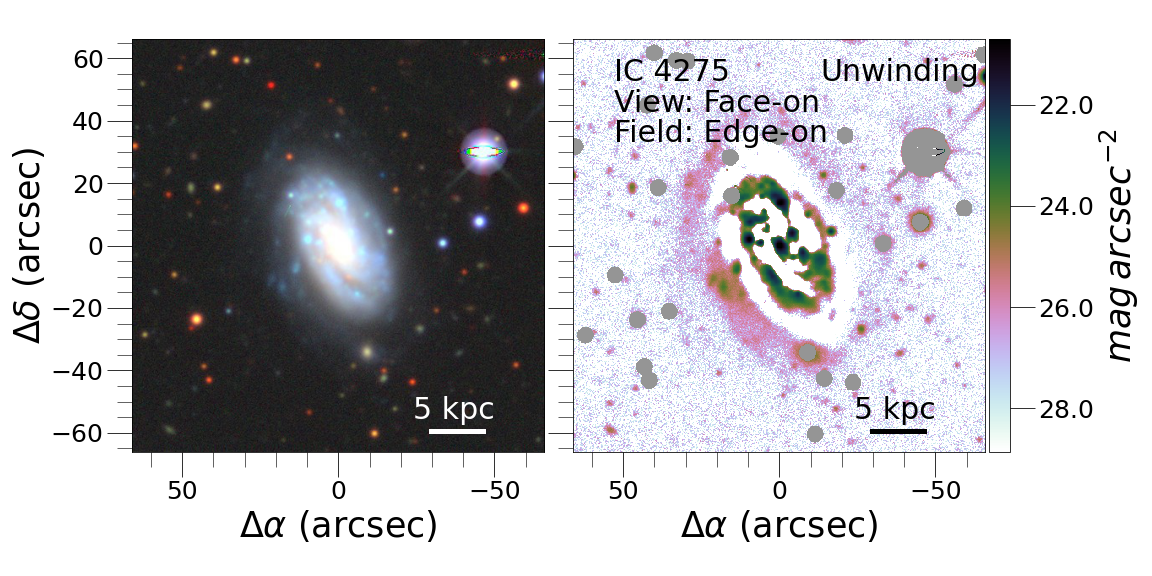}
\includegraphics*[width=0.45\textwidth]{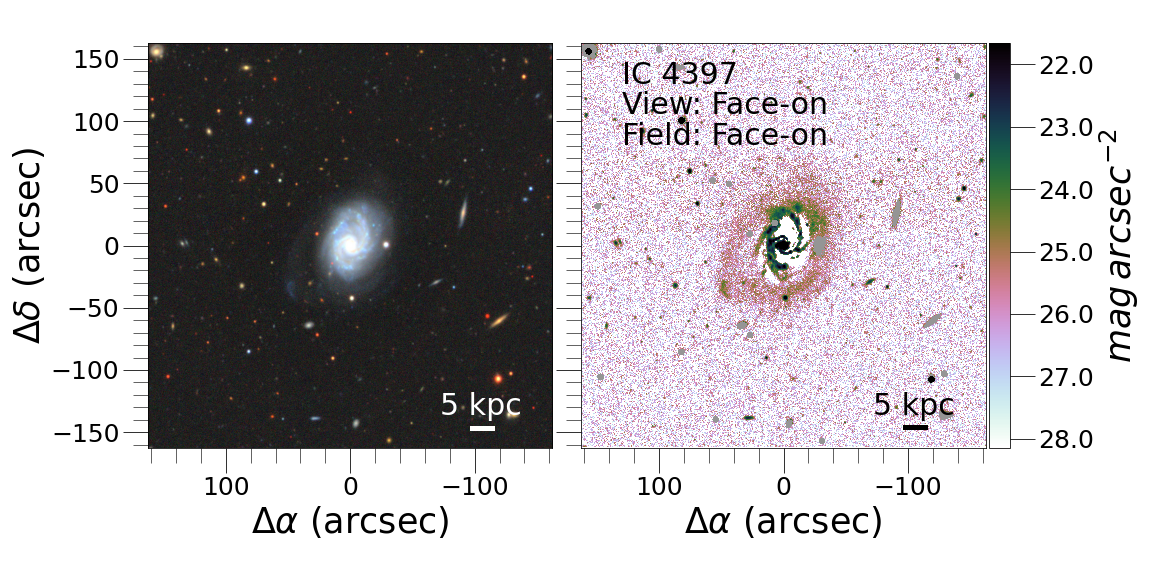}
\includegraphics*[width=0.45\textwidth]{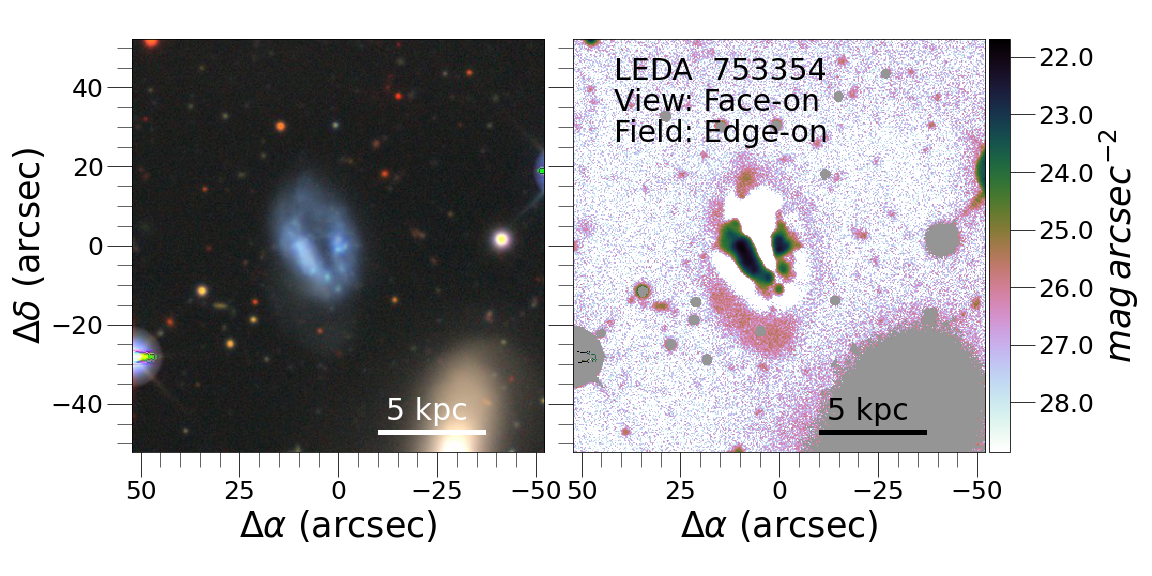}
\includegraphics*[width=0.45\textwidth]{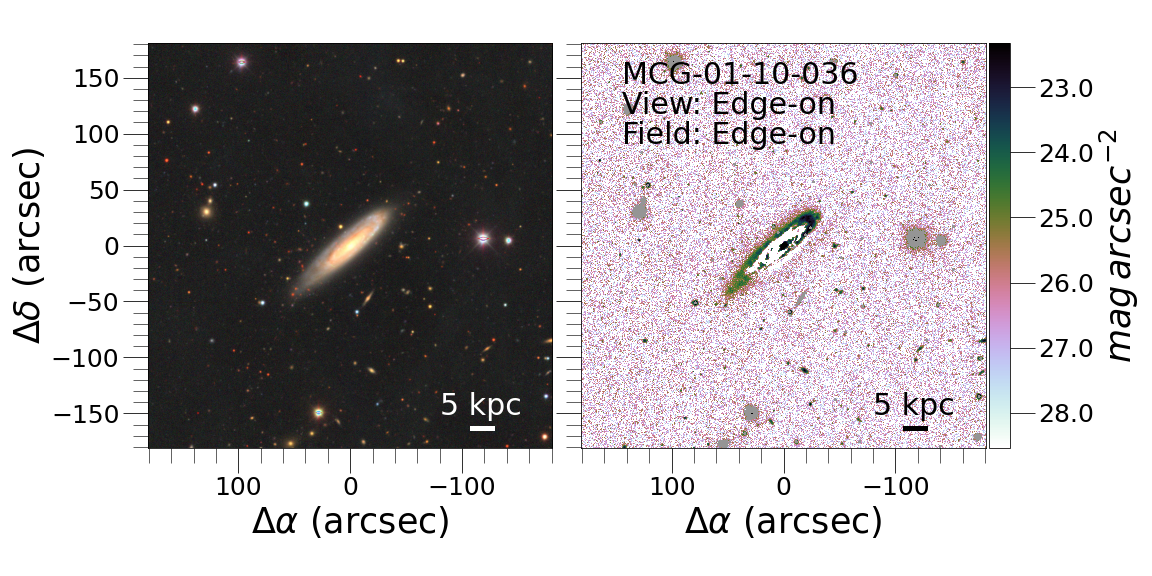}
\includegraphics*[width=0.45\textwidth]{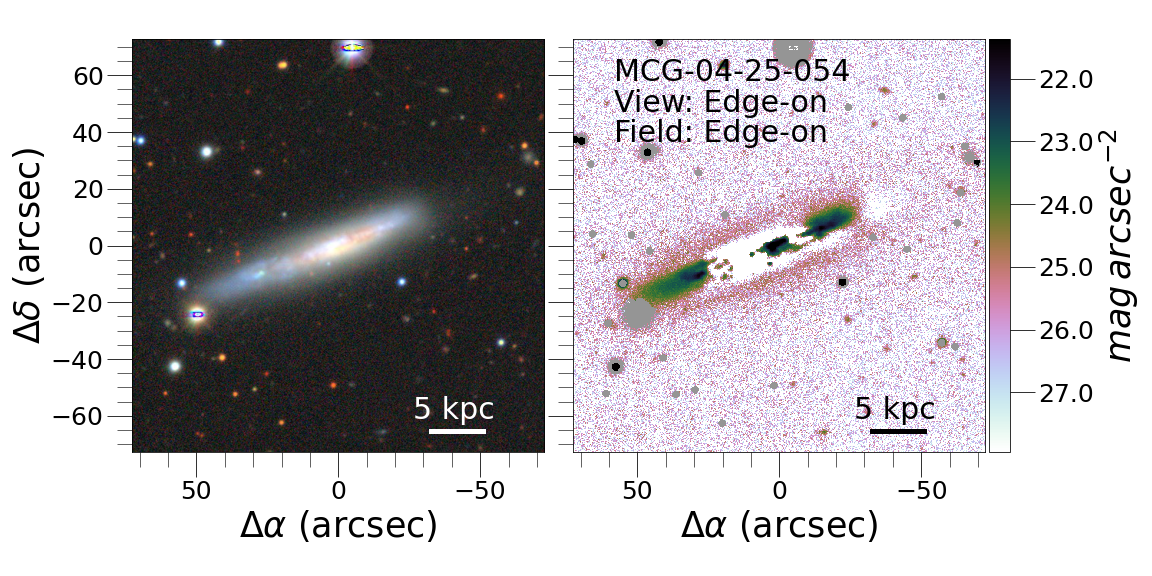}
\includegraphics*[width=0.45\textwidth]{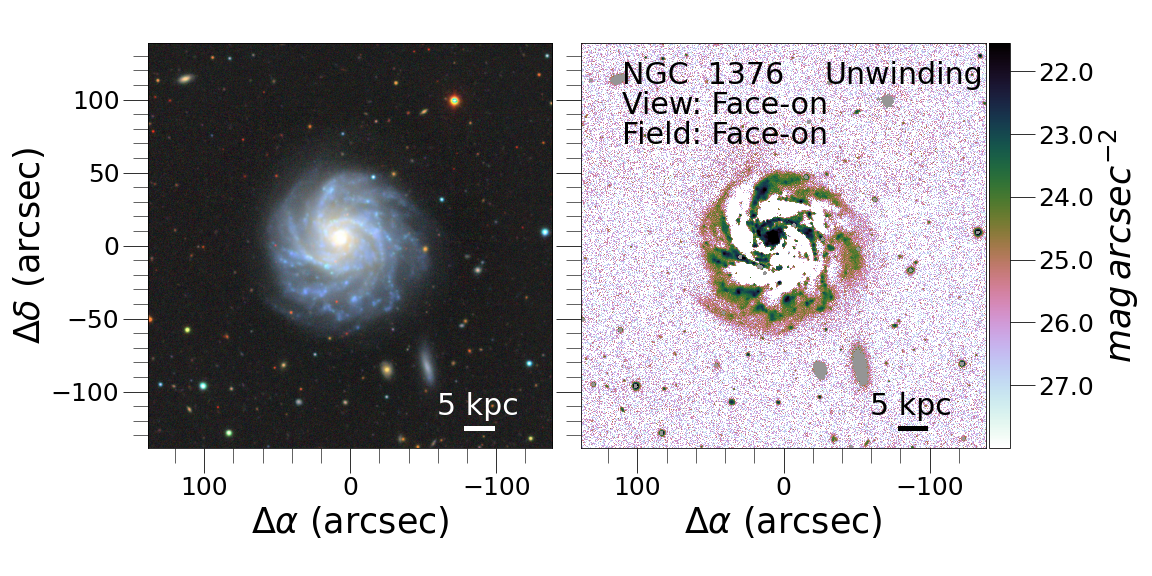}
\includegraphics*[width=0.45\textwidth]{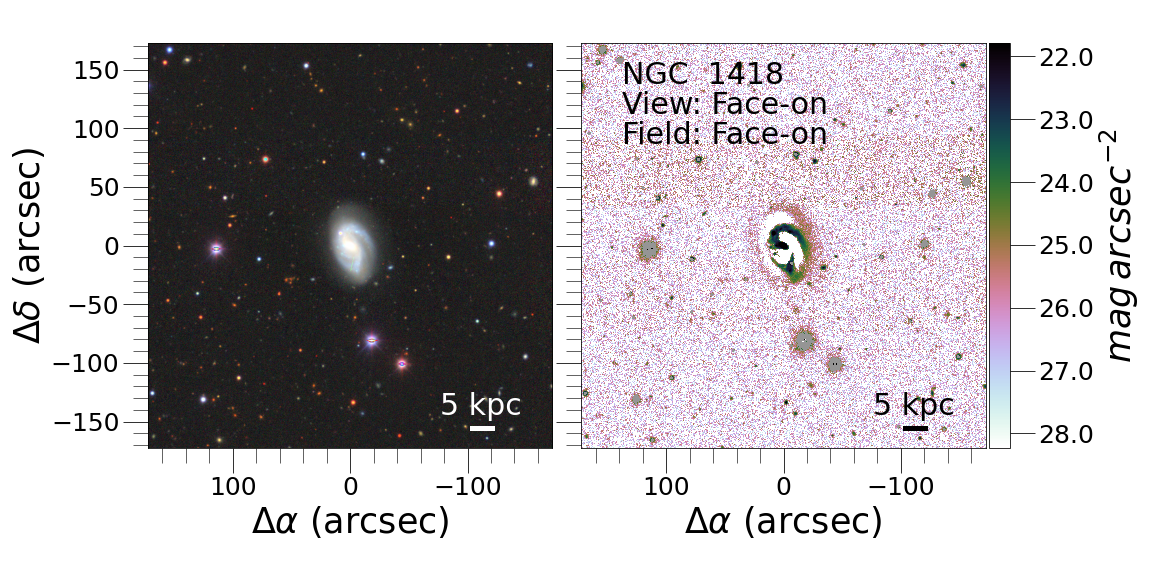}
\includegraphics*[width=0.45\textwidth]{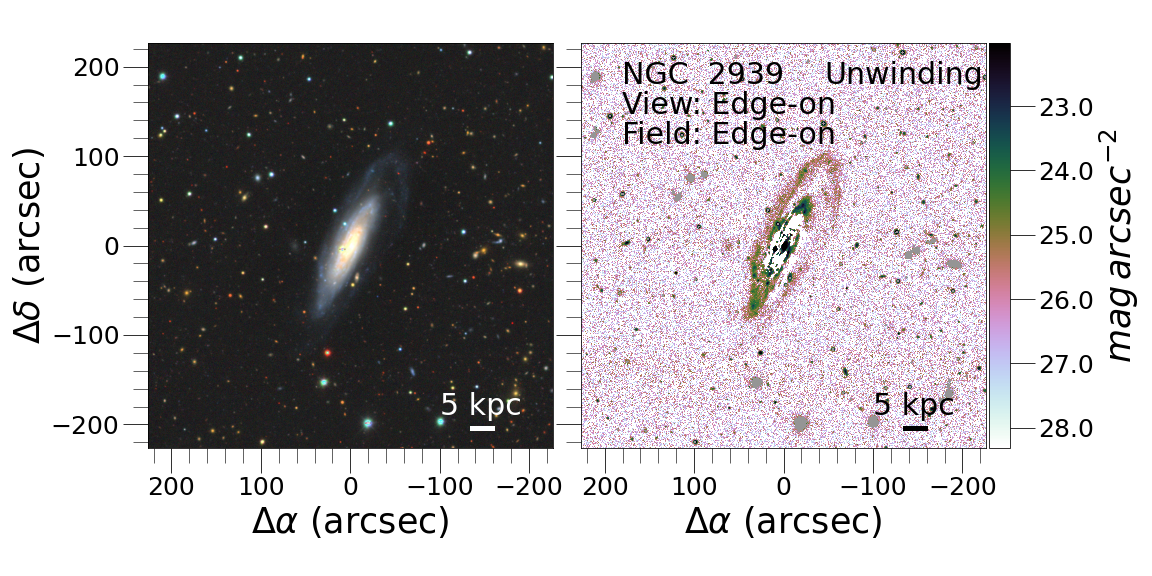}
\label{fig_model_jelly_02}
\end{figure*}

\begin{figure*}
\centering
\includegraphics*[width=0.45\textwidth]{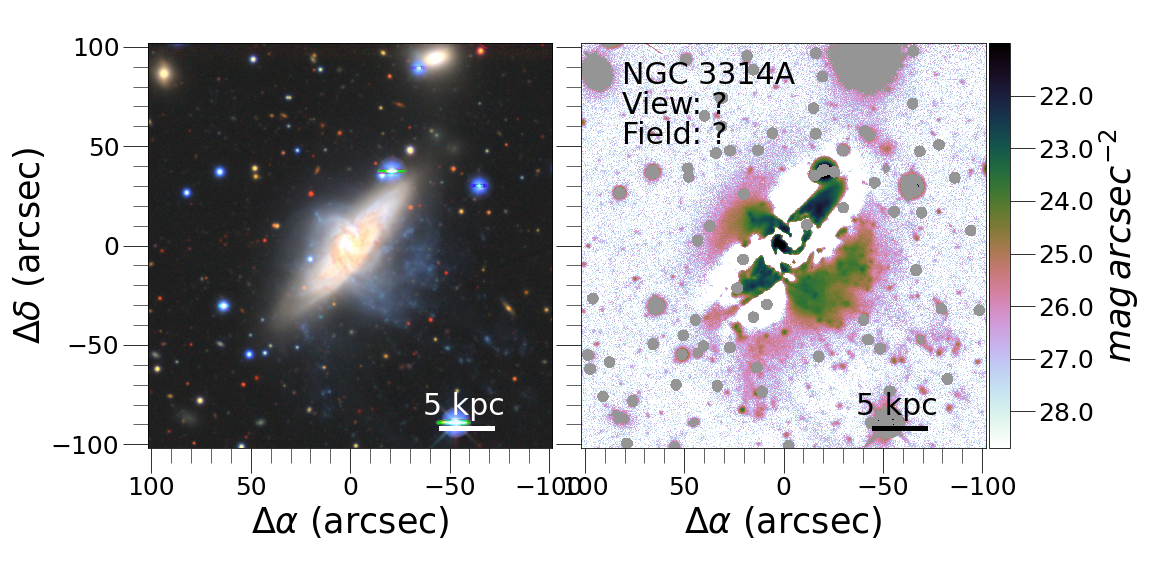}
\includegraphics*[width=0.45\textwidth]{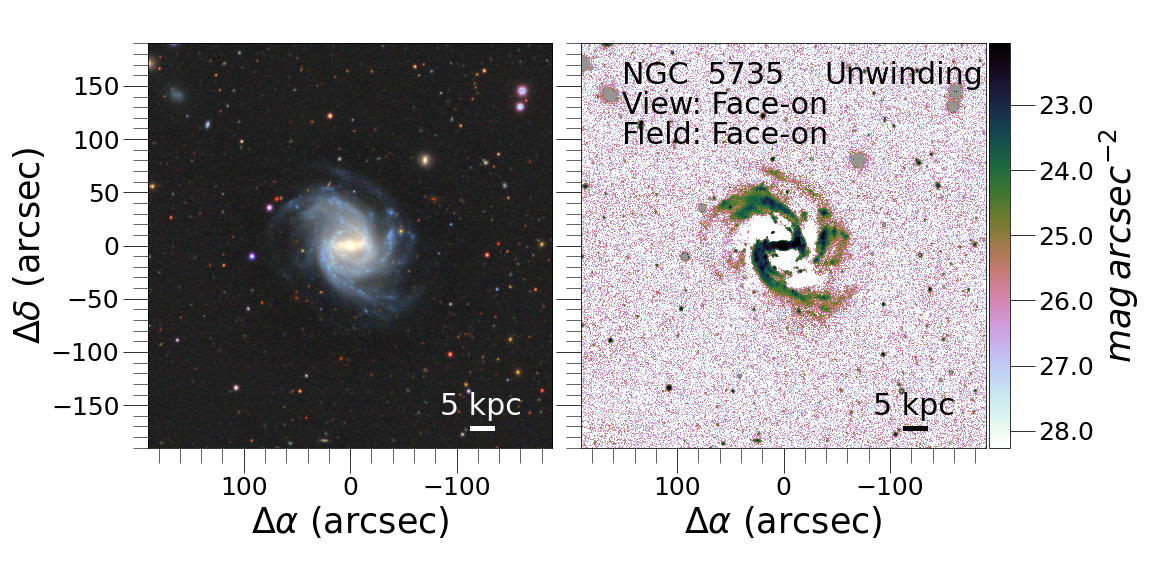}
\includegraphics*[width=0.45\textwidth]{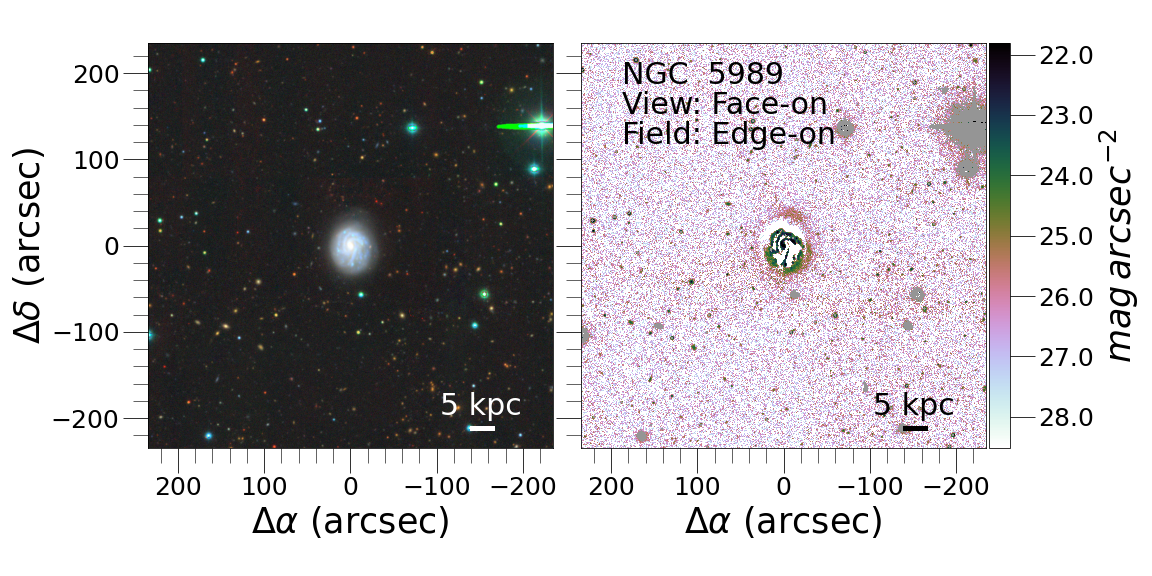}
\includegraphics*[width=0.45\textwidth]{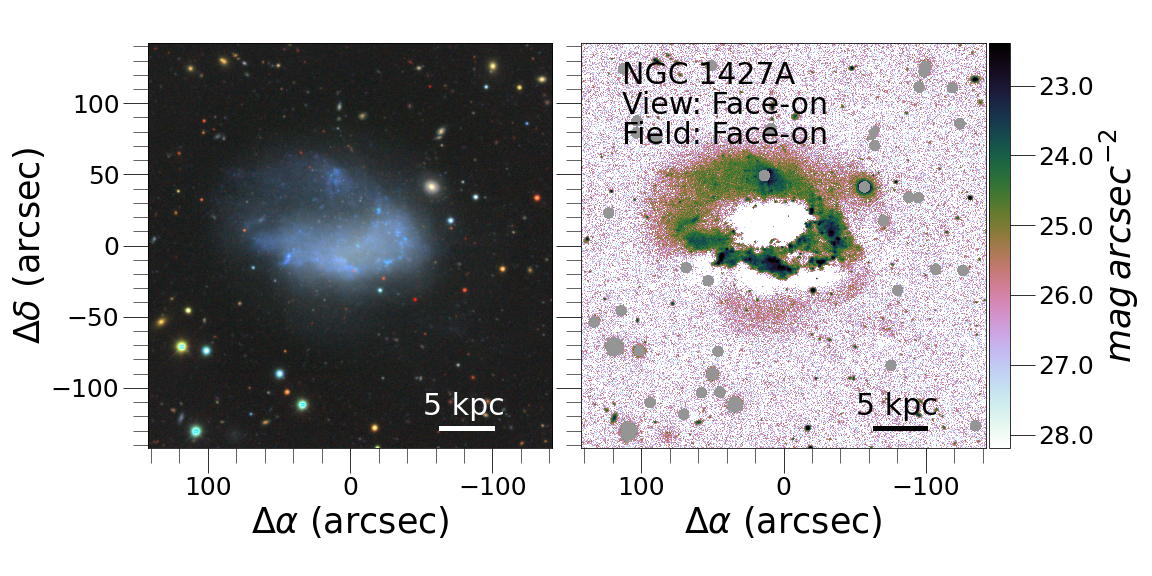}
\includegraphics*[width=0.45\textwidth]{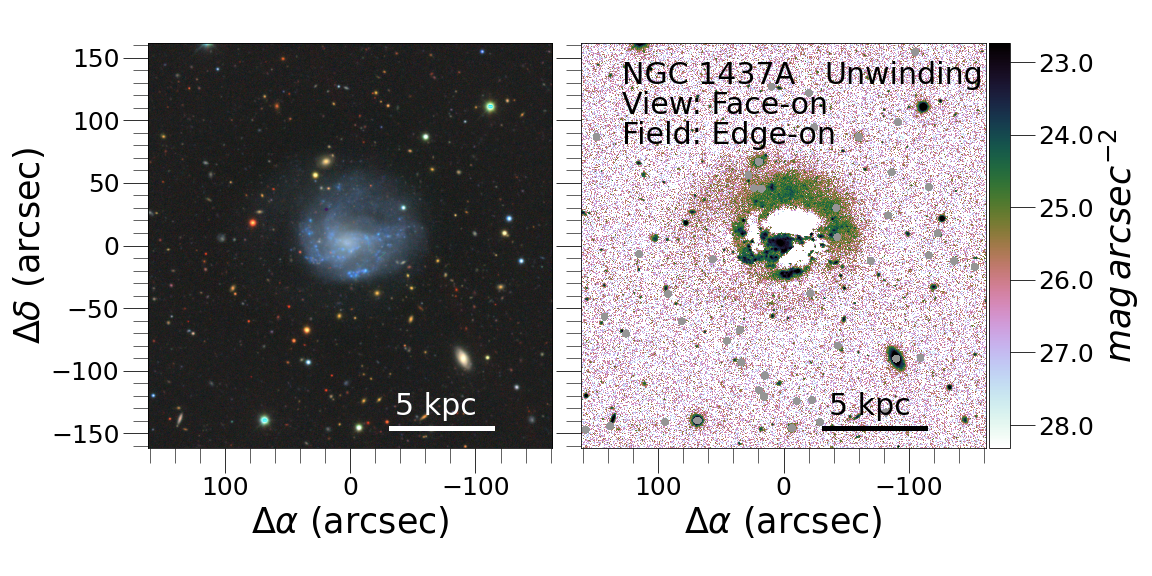}
\includegraphics*[width=0.45\textwidth]{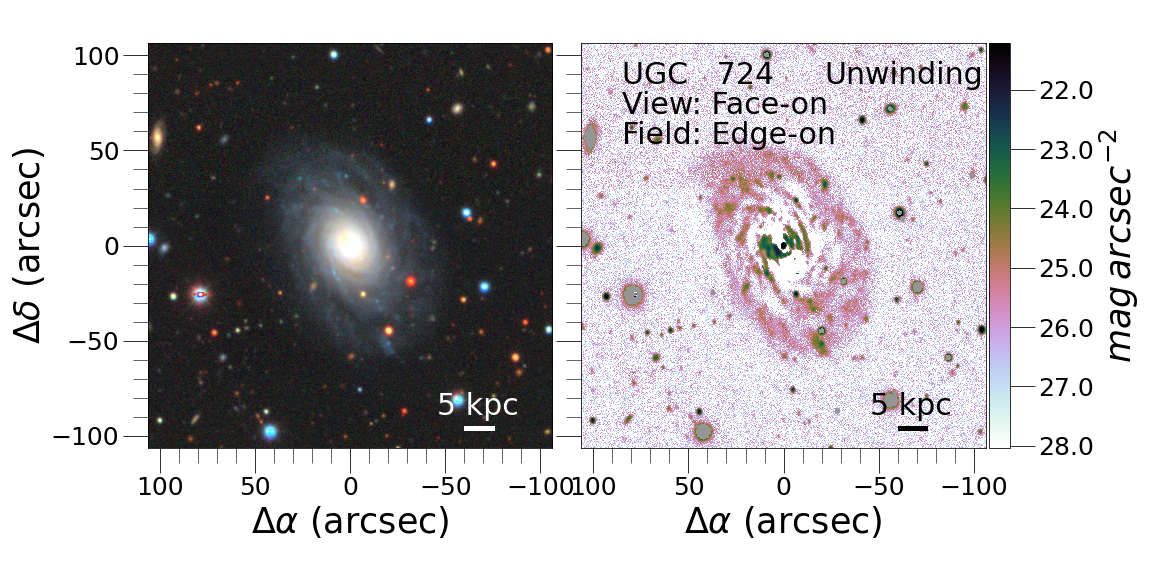}
\includegraphics*[width=0.45\textwidth]{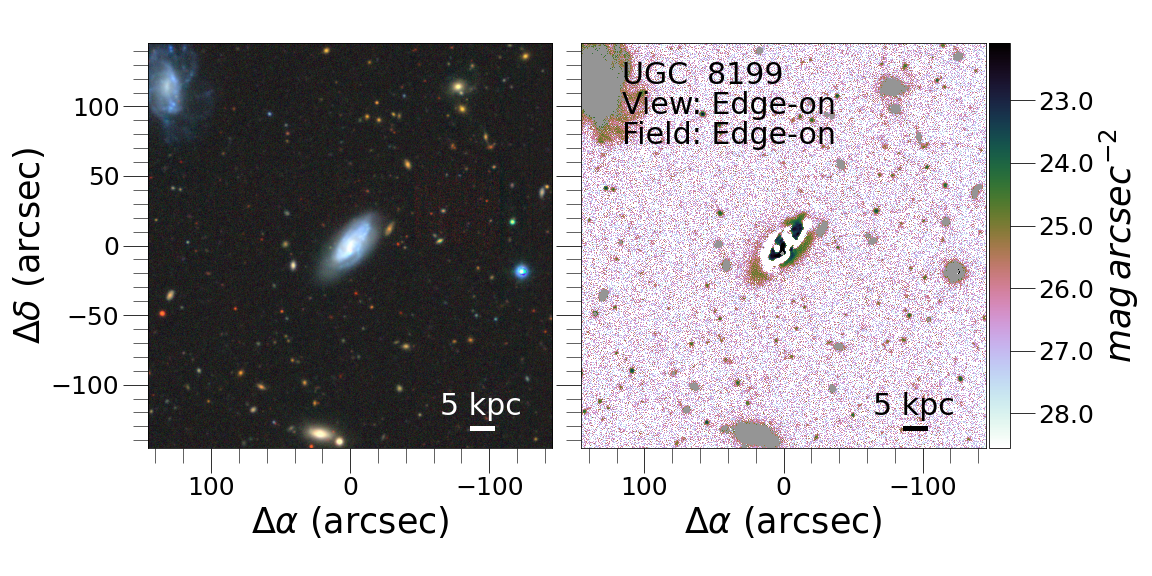}
\includegraphics*[width=0.45\textwidth]{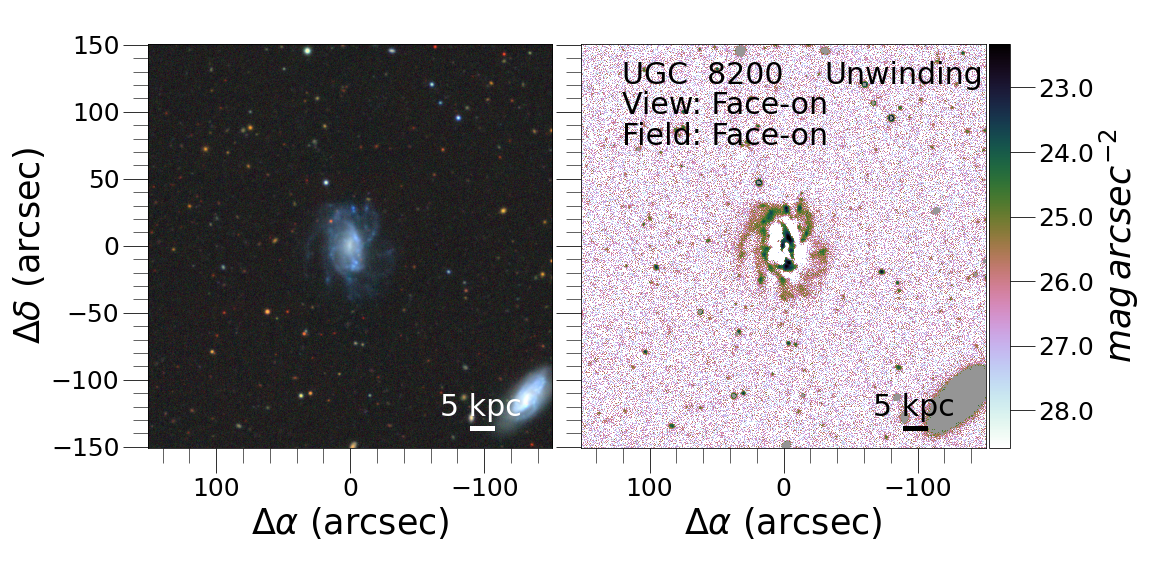}
\includegraphics*[width=0.45\textwidth]{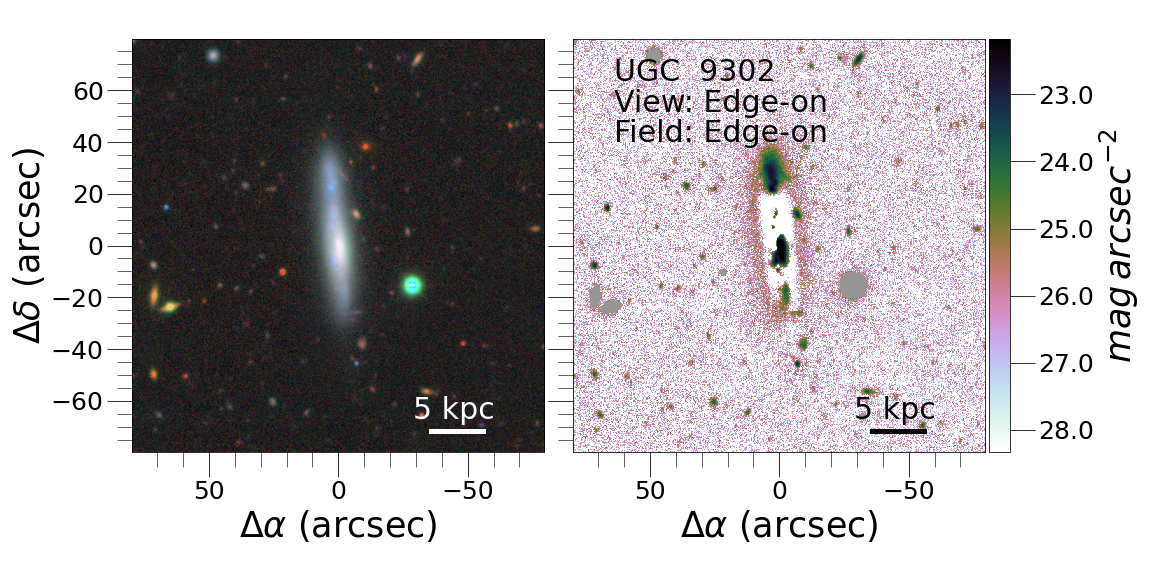}
\contcaption{}
\label{fig_model_jelly_03}
\end{figure*}

\begin{figure*}
\centering
\includegraphics*[width=0.8\textwidth]{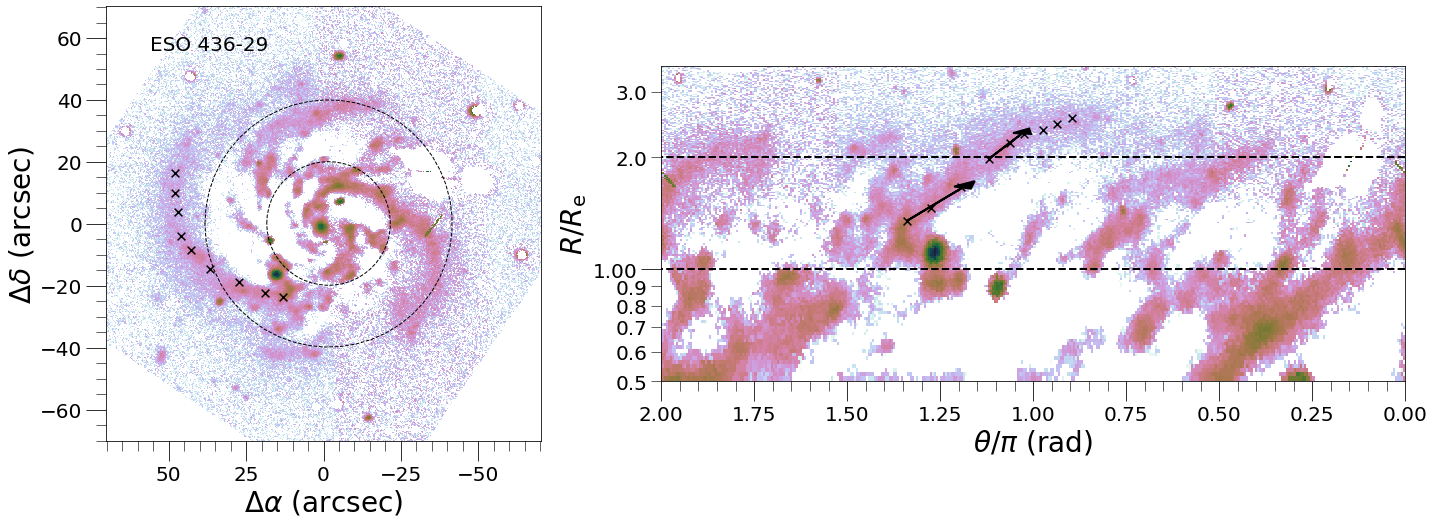}
\includegraphics*[width=0.8\textwidth]{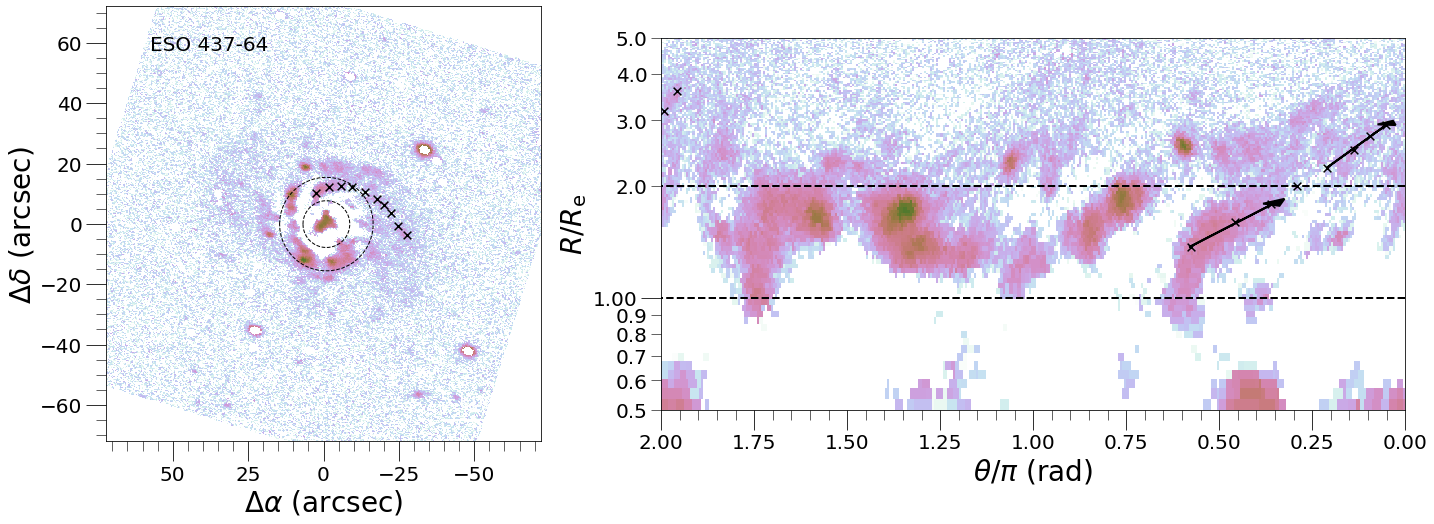}
\includegraphics*[width=0.8\textwidth]{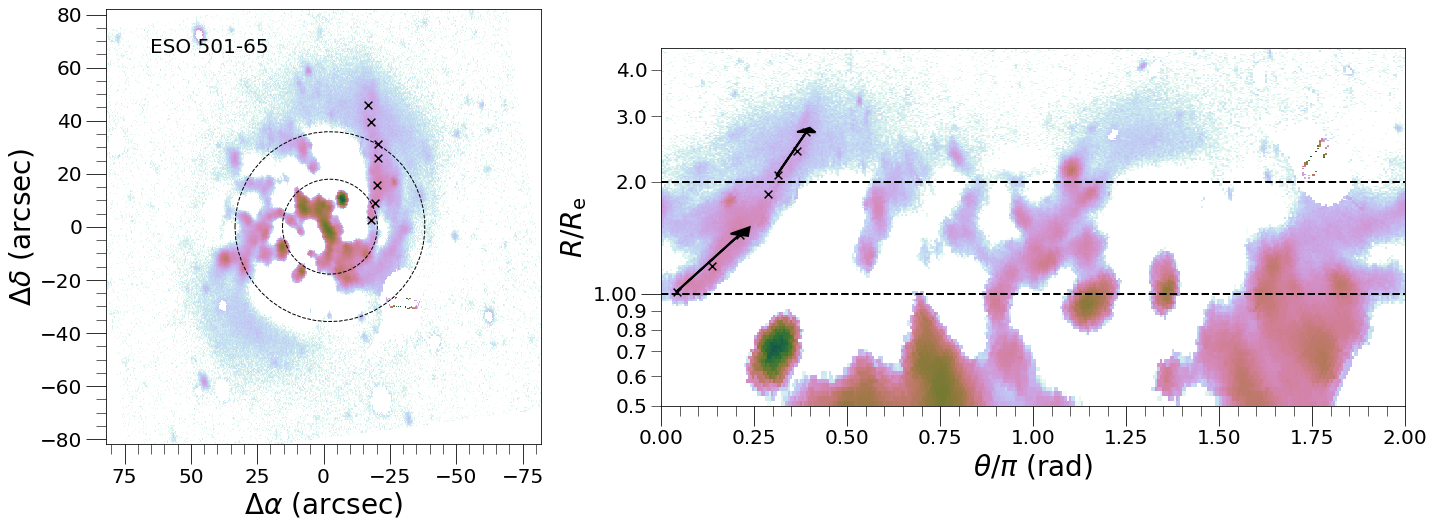}
\includegraphics*[width=0.8\textwidth]{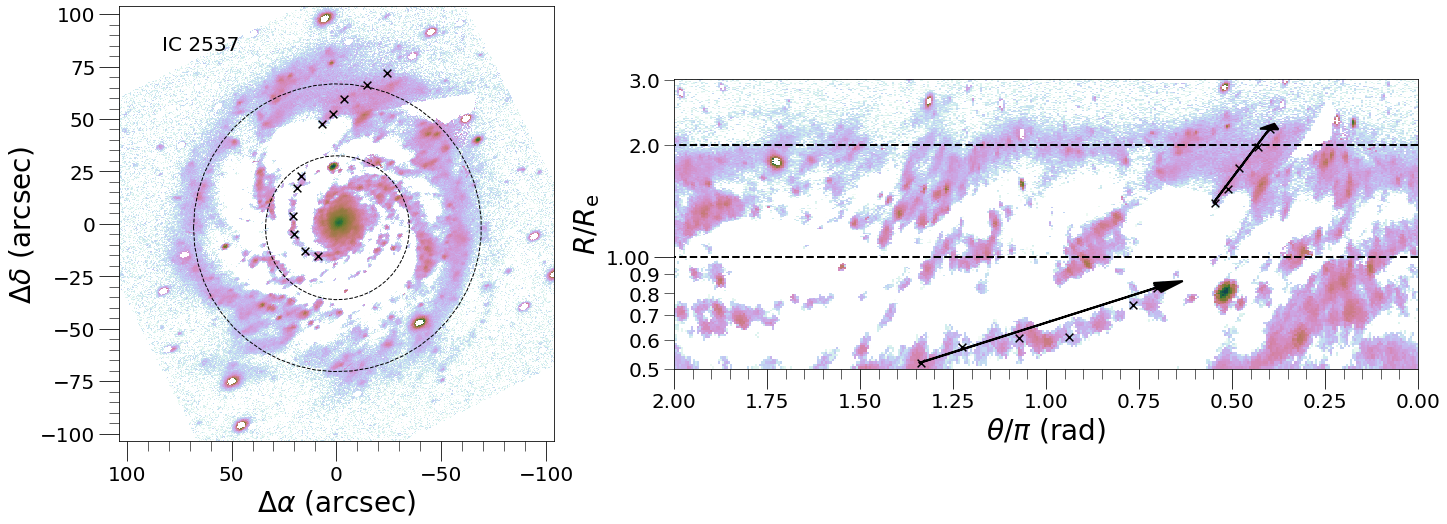}
\caption{The unwinding ram pressure galaxy candidates. Left panels contain the deprojected images {\bf in the $g$-band}, while right panels contain the radius {\it vs} $\theta$ plots. The ``X'' 's are the markers for the spiral arms for which were measured the aperture angles, while the black arrows indicate
the inner and outer ranges for the arms. The North ($\delta$) and East ($\alpha$) directions are aligned in the conventional way.}
\label{fig_unwinding_01}
\end{figure*}

\begin{figure*}
\centering

\includegraphics*[width=0.8\textwidth]{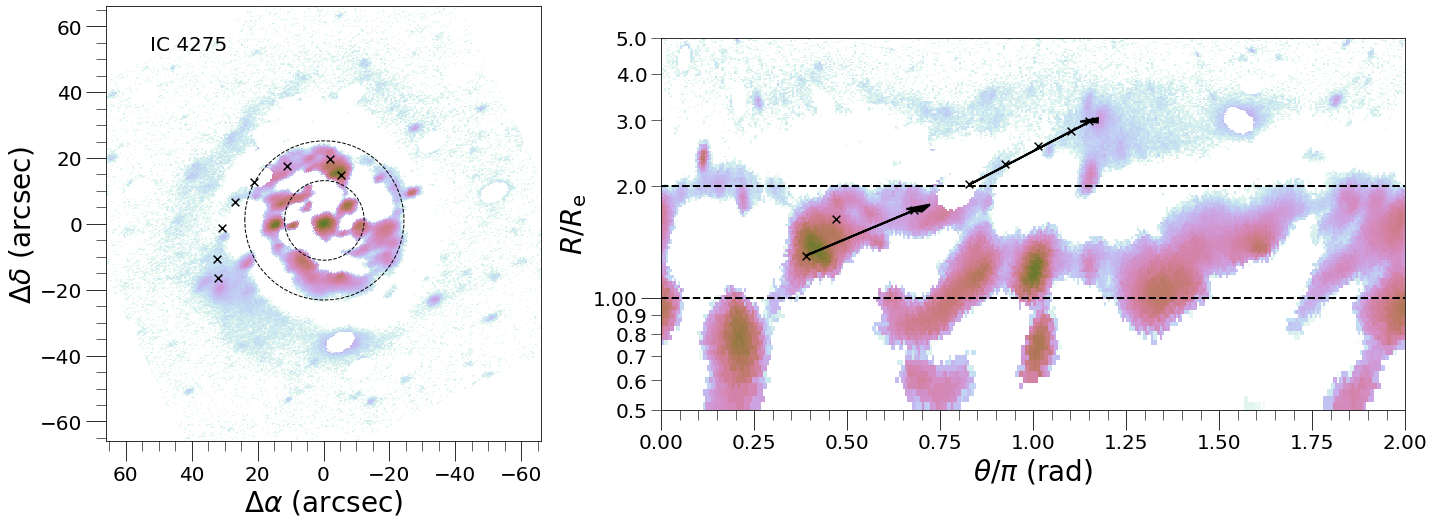}
\includegraphics*[width=0.8\textwidth]{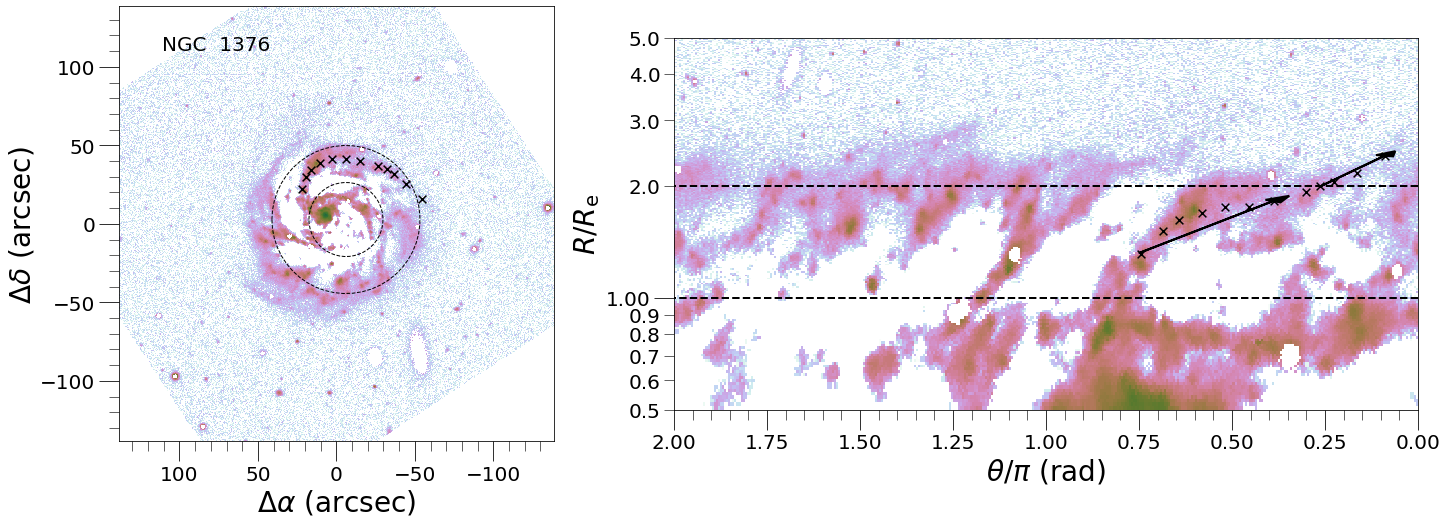}
\includegraphics*[width=0.8\textwidth]{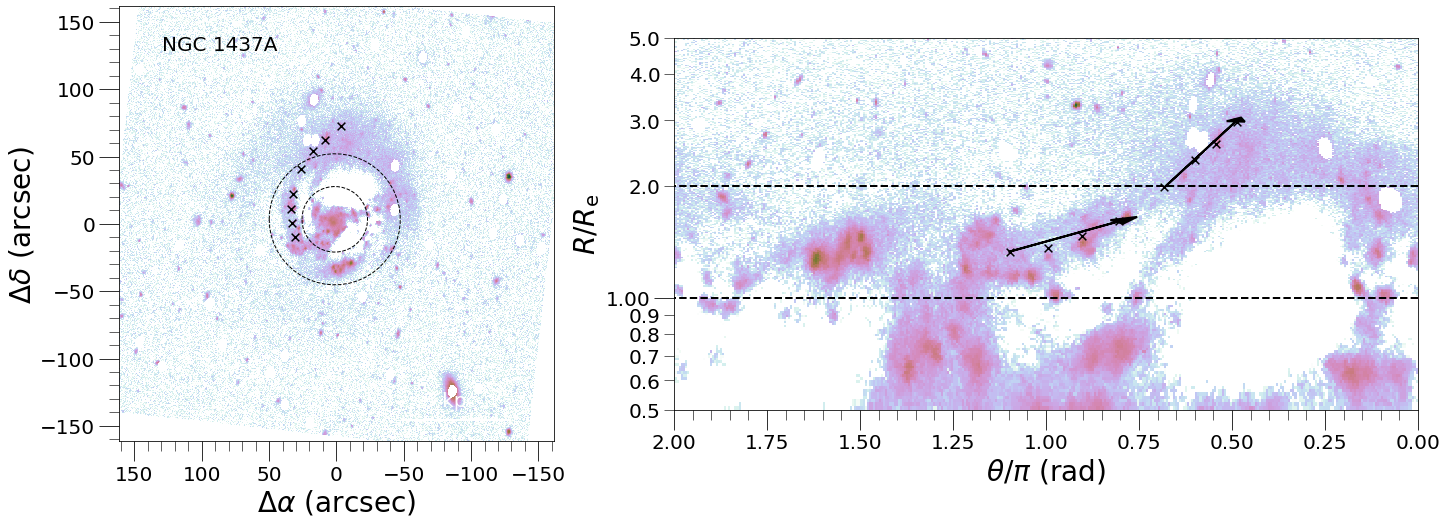}
\includegraphics*[width=0.8\textwidth]{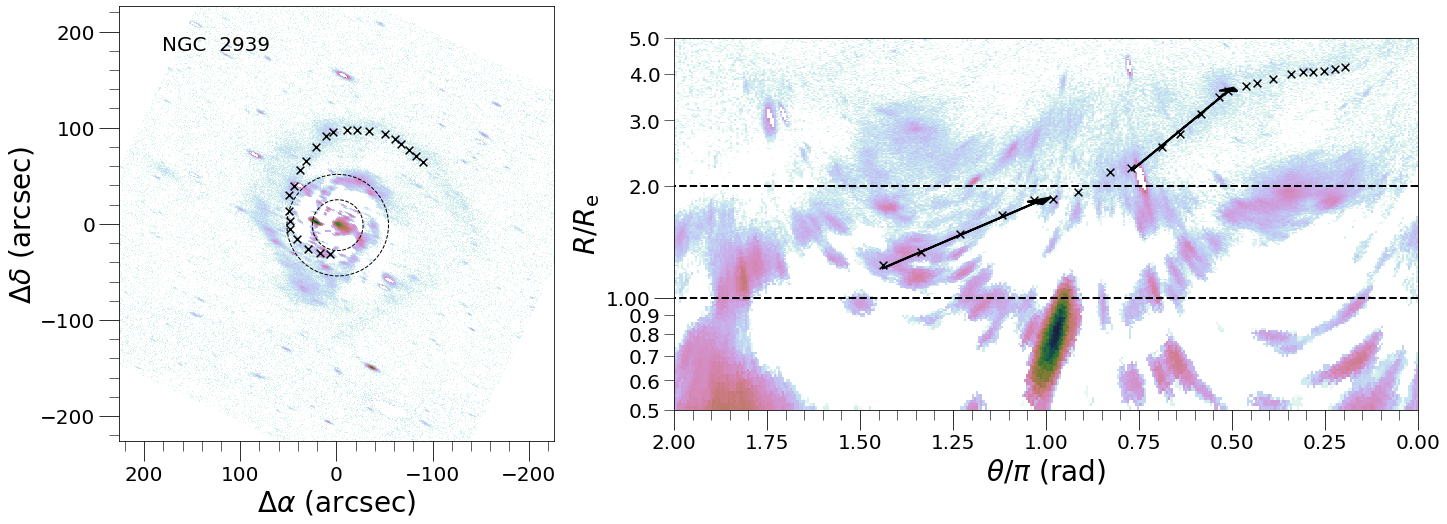}
\contcaption{}
\label{fig_unwinding_02}
\end{figure*}

\begin{figure*}
\centering
\includegraphics*[width=0.8\textwidth]{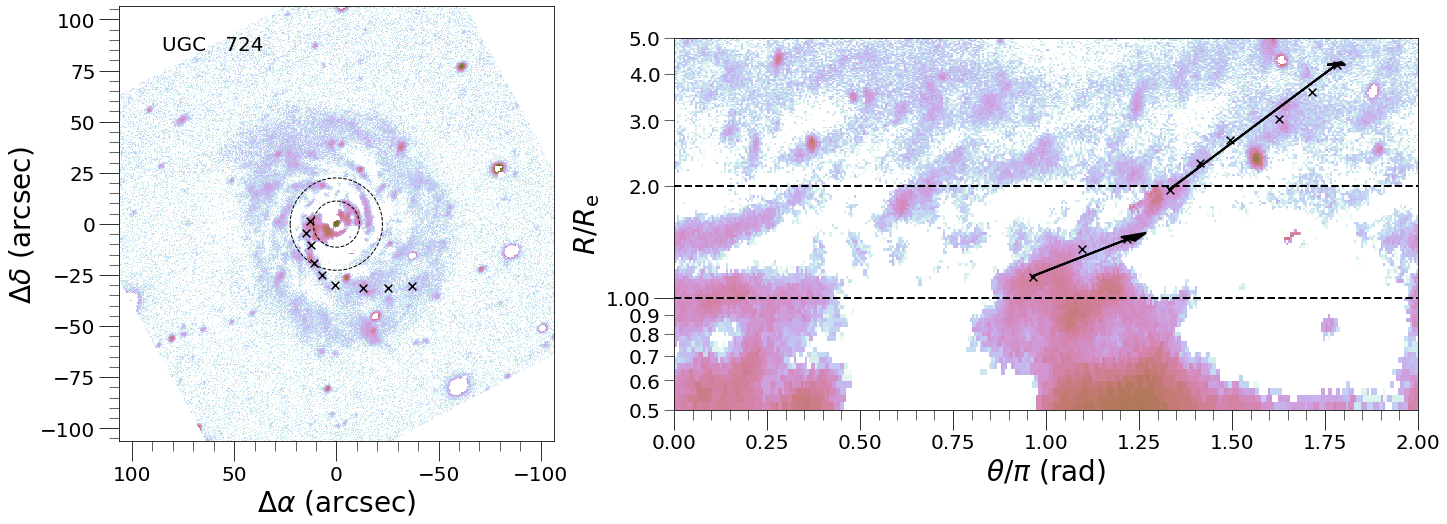}
\includegraphics*[width=0.8\textwidth]{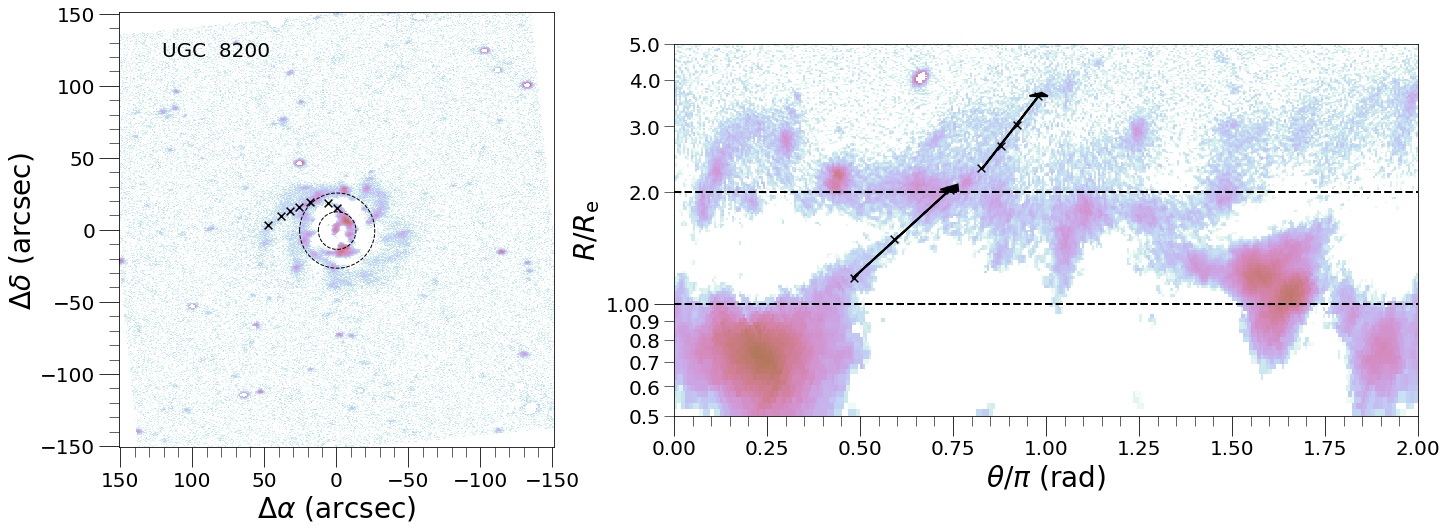}
\contcaption{}
\label{fig_unwinding_03}
\end{figure*}





\bsp	
\label{lastpage}
\end{document}